\documentclass[aps, pra, 10pt, superscriptaddress, twocolumn, floatfix]{revtex4-1}

\usepackage[nointlimits]{amsmath}
\usepackage{amsthm}
\usepackage{graphicx,nicefrac}
\usepackage{color}
\usepackage{subcaption}
\usepackage{txfonts}
\usepackage{mathtools}
\usepackage{hyperref}
\usepackage{tabularx}
\hypersetup{colorlinks=true, citecolor=blue, linkcolor=blue}
 
\usepackage{physics}
\usepackage{multirow}
\usepackage{booktabs}

\usepackage[table]{xcolor}
\usepackage{amssymb}
\usepackage{array}
\usepackage{supertabular}

\usepackage{amsfonts,amssymb}
\usepackage{array,bm,color}
\usepackage{epsfig,nomencl,revsymb4-1,upgreek,url}
\usepackage{algpseudocode}
\usepackage{xspace}
\usepackage{tabularx}
\usepackage[normalem]{ulem}
\usepackage{changes}
\usepackage{enumitem}
\usepackage{mathtools}
\usepackage{placeins}
\usepackage{quantikz}

\makeatletter

\DeclareFontFamily{OMX}{MnSymbolE}{}
\DeclareSymbolFont{MnLargeSymbols}{OMX}{MnSymbolE}{m}{n}
\SetSymbolFont{MnLargeSymbols}{bold}{OMX}{MnSymbolE}{b}{n}
\DeclareFontShape{OMX}{MnSymbolE}{m}{n}{
    <-6>  MnSymbolE5
   <6-7>  MnSymbolE6
   <7-8>  MnSymbolE7
   <8-9>  MnSymbolE8
   <9-10> MnSymbolE9
  <10-12> MnSymbolE10
  <12->   MnSymbolE12
}{}
\DeclareFontShape{OMX}{MnSymbolE}{b}{n}{
    <-6>  MnSymbolE-Bold5
   <6-7>  MnSymbolE-Bold6
   <7-8>  MnSymbolE-Bold7
   <8-9>  MnSymbolE-Bold8
   <9-10> MnSymbolE-Bold9
  <10-12> MnSymbolE-Bold10
  <12->   MnSymbolE-Bold12
}{}

\let\llangle\@undefined
\let\rrangle\@undefined
\DeclareMathDelimiter{\llangle}{\mathopen}%
                     {MnLargeSymbols}{'164}{MnLargeSymbols}{'164}
\DeclareMathDelimiter{\rrangle}{\mathclose}%
                     {MnLargeSymbols}{'171}{MnLargeSymbols}{'171}
\makeatother

\usepackage{tikz}
\usetikzlibrary{arrows}

\def\ket#1{{\left| #1 \right\rangle}}

\newcommand{\Hg}[2]{h_{#1}(#2)}
\newcommand{\Sg}[2]{s_{#1}(#2)}
\newcommand{\Cg}[3]{c_{#1#2}(#3)}
\newcommand{\Ag}[3]{a_{#1#2}(#3)}

\newcommand{\TL}{T_L}
\newcommand{\SL}{S_L}
\newcommand{\IL}{I_L}
\newcommand{\HL}{H_L}
\newcommand{\rz}{\rho_0}
\newcommand{\Md}{M}

\newcommand{\rt}{\sqrt{2}}
\newcommand{\onept}{(1+\sqrt{2})}
\newcommand{\pfac}[1]{\frac{1}{\sqrt{#1}}}

\newcommand{\expect}[1]{\langle#1\rangle}
\newcommand{\llrdiff}{\delta}
\newcommand{\Ntot}{N_\mathrm{tot.}}

\tikzset{
    vertex/.style={circle,draw,minimum size=1.5em},
    edge/.style={->,> = latex'}
}

\renewcommand{\arraystretch}{1.2} 
\makeatletter
\DeclareCaptionJustification{fulljustified}{\leftskip\z@skip\rightskip\z@skip\parfillskip\@flushglue}
\makeatother
\begin{document}

\title{Logical Gate Set Tomography}
\author{Piper C. Wysocki}
\thanks{pcwysoc@sandia.gov}
\affiliation{Quantum Performance Laboratory, Sandia National Laboratories, Livermore, CA 94550, USA}
\affiliation{Department of Physics and Astronomy, University of New Mexico, Albuquerque, NM 87106, USA}

\author{Juan Gonzalez De Mendoza}
\affiliation{Quantum Performance Laboratory, Sandia National Laboratories, Albuquerque, NM 87185, USA}
\affiliation{Department of Physics and Astronomy, University of New Mexico, Albuquerque, NM 87106, USA}

\author{Karl Mayer}
\affiliation{Quantinuum, Broomfield, Colorado 80021, USA}

\author{Natalie C. Brown}
\affiliation{Quantinuum, Broomfield, Colorado 80021, USA}

\author{Ciar\'an Ryan-Anderson}
\affiliation{Quantinuum, Broomfield, Colorado 80021, USA}

\author{Jalan A. Ziyad}
\affiliation{Quantum Performance Laboratory, Sandia National Laboratories, Albuquerque, NM 87185, USA}
\affiliation{Department of Physics and Astronomy, University of New Mexico, Albuquerque, NM 87106, USA}

\author{Jordan H. Hines}
\affiliation{Quantum Performance Laboratory, Sandia National Laboratories, Albuquerque, NM 87185, USA}

\author{Daniel G. Hothem}
\affiliation{Quantum Performance Laboratory, Sandia National Laboratories, Livermore, CA 94550, USA}

\author{Corey I. Ostrove}
\affiliation{Quantum Performance Laboratory, Sandia National Laboratories, Albuquerque, NM 87185, USA}

\author{Timothy J. Proctor}
\affiliation{Quantum Performance Laboratory, Sandia National Laboratories, Livermore, CA 94550, USA}

\author{Kevin C. Young}
\thanks{Current address: NVIDIA Corporation, 2788 San Tomas Expressway, Santa Clara, CA 95051, USA.}
\affiliation{Quantum Performance Laboratory, Sandia National Laboratories, Livermore, CA 94550, USA}

\author{Robin J. Blume-Kohout}
\affiliation{Quantum Performance Laboratory, Sandia National Laboratories, Albuquerque, NM 87185, USA}

\author{Charles H. Baldwin}
\affiliation{Quantinuum, Broomfield, Colorado 80021, USA}

\author{Kenneth M. Rudinger}
\affiliation{Quantum Performance Laboratory, Sandia National Laboratories, Albuquerque, NM 87185, USA}

\begin{abstract} 
Quantum error correction (QEC) can strongly suppress noise, but predicting whether a computation succeeds requires understanding not only how much effective noise persists after QEC, but what form that noise takes. To thoroughly characterize this noise, we introduce logical gate set tomography (``logical GST''). Logical GST produces detailed representations of noisy logical operations that can be used to derive higher-level metrics, such as gate fidelity and diamond distance, and to test for non-Markovian and coherent logical errors. We use logical GST to obtain the first full, self-consistent characterization of a universal single-qubit logical gate set, implemented in the seven-qubit Steane code using Quantinuum System Model H1, a 20-qubit trapped-ion quantum processor. In the experimental data from H1, logical GST identified predominantly logical Pauli-stochastic errors, found no evidence of logical non-Markovianity, and detected weak potential signatures of coherent logical error. By contrast, higher-shot datasets generated with Quantinuum's detailed system emulator, identify mechanisms for weak but statistically significant logical non-Markovianity and strong evidence of coherent logical error. Finally, we use logical GST to perform error attribution with the system emulator, revealing how specific physical error sources shape logical error mechanisms.
\end{abstract}

\maketitle

\section{Introduction}

The frontier of experimental quantum error correction (QEC) has transitioned from demonstrations of logical memory to the implementation of fault-tolerant logical operations, and the execution of simple algorithmic circuits on error-corrected logical qubits \cite{besedin2026lattice, rosenfeld2025magic, mayer2024benchmarking, bluvstein2026a, proctor2026benchmarking}.  QEC reduces, but does not eliminate error. So logical operations still fail, and---like physical operations---they can fail in complex nontrivial ways.  Understanding and predicting the behavior of logical operations and logical circuits requires more than just measuring logical error rates. It requires experimentally probing the structure of logical gate errors and testing the assumptions used to model them. Here, to meet this challenge, we introduce logical gate set tomography (``logical GST'') and use it to characterize logical operations using Quantinuum System Model H1, a 20-qubit trapped-ion quantum processor \cite{Pino2020, ryananderson2021realization}. 

An $\llbracket n,k,d \rrbracket$ code protects---via syndrome extraction, decoding, and frame updates or physical corrections---a $k$-qubit logical subsystem encoded in a system of $n$ physical qubits. Previous work has shown that in many circumstances, the dynamical evolution of the protected logical qubit(s) during QEC can be accurately modeled by an effective logical CPTP (completely positive trace-preserving) map \cite{iverson2020coherence,bravyi2018correcting,beale2018quantum}, i.e., a process matrix or superoperator.  This ansatz remains valid if a nontrivial logical operation is integrated into a cycle of QEC.  

It is also reasonable to model logical operations as \textit{Markovian}. A Markovian logical operation has the same effective action each time it is applied and can be represented by a fixed quantum channel regardless of where it appears within a logical circuit. This model is not always accurate, though---logical operations can be non-Markovian \cite{ceasura2022nonexponential,endo2025nonmarkovianity,ziyad2025emergent}. Certain conditions ensure that logical dynamics will be well-approximated by a Markovian model \cite{kwiatkowski2025constructing}, but those conditions do not always hold. Logical non-Markovianity can be inherited from physical-level effects such as calibration drift or leakage out of the computational subspace, and emergent non-Markovianity can be produced by QEC even when the underlying physical operations are strictly Markovian, if information from earlier rounds persists in nonlogical degrees of freedom and impacts logical dynamics in subsequent rounds \cite{endo2025nonmarkovianity,ziyad2025emergent,kwiatkowski2025constructing}. 

An even stronger assumption is often made: that errors in logical operations are Pauli-stochastic---i.e., the $k$-qubit CPTP map describing errors in the logical operation can be represented as a probabilistic mixture of Pauli $X$, $Y$, and $Z$ flips (or tensor products of these). This assumption is motivated by theoretical results showing that QEC can convert coherent physical errors into predominantly stochastic logical noise at sufficiently large code distances \cite{Greenbaum2017modeling, bravyi2018correcting, iverson2020coherence, witzel2026correcting, chang2025taming}. Early generations of fault-tolerant quantum processors, however, will use small-to-moderate code distances and hardware-tailored QEC implementations.  Under these conditions, logical operations may display non-Pauli-stochastic (e.g., coherent) logical errors \cite{bravyi2018correcting, witzel2026correcting, cheng2025emergent}.  Moreover, the theoretical results justifying Pauli-stochastic error models mostly apply to logical memories. Nontrivial logical operations and fault-tolerant circuits including both transversal and non-transversal logical operations may display unexpected effects.

\begin{figure*}[t!]
    \includegraphics[width=1\linewidth]{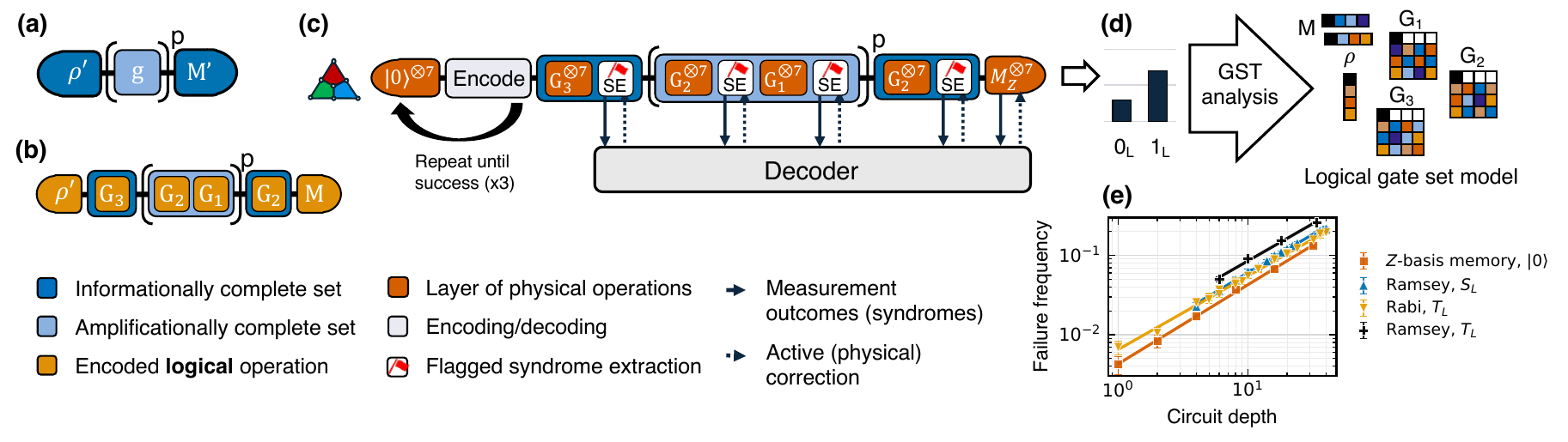}
    \caption{Schematic overview of the logical gate set tomography (``logical GST'') protocol and experimental implementation on Quantinuum H1-1, a 20-qubit trapped-ion quantum processor. Panels (a)--(c) show a hierarchy of circuit representations, from the high-level GST circuit structure, to its decomposition into logical operations, to the physical implementation of those logical operations in the Steane code. (a)  Logical GST circuit structure. Each circuit consists of an effective state preparation  $\rho'$, a short gate sequence called a \textit{germ} that is repeated $p$ times to amplify gate errors, and an effective measurement $M'$. The full protocol uses a structured list of such circuits with different germs, states, measurements, and repetition lengths. (b) Logical-operation-level implementation. Each effective state preparation, germ, and effective measurement is implemented using logical operations drawn from the logical gate set to be characterized. (c) Physical Steane-code implementation. The implementation used in this paper is identical to the one introduced in Ref. \cite{ryananderson2021realization} except for the usage of active error correction in this work. The layer of physical operations used to implement a logical gate need not be transversal; this flexibility is used for the non-transversal $T_L$ gate characterized in this work, although that implementation is not shown here. (d) Logical gate set estimation from observed circuit-outcome data. Outcome frequencies from the full GST experiment design are used (via maximum likelihood estimation) to estimate a logical gate set model, represented as Hilbert-Schmidt vectors (states and measurement effects) and Pauli transfer matrices (gates). (e) Application to circuit-class behavior. This logical gate set model can be used to explain differing behavior between different circuit classes; see Figure \ref{fig:dataset_fitting} for further details.} 
    \label{fig:schematic}
\end{figure*}

These concerns motivate the creation of logical-qubit characterization techniques that can be used to experimentally probe and estimate logical gate dynamics, and thus to test the validity of assumptions like Markovianity and stochasticity. One way to create such protocols is to ``lift'' physical-qubit benchmarking and characterization protocols to the logical level. The resulting protocols inherit the features and limitations of their physical-level counterparts. For example, logical randomized benchmarking (RB) measures a single average logical gate error rate \cite{combes2017logical, mayer2024benchmarking, lacroix2025scaling, ceasura2022nonexponential, dasu2026computing, berthusen2026experimental}, which is rarely sufficient to diagnose violations of Markovianity or stochasticity (although emergent logical non-Markovianity can sometimes be detected from the appearance of non-exponential RB decay curves \cite{ceasura2022nonexponential}). Tomographic approaches provide more detailed information. They have been used to estimate logical fidelities for logical resource-state generation \cite{daguerre2025experimental, lacroix2025scaling, besedin2026lattice, reinhold2020error-corrected} as well as to reconstruct certain logical operations \cite{zhang2026universal, besedin2026lattice, gutierrez2016errors, Fazio2025logical}. However, standard state and process tomography are not \textit{self-consistent}---they rely on known pre-calibrated measurements (and pre-calibrated input states for process tomography), and incorrect assumptions about these reference frames can corrupt tomographic estimates \cite{merkel2013selfconsistent}.  

Gate set tomography (GST) \cite{Nielsen2021gatesettomography, blume-kohout2017} 
was developed to address this self-consistency problem. Since its introduction, GST has been applied across hardware platforms, including superconducting qubits \cite{Govia2020bootstrapping, brieger2023compressive,cao2024efficient, wysocki2026detailed}, silicon quantum dots \cite{Madzik2022precision, Xue2022quantum,jones2026midcircuit}, neutral atoms \cite{rozanov2026benchmarking}, spin qubits in diamond \cite{bartling2025universal}, and trapped ions \cite{blume-kohout2017, hughes2020benchmarking, proctor2020detecting, rudinger2021experimental, piliouras2026highperformance}. We adapt GST to logical qubits, and use logical GST to experimentally and self-consistently estimate a set of \textit{logical} operations on a logical qubit encoded in the $\llbracket 7,1,3\rrbracket$ Steane code implemented on the Quantinuum H1-1 quantum processor. The logical gate set includes logical $\ket{0}$ state preparation, logical $Z$ measurement, transversal Clifford gates, and a non-transversal logical $T$ gate implemented by magic state injection. 

Our analysis of the experimental logical GST data showed no statistically significant evidence for logical non-Markovianity, and only weak evidence of coherent logical error. However, the modest size ($N_{\mathrm{shots}}=50$) of the experimental dataset limits the power of these tests---only large non-Markovianity or coherence effects would have been detected. To study these effects beyond the detection limits of the experiment, we repeated our analysis on large ($N_{\mathrm{shots}}\geq 10^4$) datasets generated using a detailed emulator of the H1-1 system, H1-1E. This analysis revealed strong evidence for coherent logical errors, and small but statistically significant evidence for logical non-Markovianity under default settings. Finally, by selectively disabling physical error sources in the emulator, we identified the primary cause of the coherent logical error as coherent physical-qubit memory errors, and determined that logical non-Markovianity can both be caused by physical leakage \textit{and} emerge from Markovian physical noise. Figure~\ref{fig:schematic}(a)-(d) gives a schematic overview of the logical GST protocol and its implementation on H1-1. Figure~\ref{fig:schematic}(e) summarizes the motivation for detailed modeling of logical qubit dynamics: logical errors that are not well approximated by simple depolarizing noise can accumulate differently in different classes of circuits, with especially pronounced effects possible for coherent and non-Markovian logical error. 

The remainder of this paper is structured as follows. In Sec.~\ref{sec:code_choice}, we describe the Steane code implementation on H1-1. In Sec.~\ref{sec:LoGST}, we introduce the logical gate set tomography protocol in detail. In Sec.~\ref{sec:exp_demo}, we describe the experiment design and present the experimental logical GST results. In Sec.~\ref{sec:testing}, we examine the evidence for logical non-Markovianity and coherent logical error, using Quantinuum's detailed system emulator to produce a higher-shot counterpart to the experimental data. In Sec.~\ref{sec:model_selection}, we use automated model selection to procedurally construct a reduced logical error model that retains only the parameters needed to predict logical circuit performance. In Sec.~\ref{sec:discussion}, we provide concluding remarks and future outlook. 

\section{Code choice and implementation} \label{sec:code_choice}
We characterize single-qubit operations on a logical qubit encoded in the $[\![7,1,3]\!]$ Steane code, which uses seven physical qubits to protect one logical qubit and can correct a single physical error \cite{Steane1996multiple, Steane1996error}. For a fault-tolerant QEC cycle, we use an adaptive flagged syndrome extraction circuit introduced in Ref.~\cite{Reichardt_2021} and realized experimentally in Ref.~\cite{ryananderson2021realization}, together with a lookup table decoder. Unlike Ref.~\cite{ryananderson2021realization}, we employ active error correction instead of Pauli frame tracking: the syndromes are decoded during the circuit execution, and the correction is applied directly to the physical qubits via real-time feedforward. This approach demonstrates the system's dynamic control capabilities and avoids additional Pauli-frame bookkeeping in the generated circuits, with comparable performance expected for the circuits considered here. When compared to Pauli-frame tracking, active error correction also has one other key difference: unlike Pauli frame tracking, which can effectively twirl logical error, active error correction has been shown to allow the coherent accumulation of logical errors \cite{witzel2026correcting}. As such, we are more likely to observe violation of the Pauli-stochastic model than if we had used Pauli-frame tracking. 

The characterized logical operations are phase ($S_L$), Hadamard ($H_L$), and $Z_{\pi/4}$ ($T_L$) gates; an idle ($I_L$) operation defined as a QEC cycle; logical state preparation in $|0\rangle_L$; and logical $Z$-basis measurement ($M^Z_L$). The gates $H_L$ and $S_L$ are Clifford gates, while $T_L$ is non-Clifford; together, these gates form a universal gate set for single-qubit operations. The Steane code admits transversal logical Clifford operations, in particular, $H_L$ and $S_L$ are implemented $H^{\otimes 7}$ and $(S^{\dagger})^{\otimes7}$, respectively. $T_L$ is applied by preparing a $\ket{T}_L$ state on an ancilla code block as described in \cite{chamberland2019fault-tolerant}, and injecting the state using a teleportation gadget followed by a QEC cycle. We use a version of the teleportation gadget that swaps the data and ancilla code blocks to be more robust to leakage and memory errors \cite{mayer2024benchmarking}. The $\ket{T}_L$ state preparation involves a flagged fault-tolerance check and requires post-selecting on the flagged measurement outcomes. This type of post-selection is still considered scalable, because the discard rate can be exponentially suppressed using repeat-until-success. In our experiments, however, we limited ourselves to one attempt for each $\ket{T}_L$ state preparation, with an measured average acceptance rate of $95.9\%$ per $T_L$ gate on H1-1. For circuits containing $T_L$ gates, only shots that are post-selected for having successfully injected all $T_L$ instances are incorporated into the logical GST dataset.
Therefore, the fitted $T_L$ operation should be interpreted as an effective trace-preserving logical operation conditioned on acceptance, rather than as the unconditional operation including discarded events.

We use a repeat-until-success procedure to initialize $|0\rangle_L$ with a maximum of three attempts; unsuccessfully initialized shots are post-selected with a rate less than one shot per million in noise-aware simulations. Logical $Z$-basis measurement $M^Z_L$ is implemented by terminating the circuit with computational-basis measurements of all data qubits; the resulting bit values are decoded and an inferred correction is applied to obtain the final logical measurement outcome \cite{ryananderson2021realization}. Logical circuits are specified using Simple Logical Representation (SLR), which allows the user to procedurally generate low-level QASM code for various QEC primitives \cite{Pecos, mayer2024benchmarking} (see Appendix \ref{app:steane} for more details). Figure~\ref{fig:schematic}(c) summarizes many of these QEC implementation details within the context of logical GST, which we will discuss next. 

\section{Logical Gate Set Tomography}\label{sec:LoGST}
We now describe the logical GST protocol. The central idea is to characterize a fault-tolerant quantum computer in terms of effective operations acting only on the logical degrees of freedom, and then to estimate those operations using gate set tomography \cite{blume-kohout2017,Nielsen2021gatesettomography}. This requires two ingredients. First, the syndrome degrees of freedom and the associated recovery information must be reduced to an effective logical description, so that preparations, measurements, and gates can be modeled as maps on the logical subsystem \cite{rahn2002exact,bravyi2018correcting,beale2018quantum,iverson2020coherence}. Second, after reducing the dynamics to effective logical operations, the standard GST protocol can be implemented the logical level.

The first subsection motivates this reduction and explains when QEC operations can be represented by effective logical maps. The following subsection then described the logical gate-set model and briefly recaps the GST experiment design and analysis procedure. 

\subsection{Effective logical maps}
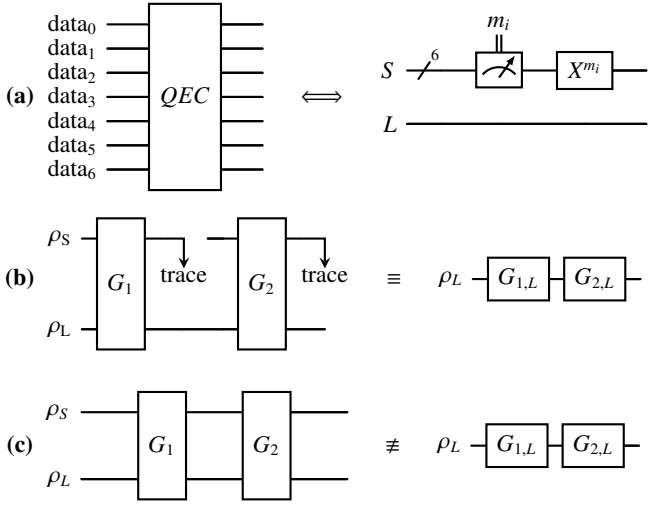
\begin{figure}[t]
\centering
\small
\textbf{(a)}
\begin{quantikz}[row sep={0.32cm,between origins}, column sep=0.55cm]
\lstick{$\mathrm{data}_0$} & \gate[7]{QEC} & \qw \\
\lstick{$\mathrm{data}_1$} & \qw           & \qw \\
\lstick{$\mathrm{data}_2$} & \qw           & \qw \\
\lstick{$\mathrm{data}_3$} & \qw           & \qw \\
\lstick{$\mathrm{data}_4$} & \qw           & \qw \\
\lstick{$\mathrm{data}_5$} & \qw           & \qw \\
\lstick{$\mathrm{data}_6$} & \qw           & \qw
\end{quantikz}
\quad
$\Longleftrightarrow$
\quad
\begin{quantikz}[row sep={0.35cm}, column sep=0.45cm]
\lstick{$S$}
    & \qwbundle{6}
    & \metercw[label style={inner sep=1pt}]{m_i}
    & \gate{X^{m_i}}
    & \qw \\
\lstick{$L$}
    & \qw
    & \qw
    & \qw
    & \qw \\
\end{quantikz}
\textbf{(b)}
\begin{quantikz}[row sep=0.35cm, column sep=0.2cm]
\lstick{$\rho_{\text{S}}$}
    & \gate[2]{G_1}
    & \trash{\text{trace}}
    & \qw
    & \gate[2]{G_2}
    & \trash{\text{trace}} \\
   \lstick{$\rho_{\text{L}}$} & \qw
    & \qw
    & \qw
    & \qw
    & \qw
\end{quantikz}
\quad
$\equiv$
\quad
\begin{quantikz}[row sep=0.35cm, column sep=0.2cm]
\lstick{$\rho_{L}$}
& \gate[2]{G_{1,L}}
    & \gate[2]{G_{2,L}}
    & \qw
\end{quantikz}

\vspace{1em}

\textbf{(c)}
\begin{quantikz}[row sep=0.35cm, column sep=0.75cm]
\lstick{$\rho_{S}$}
    & \gate[2]{G_{1}}
    & \gate[2]{G_{2}}
    & \qw \\
\lstick{$\rho_{L}$}
    & \qw
    & \qw
    & \qw
\end{quantikz}
\quad
$\not\equiv$
\quad
\begin{quantikz}[row sep=0.35cm, column sep=0.2cm]
\lstick{$\rho_{L}$}
& \gate[2]{G_{1,L}}
    & \gate[2]{G_{2,L}}
    & \qw
\end{quantikz}
\caption{\label{fig:logical_and_syndrome} Logical operations can be treated as acting \textit{only} on logical qubits only in certain circumstances. (a) The $n$-data-qubit Hilbert space can be viewed as $\mathcal H_{\mathrm{data}}\cong \mathcal H_L\otimes \mathcal H_S$,
with $k$ logical qubits and $n-k$ syndrome qubits. For the
$[\![7,1,3]\!]$ Steane code, this gives one logical qubit and six syndrome
qubits. In this picture, ideal syndrome extraction and recovery can be viewed as returning the syndrome subsystem to the trivial syndrome sector. (b) If the syndrome subsystem is correctly reset after each implemented logical gate, it can be traced out, leaving an effective operation on the logical subsystem alone. (c) If the syndrome subsystem retains information between gates, the syndrome subsystem acts as a memory. The logical-only description is then not generally exact, although it can remain approximately valid when the memory effects are small.} 
\end{figure}

For a stabilizer code, the data-qubit Hilbert space can be factored into logical and syndrome subsystems \cite{ziyad2025emergent}. Figure~\ref{fig:logical_and_syndrome}(a) shows this decomposition for the $[\![7,1,3]\!]$ Steane code, where the seven data qubits are identified with one logical qubit and six syndrome qubit degrees of freedom.

Syndrome measurements produce classical outcomes, and the decoder uses these outcomes to determine the recovery operation. For a single implemented operation, these outcomes define different branches of the experiment. We can define an average logical channel by marginalizing over the syndrome records \cite{rahn2002exact, bravyi2018correcting}. As long as the syndrome subsystem is reset after recovery and can be traced out, previous work has established that the behavior of logical qubits during quantum error correction can be well represented using an effective map acting solely on the logical degree of freedom \cite{rahn2002exact, beale2018quantum, bravyi2018correcting,iverson2020coherence}. 

By the same reasoning, we can \textit{also} represent logical gates which act non-trivially on the logical degree of freedom as effective maps on the logical subsystem under the same assumptions [Fig.~\ref{fig:logical_and_syndrome}(b)]. Operationally, we define a logical gate as the full fault-tolerant implementation of a target logical operation on the encoded state. As such, ideal syndrome extraction and recovery should reset the syndrome qubits after each logical gate. To achieve fault-tolerance, this implementation may, depending on the operation, include physical gates, ancillae preparation, measurements, post-selection, syndrome extraction, decoding, and active corrections or Pauli-frame updates. For example, in the seven-qubit Steane code, a transversal logical Hadamard can be implemented fault-tolerantly by applying Hadamard gates to each physical qubit, followed by syndrome extraction, decoding, and a decoder-inferred correction or Pauli-frame update. If syndrome information instead persists between gates, as in Fig.~\ref{fig:logical_and_syndrome}(c), the syndrome subsystem can act as a memory, and it is no longer possible to represent the logical gate acting only on the logical qubit. If memory effects are small, the description of the logical gate acting only on the logical subsystem may remain approximately true \cite{kwiatkowski2025constructing}. 

\subsection{Protocol}
We now introduce the logical GST protocol, which adapts standard physical-qubit gate set tomography to encoded operations rather than native physical operations. Its aim is to estimate a model---a collection of Pauli transfer matrices, density matrices, and POVM effects---for a \textit{logical} gate set $\mathcal{G}_L$. We define $\mathcal{G}_L$ as a set of logical operations including one or more state preparations, measurements, and gates acting on the logical degree of freedom. Logical state preparation and logical measurement are defined analogously to our definition of a logical gate in the previous section. Logical state preparation refers to the full process used to initialize a logical state, which may involve some combination of an encoding circuit \cite{ryananderson2021realization}, one or more fault-tolerant QEC cycles \cite{lacroix2025scaling}, and a repeat-until-success procedure \cite{ryananderson2021realization}. Likewise, logical measurement describes a multi-step process: the physical qubits are measured, and decoder-inferred corrections are used to assign logical measurement outcomes. 

Mathematically, the effective state of a $k$-logical-qubit register is represented by a density matrix $\rho$ acting on the $D$-dimensional logical Hilbert space, where $D=2^k$.  We represent the effective logical measurement using a positive operator-valued measure (POVM) $M=\{E_i:\ i=1\ldots m\}$, where each $E_i\geq0$ is a $D\times D$ matrix called an \textit{effect}. Each logical gate $G_i$ is represented as a fixed completely positive, trace-preserving (CPTP) linear map which acts as $G_i:\rho\rightarrow G_i[\rho]$. We use the Pauli transfer matrix representation that is standard in quantum tomography, in which density matrices and effects are vectorized and gates act as matrices on vectorized states, all represented in the Pauli basis \cite{Nielsen2021gatesettomography}. In our model, $G_i$ does not depend on the observed syndrome or, for that matter, any other context information (e.g., the time of day, preceding gates, or position in the circuit). This is the familiar Markovian assumption from physical-qubit characterization \cite{Nielsen2021gatesettomography, ziyad2025emergent}, and, as we see in Sec.~\ref{sec:non_markovian}, can be explicitly tested using statistical model validation techniques \cite{Nielsen2021gatesettomography}. This assumption can be violated by the memory effects shown in Fig~\ref{fig:logical_and_syndrome}(c), or by non-Markovian processes inherited from the physical level, e.g., calibration drift leading to \textit{logical} gate drift.

\begin{figure*}[ht!]
    \centering
    \includegraphics[width=1.0\linewidth]{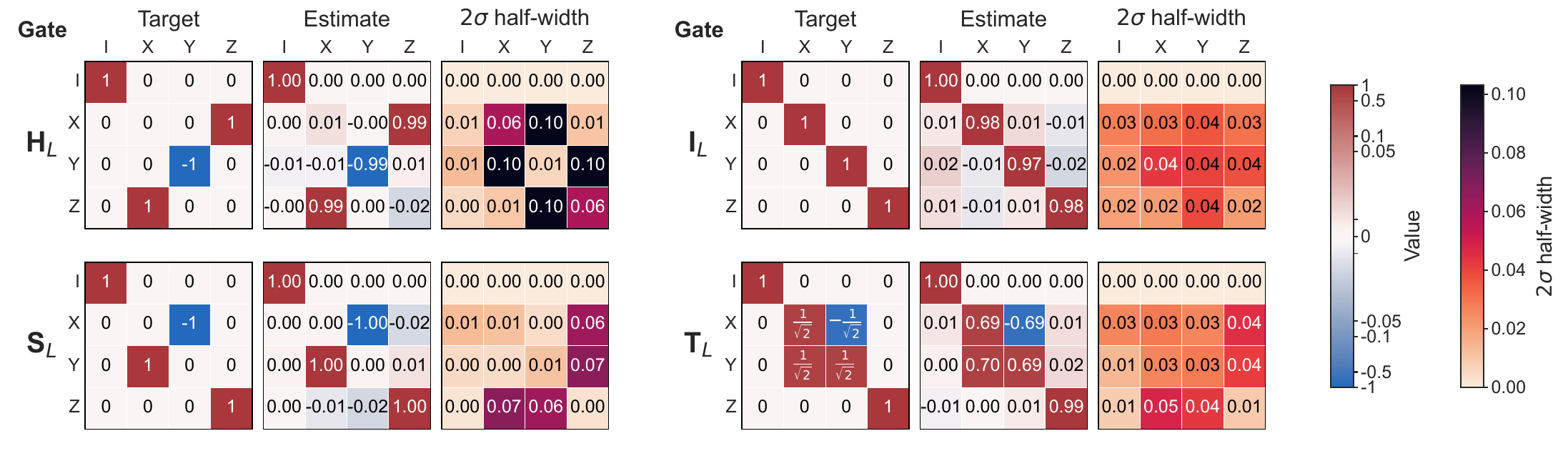}
    \caption{Pauli transfer matrices with $2\sigma$ half-widths for the logical Hadamard ($H_L$), phase ($S_L$), identity ($I_L$), and $Z_{\pi/4}$ ($T_L$) gates, estimated using logical GST on a Steane-$[\![7,1,3]\!]$-encoded qubit implemented on H1-1. Pauli transfer matrix elements are colored using a symmetric logarithmic scale with a linear region near zero to make deviations from the target operation more visible.}
    \label{fig:ptms}
\end{figure*}

The logical GST experiment design and analysis procedure proceed identically to standard physical-qubit GST. To estimate $\mathcal{G}_L$, we use a structured list of circuits each with a common form. As shown in Fig.~\ref{fig:schematic}(a), each GST circuit consists of an effective logical state preparation $\rho'$, a logical gate sequence, and an effective logical measurement $M'$. The logical gate sequence is formed by selecting a short germ $g$ and repeating it $p$ times, thereby amplifying the effects of gate errors. The germs are chosen to collectively amplify all amplifiable (i.e., non-gauge \cite{Nielsen2021gatesettomography}) gate parameters and to achieve Heisenberg-limited precision on estimates. Fiducial circuits $\{F^{\mathrm{prep}}_j\}$ and $\{F^{\mathrm{meas}}_k\}$ are used to generate the effective preparations $\{\rho_j'\}$ and measurements $\{M_k'\}$ [Fig.~\ref{fig:schematic}(b)]. These fiducials are short sequences of logical gates drawn from $\mathcal{G}$ and are chosen so that the resulting effective preparations and measurements are informationally complete. Usage of fiducial circuits is necessary because most QEC implementations only directly implement one or two logical state preparations and measurements, much as most physical quantum processors only offer one native state preparation and measurement \cite{Nielsen2021gatesettomography,ostrove2023near}. The experiment design includes all circuits of the form $F_j^{\mathrm{prep}}(g_\ell)^pF_k^{\mathrm{meas}}$. Setting $p=0$ gives circuits of the form $F_j^{\mathrm{prep}}F_k^{\mathrm{meas}}$, which characterize the effective null operation. These circuits allow state-preparation and measurement errors to be modeled self-consistently rather than attributed to the logical gates. 

Finally, as shown in Fig.~\ref{fig:schematic}(d), the observed outcome frequencies are used to fit a logical gate set model using maximum likelihood estimation, implemented in the Python package \texttt{pyGSTi} \cite{osti_code-28250, Nielsen2020}. As in standard physical-qubit GST, the fitted gate set is only defined up to a gauge freedom and all estimates and gauge-variant metrics are reported after fixing a gauge relative to the target gate set. 

\section{Experimental demonstration on Quantinuum H1-1} \label{sec:exp_demo}
Experiments were run on Quantinuum H1-1, a 20-qubit trapped-ion quantum computer with five gate zones, all-to-all connectivity, high-fidelity gate operations, low crosstalk mid-circuit measurement and reset, and leakage repump~\cite{Pino2020, ryananderson2021realization}. See Appendix~\ref{appendix:h1-1} for more detailed hardware information. These capabilities enable straightforward implementation of many low-distance QEC codes and make H1-1 an ideal testbed for early QEC studies. Experimental cost is driven both by the number of logical GST circuits and by the cost of implementing specific logical operations. In our implementation, each fault-tolerant QEC cycle required 50 physical two-qubit entangling gates and 12 mid-circuit measurements, so even relatively short logical GST circuits corresponded to substantial physical circuit complexity. The non-Clifford gate $T_L$ is especially costly as it is implemented using a teleportation gadget, that requires a second logical qubit, additional physical operations, and post-selection, leading to shot loss.

We therefore compressed the logical GST experiment design in several distinct ways. The final experiment used the smallest number of fiducials that is still informationally complete, and each fiducial sequence contained only gates that could be implemented transversally. Given this fiducial set, we then used fiducial pair reduction to reduce the number of preparation/measurement combinations used to probe each germ, omitting combinations expected to contribute little information about the gate-set model parameters in order to reduce the total number of circuits in the logical GST experiment~\cite{rudinger2023two,ostrove2023near}.

Within this compressed design, we used a maximum germ power of $p_{\max}=32$. Although this is modest by the standards of some physical-qubit GST experiments, it corresponds to relatively deep circuits at the logical level given present capabilities, especially for germs containing the non-transversal $T_L$ gate. Because each $T_L$ gate requires additional physical operations and post-selection, repeated $T_L$-containing germs quickly become among the most demanding circuits in the experiment. We therefore limited the number of repetitions of such germs so that no circuit contained more than eight $T_L$ gates, bounding the cost of $T_L$-containing circuits without removing them from the GST design. We also chose to prioritize characterization of the non-trivial, i.e., non-idle, gates, which we believed were the most interesting and least studied. As such, for the idle germ at germ powers $p>1$, we retained only the $\{|0\rangle, |+\rangle\}$ preparation and measurement fiducials. The definite-outcome circuits retained after this choice correspond to $X$- and $Z$-basis memory experiments.

The final implementation-aware experiment design contained $N_{\mathrm{circuits}}=158$ circuits with $N_{\mathrm{shots}} \approx 50$ retained shots per circuit; for circuits containing $T_L$ gates, additional attempts were used to compensate for post-selection. To check for substantial temporal drift, two circuits from the design were re-executed later in the run; their outcomes were statistically consistent with the original executions. Although the sample count is modest by GST standards, the physical circuits were substantial: circuits contained an average of 11 QEC cycles and as many as 38, corresponding to as many as 1900 physical two-qubit gates and 456 mid-circuit measurements from QEC cycles alone before accounting for additional overhead from non-transversal $T_L$ gates. By sample count, the populated dataset is roughly $100\times$ smaller than a typical physical-qubit GST experiment with the same gate set and maximum germ power, which might use $N_{\mathrm{circuits}}\approx10^3$ circuits with $10^3\gtrsim N_{\mathrm{shots}}\gtrsim10^2$. Hessian-based 95\% ($2\sigma$) confidence intervals reported below reflect this sample size; we interpret the reconstructed gate-set model at that resolution. 

Figure~\ref{fig:ptms} shows the estimated Pauli transfer matrices for the logical gates, obtained from a CPTP-constrained logical GST model. To summarize logical performance, we use the estimated gate-set model to extract the error metrics reported in Table~\ref{tab:gate_metrics}. Precise conventions for the initialization and measurement entries are given in the table caption. For gates, we report the entanglement infidelity $\epsilon_F=1-F_e$ and the diamond-distance error $\epsilon_\diamond=\frac{1}{2}\|\hat{G}-\bar{G}\|_\diamond$, where $\hat{G}$ is the estimated operation and $\bar{G}$ is the target operation~\cite{hashim2025practical}. For logical initialization, $\epsilon_F$ denotes state infidelity. At 95\% confidence, the non-idle logical Clifford gates and logical measurement have $\epsilon_F \leq 1.5\%$ and $\epsilon_\diamond \leq 4\%$. At the same confidence level, the non-Clifford $T_L$ gate has $\epsilon_F \leq 2.7\%$ and $\epsilon_\diamond \leq 6\%$, and logical initialization has $\epsilon_F \leq 2.3\%$. The transversal logical operations $H_L$ and $S_L$ have comparable performance, as expected, with error metrics that agree within the quoted 95\% confidence intervals.

\begin{table}[t!]
\centering
\small
\begin{tabular}{lcc}
Gate& \shortstack{\\ \textbf{Infidelity $\epsilon_F$  (\%)}}& \shortstack{\textbf{Diamond distance}\\ \textbf{error $\epsilon_\diamond$(\%)}}\\
\hline
\hline
Hadamard ($H_L$)
& $0.7 \pm 0.4$
& $1.3 \pm 1.9$
\\

Phase ($S_L$)
& $0.2 \pm 0.4$
& $1.3 \pm 2.5$
\\

Idle ($I_L$)
& $1.7 \pm 9.2$
& $3.2 \pm 16.8$
\\

$T_L$
& $1.2 \pm 1.5$
& $2.1 \pm 3.9$
\\
Initialization ($|0\rangle_L$)& $0.0 \pm 2.3$
& --
\\
Measurement ($M^Z_L$)& $0.1 \pm 1.4$
& $0.7 \pm 2.1$

\\
\hline

\end{tabular}
\caption{\label{tab:gate_metrics}Error metrics for logical gates from logical GST experiment on H1-1. Entanglement infidelity is given by $\epsilon_F = 1-F_e$ where $F_e$ is the entanglement fidelity; diamond distance error is defined as $\epsilon_\diamond =\frac{1}{2}||\hat{G}-\bar{G}||_\diamond$ where $\hat{G}$ is the estimated gate and $\bar{G}$ is the target gate \cite{hashim2025practical}. Diamond distance error and entanglement infidelity will have the same value in a Pauli-stochastic noise model. For initialization and measurement, analogous error quantities are reported for the associated state or POVM map. For the logical initialization $|0\rangle_L$, the infidelity is the state fidelity. For the logical terminating measurement $M^Z_L$, the reported infidelity is the entanglement infidelity of the POVM map, which to leading order coincides with readout infidelity \cite{hashim2025practical}, while the diamond-distance error is the diamond distance between POVM maps. The error bars represent $95\%$ ($2\sigma$) confidence intervals. The large error bars on $I_L$ are due to removing many of the circuits containing idle operations to condense the experiment design.} 
\end{table}

\section{Logical error inference}\label{sec:testing}
We next examine both the H1-1 hardware data and simulated data generated with H1-1E for evidence of logical non-Markovianity and of logical errors beyond a Pauli-stochastic model, and assess whether any detected non-Pauli errors are coherent. H1-1E is a full-stack device model that incorporates characterization- and theory-informed error modeling, as well as compilation and ion transport information used to implement circuits on the device \cite{emulator_docs}. The main results of this section are summarized in Fig.~\ref{fig:overrotation_ramsey}. In particular, the experimental data from H1-1 did not show any significant evidence of logical non-Markovianity or inconsistency with a Pauli-stochastic logical error model [Fig.~\ref{fig:overrotation_ramsey}(a)]. However, data generated with H1-1E under certain settings showed small but statistically significant evidence of logical non-Markovianity [Fig.~\ref{fig:overrotation_ramsey}(a)-(b)] and logical coherent errors [Fig.~\ref{fig:overrotation_ramsey}(c)-(d)]. 

The H1-1E noise model includes one- and two-qubit gate errors, measurement bit flips, initialization errors, crosstalk, leakage, and memory errors. Gate errors are modeled primarily as Pauli stochastic error channels with a smaller leakage component due to spontaneous emission. Memory error is treated as a coherent phase error accumulated during transport and cooling \cite{moses2023a,ransford2026a}. Both leakage and memory error have previously been shown to significantly affect trapped-ion QEC performance \cite{ryananderson2021realization}. For exact H1-1E noise model settings, refer to Appendix~\ref{app:noise_model}. We use H1-1E to study how non-Markovian and coherent physical errors in the QEC implementation are reflected in the effective logical error. In H1-1E, leakage is the only non-Markovian physical error, and memory error is the only modeled source of coherent physical error. By independently disabling these mechanisms we test whether logical non-Markovianity and coherent logical error are inherited from corresponding physical errors, are suppressed by QEC, or arise from the remaining physical error processes. 

Using high-performance computing resources and state-vector simulation, we sample four simulated datasets with H1-1E, using the same experiment design ($N_{\mathrm{circuits}}=931$ circuits with $N_{\mathrm{shots}}=14100$) but independently enabling or disabling leakage and memory error. One dataset uses the default emulator settings with both mechanisms enabled. The other three disable memory error, leakage, or both. Since the experiments on H1-1 used the leakage repump feature, we conjecture that the H1-1E leakage ``off" and memory error ``on" setting is the most accurate model for H1-1. However, no emulator is perfect and H1-1E may not fully capture certain errors or may be overly conservative in estimating others. When both leakage and memory error are disabled, the emulator still includes stochastic gate errors, SPAM errors, and crosstalk. The maximum germ power is $p_{max}=32$, matching the hardware study. For the data generated with H1-1E, we expanded our fiducial states and measurements to the six Pauli eigenstates and did not apply fiducial-pair reduction. Appendix~\ref{app:simulated_data} reports error metrics and post-selection acceptance rates for each simulated dataset. In the following subsections, we first examine the evidence for logical non-Markovianity in these datasets, beginning with the hardware data, then proceed to the simulated data, and then turn to evidence for logical coherent error. 

\subsection{Testing for logical non-Markovianity}\label{sec:non_markovian}
To assess whether the effective logical dynamics are consistent with a Markovian description, we test whether the CPTP-constrained gate-set model that was first fit to the experimental data in Sec.~\ref{sec:LoGST}. We refer to this model as the ``CPTP'' model is \textit{valid}, i.e. consistent with the data to which it was fitted. Because this model can represent arbitrary Markovian logical dynamics, significant violation of the CPTP model is evidence for logical non-Markovianity. We quantify violation of this candidate model by comparing the log-likelihood ratio statistic $\lambda = 2\Delta \log{\mathcal{L}}$ between the candidate model (e.g. CPTP) and a known statistically valid model. Per Wilks' theorem, $\lambda$ is expected to follow a $\chi^2_\kappa$ distribution when the candidate model is statistically valid \cite{wilksthm}. We therefore quantify model violation by
\begin{equation}
    N_{\sigma}
   =  \frac{\lambda-\kappa}{\sqrt{2\kappa}},
   \label{eq:n_sigma}
\end{equation}
where $\kappa$ is the difference between the number of degrees of freedom in the dataset and the number of observable model parameters. $N_\sigma \gg 1 $ corresponds to statistically significant evidence of model violation. However, $N_\sigma$ is not a measure of the effect size: with sufficient data, a very small departure from the model can produce a large $N_\sigma$. Wildcard error was introduced to quantify unmodeled effect size \cite{blumekohout2020wildcard}. We fit a single parameter wildcard model, where $\alpha$ is the fraction of each model operation's diamond distance error required as unmodeled error in order to reconcile the model with the data. If $\alpha \ll 1$, then we can assume the non-Markovian behavior is negligible and the fitted gate set model captures the majority of the observed errors \cite{rudinger2021experimental, blumekohout2020wildcard, hashim2023benchmarking}. 

\begin{figure*}
    \centering
    \includegraphics[width=1\linewidth]{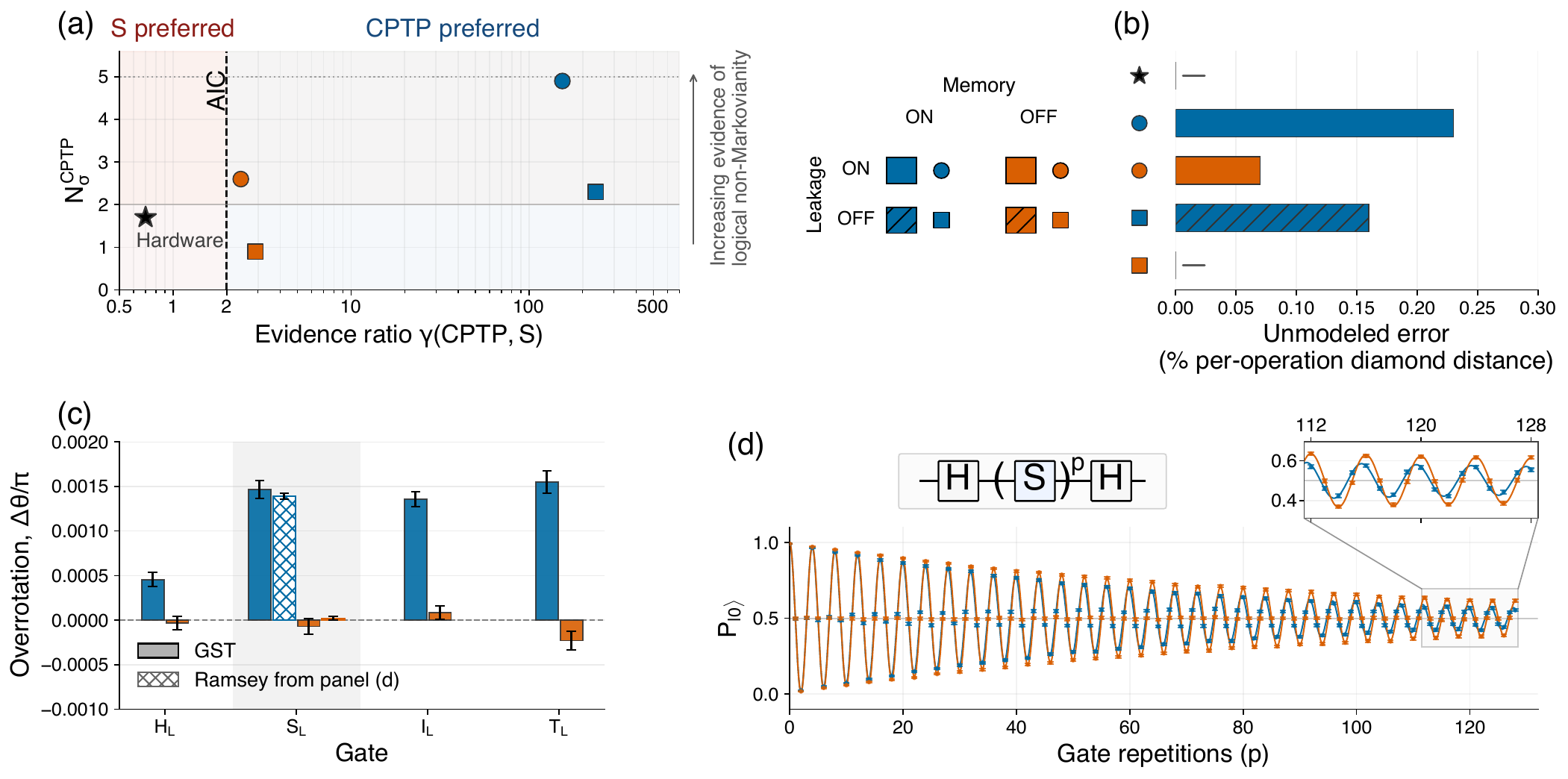}
    \caption{Quantifying coherent logical error in simulated data from Quantinuum's detailed system emulator. (a) Summary of model agreement quantified by model violation $N_\sigma$ [Eq.~\eqref{eq:n_sigma}] and evidence ratio $\gamma(\mathrm{CPTP}, \mathrm{S})$ [Eq.~\eqref{eq:evidence_ratio}] for the hardware data and four datasets simulated on H1-1E. The Akaike information criterion (AIC) is shown by a dashed line at $\gamma = 2$. Memory settings
are encoded by color, while leakage settings are encoded by hatching and marker shape. (b) Unmodeled error estimated using a single-parameter wildcard model shown as a percentage of per-operation diamond distance for all datasets. For exact numerical values for quantities plotted in panels (a) and (b), see Table~\ref{tab:nsigma_comparison} in Appendix~\ref{app:simulated_data}. (c) Over-rotation in units of $\pi$ for each logical gate for memory error enabled and disabled. Over-rotation is defined as $\Delta\theta=\theta_{\mathrm{est}}-\theta_{\mathrm{targ}}$ where $\theta_{\mathrm{est}}$ and $\theta_{\mathrm{targ}}$ are the estimated and target rotation angle, respectively. Solid bars show the over-rotation estimates extracted from the CPTP-constrained gate-set model, while cross-hatched bars show the over-rotation estimated using the $S_L$-gate Ramsey experiment shown in panel (d). (d) Ramsey experiment used to estimate S-gate over-rotation. Each marker on the plot corresponds to the $0$-outcome probability for a circuit with $p$ repetitions of the $S_L$ gate. For both (c) and (d), error bars represent $95 \%$ ($2\sigma$) confidence intervals.}
    \label{fig:overrotation_ramsey}
\end{figure*}

Figure~\ref{fig:overrotation_ramsey}(a) shows the $N_\sigma^{\mathrm{CPTP}}$ values for each dataset on the y-axis. In the experimental data, the violation of the CPTP model is small, with $N_\sigma^{\mathrm{CPTP}} < 2$ (black star), indicating no statistically significant evidence of logical non-Markovianity. By contrast, the default emulator configuration, with both leakage and memory errors enabled, gives $N_\sigma^{\mathrm{CPTP}} \approx 5$ (blue circle). Disabling leakage while retaining memory error reduces the violation, although it remains statistically significant at $N_\sigma^{\mathrm{CPTP}} \approx 2.3$ (blue square). Enabling leakage while disabling memory error also produces a smaller but non-negligible model violation, $N_\sigma^{\mathrm{CPTP}} \approx 2.6$. Only when both error modes are disabled does the statistic fall to $N_\sigma^{\mathrm{CPTP}} \approx 1$, consistent with no significant logical non-Markovianity (orange square). Jointly, these results suggest that non-Markovianity can both persist from physical leakage at the logical level and potentially emerge from Markovian error sources, such as coherent memory error. 

However, the unmodeled effect size of deviations from the fixed Markovian gate-set model are relatively small. The wildcard estimates in Fig.~\ref{fig:overrotation_ramsey}(b) show that the required unmodeled error accounts for less than $1\%$ of the per-operation diamond-distance error. As is the case in Refs.~\cite{kwiatkowski2025constructing,ziyad2025emergent}, even though logical non-Markovianity is detectable with sufficiently sensitive datasets, the effect size is sufficiently small that 
an approximate Markovian gate set model captures the dominant logical error dynamics. 

We also examine how sensitivity to non-Markovian dynamics in the simulated datasets varies with circuit depth.  A full accounting of this investigation is provided in Appendix~\ref{app:non_markovianity}. Interestingly, we find that when both leakage and memory errors are turned off, non-Markovian dynamics \textit{are} detected at low circuit depth, but \textit{not} at higher ones, indicating the potential presence of transient emergent logical non-Markovian noise. The exact interplay between leakage and memory errors (along with the background errors that were not toggled) and their contributions to logical non-Markovianity remain a rich subject for further study.

\subsection{Testing a logical Pauli-stochastic model}\label{sec:pauli_stochastic_model}
Having shown that all datasets are consistent with only small deviations from a Markovian model, we next ask whether an effective model containing only Pauli-stochastic errors is sufficient to predict the experimental and simulated data. We can fit such a model, which we refer to as the ``S'' model, using the error generator framework outlined in Ref.~\cite{blume-kohout2022a}, and once again test whether there is evidence of out-of-model effects using model violation. In the experimental dataset, we observe no evidence of errors that are not Pauli-stochastic ($N^{S}_\sigma <1$). For the simulated data, however, looking at $N^{S}_\sigma$ by itself makes less sense because in three of the four datasets the CPTP model fails to fit the data ($N^{CPTP}_\sigma >2$).  We therefore would expect the S model to perform worse (as measured by $N_\sigma$) than CPTP, even if the remaining Markovian errors are all Pauli stochastic.  This is because maximum likelihood estimation will use the extra parameters in the CPTP model to try to capture the effects of the non-Markovian dynamics.

This motivates asking a slightly different question: does the S model predict the data as well as or better than the CPTP model, while properly accounting for the fact that CPTP will superficially look better due to its access to extra parameters? To answer this question, we use the \textit{evidence ratio} \cite{rudinger2021experimental, Nielsen2021efficient, PORTET2020111}, which compares the predictive power of a smaller model $\mathcal{M}_A$ and a larger model $\mathcal{M}_B$, 

\begin{equation}
    \gamma(\mathcal{M}_A, \mathcal{M}_B) = \frac{(\lambda_A-\lambda_B)}{(\kappa_B-\kappa_A)}.
    \label{eq:evidence_ratio}
\end{equation}
As in previous work, we use the Akaike information criterion (AIC) to interpret the evidence ratio. The AIC states that a smaller model $\mathcal{M}_A$ has greater predictive accuracy iff $\gamma(\mathcal{M}_A, \mathcal{M}_B) < 2$\footnote{The AIC assumes that each ``useless'' parameter contributes one unit of log-likelihood on average. However, when there is evidence of model violation, we can expect the data to be ``overdispersed'' and a ``useless'' parameter will instead on average contribute more than one unit. Approximating a corrected threshold using the method in Ref.~\cite{Stemp2024scalable}, we find that the new threshold differs by less than $5\%$ from the standard AIC threshold for all datasets and does not substantively change any of the reported results. As such, we choose to retain the standard threshold for simplicity.}. It is worth noting that AIC explicitly guards against overfitting by requiring the larger model's improvement in fit to exceed the penalty associated with its additional parameters. The process of constructing different candidate models, e.g. S and CPTP, and comparing them by computing some ``score" such as $N_\sigma$ or $\gamma$ is known as \emph{model selection}. 

For the experimental data, there is no evidence against the smaller model [$\gamma(\mathrm{CPTP},\mathrm{S}) < 1$]. As such, the data can be well predicted by a Pauli-stochastic model. However, sectorwise comparisons where an entire error generator sectors are removed (e.g., between the CPTP and S models) can sometimes be too coarse, obscuring single terms which may improve the model. So far we have only considered sectorwise candidate models. In order to probe the statistical significance of individual errors we can instead construct candidate models which differ by a single parameter. This approach is hindered by gauge freedom \cite{Nielsen2021gatesettomography, sinha2026efficient}. The model parameterizations used so far, based on Ref. \cite{blume-kohout2022a}, contain infinitely many different gate sets that are gauge-equivalent and hence physically indistinguishable. Thus, we pair this fine-grained model selection approach with a first-order gauge-invariant (FOGI) gate set error parameterization \cite{FOGI} which allows us to instead probe individual \emph{observable} coherent errors. More specifically, these errors quantify (1) the over-rotation of any individual logical gate, and (2) the axis-misalignment between pairs of logical gates. For more details about FOGI parameters see Appendix \ref{app:model_selection}.

For the experimental data, we start with the S model, which we showed earlier provides a good fit ($N_\sigma^\mathrm{S}\approx1$). We then iteratively add individual FOGI coherent error parameters and compute their corresponding evidence ratio to obtain the amount of log-likelihood gained by each coherent degree of freedom. This procedure revealed three coherent FOGI error parameters to be (weakly) statistically significant in isolation  ($\gamma \approx 3,3,3.6$), and one pair of these FOGI error parameters was found to be (weakly) statistically significant in combination ($\gamma\approx 2.11$). All of these FOGI error parameters correspond to axis-misalignment errors between $T_L-H_L$ and $T_L-S_L$ (see appendix Appendix \ref{app:model_selection} for more details). Given there is no evidence of violation of the Pauli-stochastic model ($N_\sigma^{\mathrm{S}} \approx 1$), this result is best interpreted as evidence that including a small number of coherent parameters makes it possible to \textit{better} reproduce the data (without overfitting), rather than evidence that coherent parameters are \textit{necessary} to reproduce the data.     

We now turn to the simulated data. Unsurprisingly, when memory error is enabled (which introduces coherent physical errors), the smaller Pauli-stochastic model is unambiguously rejected by the AIC [we observed $\gamma(\mathrm{CPTP},\mathrm{S}) > 100$ for both leakage settings], indicating strong evidence for non-Pauli-stochastic errors. However, when memory error is disabled, the AIC only weakly rejects the S model [$\gamma(\mathrm{CPTP},\mathrm{S}) > 2.4$ with leakage enabled and $\gamma(\mathrm{CPTP},\mathrm{S}) > 2.6$ with leakage disabled]. This suggests memory error is the primary source of violation of logical non-Pauli-stochastic errors. As memory error has a coherent $Z$ component, we expect departures from the Pauli-stochastic model to appear primarily as logical coherent error, in particular over- or under-rotations of each of the logical gates. To test this hypothesis, we extract estimates of the over-rotation from the estimated Pauli transfer matrices. Figure~\ref{fig:overrotation_ramsey}(c) plots over-rotation $\Delta\theta$, the difference between the target and estimated rotation angle, in units of $\pi$ for each of the logical gates. Blue (orange) bars show memory error enabled (disabled). In each case, leakage is enabled. Negative values indicate an under-rotation. 

When memory error is included, we observe coherent over-rotations of comparable size for $S_L$, $I_L$, and $T_L$. The fitted values are $\Delta\theta = (0.0015 \pm 0.0001)\pi$ for $S_L$, $\Delta\theta = (0.0014 \pm 0.0001)\pi$ for $I_L$, and $\Delta\theta = (0.0016 \pm 0.0001)\pi$ for the non-Clifford $T_L$ gate. Despite the significant implementation difference for $T_L$, its fitted over-rotation is similar to those of $S_L$ and $I_L$. In contrast, $H_L$ has a notably smaller fitted over-rotation, $\Delta\theta = (0.0005 \pm 0.0001)\pi$.

Denoting the approximate observed logical over-rotation for $S_L$, $I_L$, and $T_L$ by $\phi_L$, we observe that the observed size of the logical over-rotation is $\phi_L/(\sqrt{2})^3$ for $H_L$. This is consistent with a physical coherent $Z$ error mechanism (e.g., memory error) of fixed size leading to an over-rotation of size $\phi$ for the physical $S$, $I$, and $T$ gates and size $\phi/\sqrt{2}$ for the physical $H$ gate. The smaller size for $H$ is expected because $H$ ideally rotates around the $X+Z$ axis, which is only partially aligned with the $Z$ error. Refs.~\cite{gutierrez2016errors, beale2018quantum} predict that physical over-rotations of size $\phi$ yield logical over-rotations of size $\phi_L=\mathcal{O}(\phi^3)$ at distance 3, which is consistent with what we see here. However, it is somewhat surprising that we see the same scaling for the non-Clifford $T_L$ gate, which has a significantly more complex implementation than the Clifford gates via state injection from an auxiliary logical qubit. 

Turning off the memory error leaves no significant over-rotation on the transversal gates [solid orange bars, Fig.~\ref{fig:overrotation_ramsey}(c)], supporting the interpretation that their coherent logical error originates from the physical memory error. However, the non-transversal gate still exhibits a coherent under-rotation $\Delta\theta=(-0.0002\pm 0.0001)\pi$, despite the presence of no other coherent physical error sources in the H1-1E simulation. We hypothesize this is due to the logical measurement and conditional $S_L\cdot X_L$ gate that are part of the teleportation gadget. If an uncorrectable $X$-type error precedes the logical measurement, then the measurement outcome will be decoded incorrectly and $S_L\cdot X_L$ will be erroneously applied or not applied, resulting in a stochastic Clifford error channel. Unlike a stochastic Pauli error channel, a stochastic Clifford channel can have a coherent error component, as shown in Appendix~\ref{sec: appendix Kevin's argument}.

To provide an independent probe of logical coherent error, we compare the logical GST over-rotation estimate for the $S_L$ gate with an estimate [Fig.~\ref{fig:overrotation_ramsey}(c)] obtained using a logical Ramsey experiment performed on the emulator with both memory error settings and leakage enabled [Fig.~\ref{fig:overrotation_ramsey}(d)]. Each circuit in the logical Ramsey experiment consists of preparation in $|+\rangle_L$ and logical $X$-basis measurement implemented using logical Hadamards, with linearly spaced numbers of repeated $S_L$ gates inserted between preparation and measurement, up to $p_{max}=128$ gate repetitions. Both logical Ramsey and logical GST experiments were performed with $N_{\mathrm{shots}}=14100$ shots per circuit. The inset in Fig.~\ref{fig:overrotation_ramsey}(d) shows that the Ramsey data exhibits a phase shift characteristic of a coherent over-rotation when memory error is enabled, but not when it is disabled. Fitting a decaying sinusoid to each dataset, we extract phase shifts of $\Delta\theta = (0.00139 \pm 0.00003)\pi$ for memory error enabled and $\Delta\theta = (0.00002 \pm 0.00007)\pi$ for memory error disabled. For both memory error settings, the $S_L$-gate Ramsey estimates [cross-hatched white bars, Fig.~\ref{fig:overrotation_ramsey}(c)] are consistent with the logical GST estimates [solid bars, Fig.~\ref{fig:overrotation_ramsey}(c)].

In datasets generated using H1-1E, the logical GST estimates show over-rotations on all logical gates, and the logical Ramsey experiment independently resolves the $S_L$-gate over-rotation. Furthermore, both analyses point towards memory error as the primary source of logical coherent error. Since this physical error source is almost certainly \textit{also} an appreciable error source on on H1-1, the emulator provides a concrete mechanism by which coherent physical noise can produce coherent logical error in the experiment. Additionally, while the H1-1 results are consistent with a Pauli-stochastic error model, model selection found that a small number of coherent parameters improve the fit without evidence of overfitting. Although these coherent parameters describe axis-misalignment, not over-rotation, they still provide evidence that logical coherent error mechanisms provide additional causal predictive power. Together, these results suggest that had we been able to take more samples, we may have observed conclusive evidence for logical coherent errors on H1-1.

\section{Automated model selection}\label{sec:model_selection}

\begin{figure*}
    \centering
    \includegraphics[width=1\linewidth]{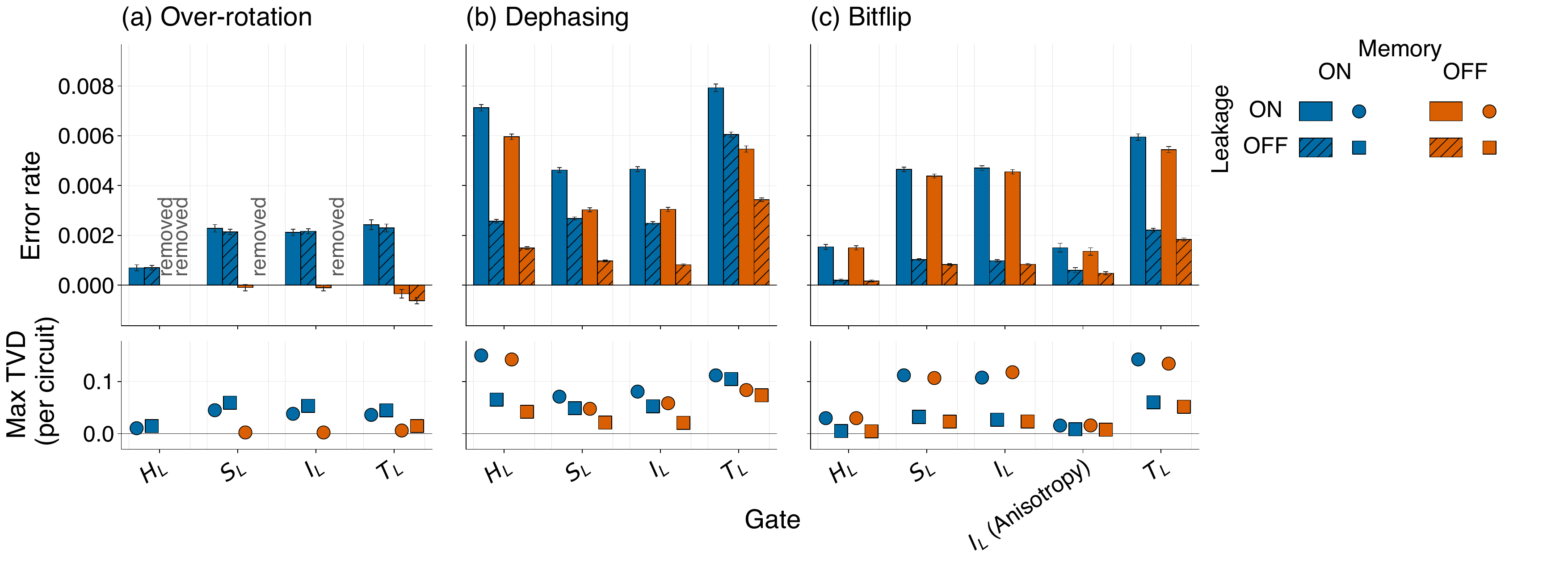}
    \caption{Automated model selection (AMS) results for H1-1E simulated datasets under the four combinations of memory and leakage error settings. Shown are the 13 FOGI parameters in the resulting reduced models with the largest maximum per-circuit TVD influence, grouped by qualitative error type: (a) on-axis over-rotations, (b) dephasing errors, and (c) generalized bitflips. Within each labeled panel, the upper plot shows the reduced-model absolute error-rate estimate, with error bars showing $2\sigma$ confidence intervals, and the lower plot shows the corresponding maximum per-circuit TVD. Memory settings are encoded by color, while leakage settings are encoded by hatching and marker shape. By imperfect analogy to the physical case, the logical idle $I_L$ is treated as an ideal $Z_L(\theta)$ operation with $\theta = 0$. For the idle, bitflip anisotropy is also observable and quantifies the asymmetry between the $X$- and $Y$-dephasing rates. The AMS over-rotation estimates are qualitatively consistent with those obtained from the CPTP model [Fig.~\ref{fig:overrotation_ramsey}(c)]. Additional details on the model selection algorithm and FOGI parameterization are given in Appendix~\ref{app:model_selection}, which also includes a plot of the FOGIs not shown here (Fig.~\ref{fig:AMS_remaining_params}).}
    \label{fig:AMS}
\end{figure*}

In this work we have considered a variety of logical gate set error models with different parameterizations to fit the empirical data. Underlying this is the goal to make inferences about the logical dynamics that produced the data. To do such inference we follow the principle of parsimony, i.e. we aim to find a model with the least number of parameters which adequately matches the data. For any given data set it is the case that even unnecessary parameters can fit statistical noise (overfitting), and thus improve the apparent quality of a statistical fit. These spurious parameters reduce the predictive power of the model with respect to future data, however. 
 
In the case of the experimental logical GST data we found that the logical dynamics are well-modeled by a few-parameter Pauli-stochastic model. For the simulated data, however, we found nontrivial value from using larger parameterizations (e.g. ``CPTP"). The jump from Pauli-stochastic to CPTP is a large one, so to identify the smaller subset of parameters actually needed, we implement an \emph{automated} version of model selection. Automated model selection (AMS) explores the space of candidate noise models by following a greedy tree-based search heuristic to obtain a parsimonious model. As argued in Sec. \ref{sec:pauli_stochastic_model}, gauge-freedom can cause the AMS heuristic to get stuck in local minima, so we use a FOGI-parameterized model. In addition, the FOGI representation allows for easier interpretation of the errors in the simulated datasets, as described in Appendix \ref{app:model_selection}. A basis of FOGI parameters for the logical gate set used here is given in Tables~\ref{tab:FOGI-error-generator-combinations-short} and \ref{tab:FOGI-error-generator-combinations-long}. To perform AMS, we construct a tree of candidate reduced models corresponding to the iterative removal of a single parameter at a time, and perform a greedy breadth-first search choosing the reduced model at each step which is best using the AIC. This process continues until none of the candidate reduced models are accepted by the AIC. 

\begin{figure*}
    \centering
    \includegraphics[width=1\linewidth]{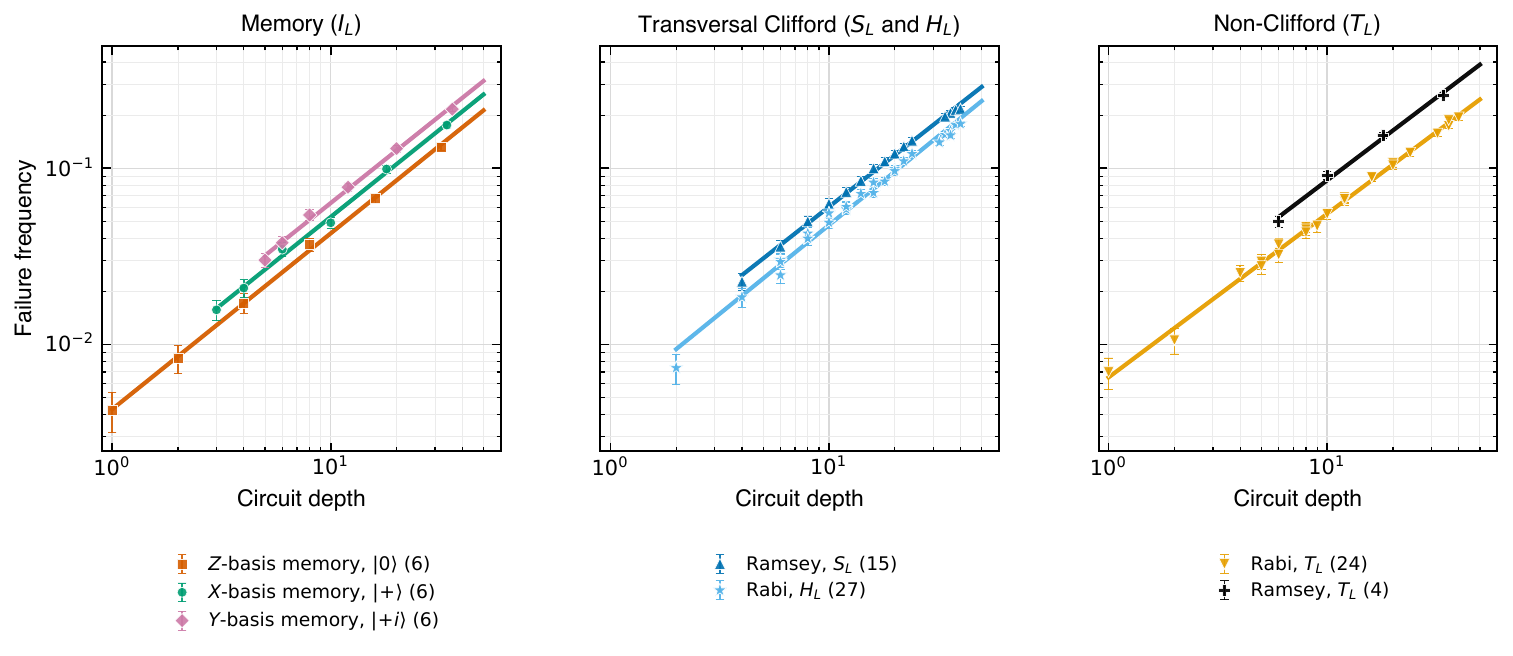}
    \caption{Failure frequency as a function of circuit depth for all definite-outcome circuits in the logical GST simulations using the default (“ON/ON”) emulator settings. Shown are (a) memory experiments, (b) transversal single-qubit Clifford Ramsey and Rabi experiments, and (c) non-Clifford $T_L$ Ramsey and Rabi experiments. Line fits are intended to guide the eye. The circuit classes show broadly similar depth scaling, but with systematic offsets in correct-outcome frequency at fixed depth, indicating class-dependent susceptibility to logical failure due to coherent and biased Pauli logical error. Small differences in slope are also present, suggesting modest variation in error accumulation with depth.} 
    \label{fig:dataset_fitting}
\end{figure*}

In Figure~\ref{fig:AMS}, we present the reduced models found by the AMS algorithm on all four simulated datasets side by side. For every emulator configuration, AMS removes at least 22 out of 43 parameters, with 15 parameters removed in all four cases. Thus, although the full model allows many possible logical error mechanisms, only a relatively small subset is needed to explain the simulated logical GST data. Most of the retained parameters are also simple. They mainly describe errors associated with individual gates, rather than SPAM errors or errors with support on more than one gate, e.g. axis misalignment error. Appendix~\ref{app:model_selection} gives more details on the classification of FOGI quantities. For readability, Figure~\ref{fig:AMS} only displays the 13 parameters with the greatest impact on the total variation distance (TVD) between the model's predicted outcome probabilities and the data; the remaining parameters are plotted in Appendix Fig.~\ref{fig:AMS_remaining_params} in Appendix~\ref{app:model_selection}. These 13 parameters can be categorized into two stochastic error sources, generalized bitflips (with respect to a gate's eigenstates\footnote{A generalized bitflip error is the analogue of an ordinary bit flip, but defined with respect to the eigenstates of the gate being studied. Thus, for a $Z$ gate it mixes $|0\rangle$ and $|1\rangle$, while for an $X$ gate it mixes $|+\rangle$ and $|-\rangle$ instead.}) and dephasing, and a single coherent error source, over-rotation. For the purposes of this classification, we treat the logical idle $I_L$ as $Z_L(\theta)$ where $\theta=0$. Additionally, for $I_L$ we include the bitflip anisotropy, which measures the difference between the $X$ and $Y$ dephasing rates. We do not report AMS on the experimental data as the low shot count limits the ability to reliably resolve between many fine-grained candidate models.

The AMS results reveal an unsurprising correlation between leakage and both Pauli-stochastic logical error sources, with a significant increase in the rate of generalized bitflips and dephasing when leakage is enabled. Furthermore, they corroborate the finding from the previous section connecting memory errors to logical coherent errors. First, we see that without memory error most coherent error parameters (over-rotation) are removed. This indicates that memory error is the primary source of violation of the Pauli-stochastic model. Second, the AMS results confirm that the parameter corresponding to over-rotation on $T_L$ is significant even when memory error is disabled. 

This reduced modeling is useful because predicting the behavior of generic logical circuits requires understanding which logical error mechanisms are actually present. Logical GST probes a broad range of circuit behaviors, with different circuits sensitive to different error mechanisms. 
Indeed, the definite-outcome circuits in Fig.~\ref{fig:dataset_fitting} show distinctly different failure frequencies across circuit classes, with effects becoming particularly pronounced at larger depths, although the overall scaling is simlar in each case. This behavior reflects the effects of coherent and biased Pauli logical error. AMS then compresses this information into the most parsimonious model that still predicts the data, identifying the parameters that are actually necessary. In the absence of model violation, the retained parameters isolate the logical error mechanisms present in the system.

\section{Discussion} \label{sec:discussion}
In this work, we establish logical gate set tomography as a tool for constructing detailed, predictive models of logical operations. As QEC experiments move from logical memory demonstrations to those of logical gates and small algorithmic circuits of increasing code distance, these models will become increasingly important for identifying and mitigating the features of logical noise that limit performance of fault-tolerant quantum processors. Figure~\ref{fig:dataset_fitting} illustrates this point: different logical circuit classes can exhibit diverging performance because they are sensitive to different effective error mechanisms, such as over-rotation or dephasing at the logical level. 

Applying logical GST to H1-1 produced Pauli-transfer matrix representations of a set of universal single-qubit gates. From further analysis, we found no evidence for logical non-Markovianity and evidence that logical coherent error terms may have causal explanatory power, although these conclusions are limited by the statistical resolution of the data, as expected from the small total number of logical shots. In contrast, higher-shot simulated data from Quantinuum's detailed system emulator resolved strong evidence of coherent logical error and weak evidence of logical non-Markovianity. Furthermore, these results illustrate that coherent and non-Markovian logical noise need not arise from corresponding physical-qubit error mechanisms, but can emerge from the QEC implementation itself.

It is worth noting that that these results reflect one qubit modality and code implementation. Changing any of these factors may change the observed logical error dynamics. In particular, passive error correction through Pauli-frame updates could mitigate the constructive build-up of coherent physical errors and spatial or temporal correlations more than active error correction, so a comparable implementation using passive error correction might observe reduced coherent logical error or non-Markovianity \cite{witzel2026correcting}. Furthermore, code distance can greatly shape logical error character. In particular, distance 3 is subject to finite-size effects that make logical coherent error more likely than even slightly higher distance implementations \cite{beale2018quantum, bravyi2018correcting}. 

This dependence on implementation motivates applying logical GST more broadly.  Much as QCVV protocols implemented on physical qubits are designed to be hardware-agnostic, logical GST is intended to be agnostic to both the platform choice and the chosen QEC architecture, including the code choice, code distance, logical gate implementation, and decoder. This portability makes it a useful tool for studying when, how, and to what extent deviations from the logical Pauli-stochastic noise model arise. By applying logical GST to two logical qubits, it would also be possible to characterize a universal logical gate set and enable detailed prediction of near-term algorithmic performance. 

\section*{AI Usage Statement}
During the preparation of this manuscript, the authors used AI-based tools to assist in generating code for analysis and visualization and to support manuscript outlining, drafting, and prose development. The authors critically reviewed, validated, and edited all AI-assisted outputs, including code and text, and take full responsibility for the accuracy, integrity, and final content of the work.

\section*{Acknowledgments}
The authors would like to thank Janine Bennett and Setso Metodi for programmatic assistance; and Dan Gresh, Kevin Gilmore, Nathan Hewitt, Craig Holliman, Alistair Milne, and Nikhil Kotibhaskar for assistance running circuits on Quantinuum H1-1; and Erik Nielsen, Setso Metodi, Mario Morford-Oberst, Julie Campos, Aliza Siddiqui, and Stefan Seritan for early conversations influential to the development of logical GST. This research used resources of the Oak Ridge Leadership Computing Facility at the Oak Ridge National Laboratory, which is supported by the Advanced Scientific Computing Research programs in the Office of Science of the U.S. Department of Energy under Contract No. DE-AC05-00OR22725. The research is based upon work supported in part by the Office of the Director of National Intelligence (ODNI), Intelligence Advanced Research Projects Activity (IARPA), specifically the ELQ program, and the Laboratory Directed Research and Development program at Sandia National Laboratories. Sandia National Laboratories is a multimission laboratory managed and operated by National Technology \& Engineering Solutions of Sandia, LLC, a wholly owned subsidiary of Honeywell International Inc., for the U.S. Department of Energy’s National Nuclear Security Administration under contract DE-NA0003525. This paper describes objective technical results and analysis. The views and conclusions contained herein are those of the authors and should not be interpreted as necessarily representing the official policies or endorsements, either expressed or implied, of the ODNI, IARPA, or the U.S. Government. The U.S. Government is authorized to reproduce and distribute reprints for Governmental purposes notwithstanding any copyright annotation thereon.

\section*{Author Contributions}
P.W. led manuscript preparation, drafted the main text, and synthesized, integrated, and refined contributions from the broader author team, including appendix contributions from C.B., R.B.-K., N.B., J.H., J.M.G., K.M., K.R., and J.Z. into the final manuscript. K.R. conceived of the original research project. P.W., K.R., C.O., K.Y., J.H., and R.B.-K. developed and refined the scientific framing, research questions, interpretation of results, and manuscript narrative. P.W., K.R., K.M., C.B., and R.B.-K. jointly developed the logical GST protocol and its adaptation to hardware. C.R.-A., C.B., P.W., and K.R. implemented the code for circuit generation and verified its correctness. C.B. and K.M. executed experiments on the quantum computer and collected data. N.B. ran high-performance-computing-based simulations. P.W. led the analysis of experimental and simulated data, with supplemental analyses by J.G.M., J.H., J.Z., T.P., K.R, and K.Y. D.H. and K.R. secured access to quantum-computing resources through proposal development. K.R. acquired funding and provided overall project supervision. K.R. supervised P.W. and J.Z.; C.O. supervised J.M.G.; R.B.-K. served as formal advisor of record for P.W., J.Z., and J.M.G. All authors reviewed and edited the manuscript. 
\FloatBarrier
\bibliographystyle{apsrev4-1}
\bibliography{LoGST}

@article{besedin2026lattice,
	author = {Besedin, Ilya and Kerschbaum, Michael and Knoll, Jonathan and Hesner, Ian and B{\"o}deker, Lukas and Colmenarez, Luis and Hofele, Luca and Lacroix, Nathan and Hellings, Christoph and Swiadek, Fran{\c c}ois and Flasby, Alexander and Bahrami Panah, Mohsen and Colao Zanuz, Dante and M{\"u}ller, Markus and Wallraff, Andreas},
	date = {2026/02/01},
	doi = {10.1038/s41567-025-03090-6},
	id = {Besedin2026},
	isbn = {1745-2481},
	journal = {Nature Physics},
	number = {2},
	pages = {189--194},
	title = {Lattice surgery realized on two distance-three repetition codes with superconducting qubits},
	volume = {22},
	year = {2026}}

@misc{rosenfeld2025magic,
      title={Magic state cultivation on a superconducting quantum processor}, 
      author={Emma Rosenfeld and others},
      year={2025},
      eprint={2512.13908},
      archivePrefix={arXiv},
      primaryClass={quant-ph},
      url={https://arxiv.org/abs/2512.13908}, 
}

@misc{mayer2024benchmarking,
      title={Benchmarking logical three-qubit quantum Fourier transform encoded in the Steane code on a trapped-ion quantum computer}, 
      author={Karl Mayer and others},
      year={2024},
      eprint={2404.08616},
      archivePrefix={arXiv},
      primaryClass={quant-ph} 
}

@article{bluvstein2026a,
	author = {Bluvstein, Dolev and others},
	date = {2026/01/01},
	doi = {10.1038/s41586-025-09848-5},
	id = {Bluvstein2026},
	isbn = {1476-4687},
	journal = {Nature},
	number = {8095},
	pages = {39--46},
	title = {A fault-tolerant neutral-atom architecture for universal quantum computation},
	volume = {649},
	year = {2026}}

@article{Pino2020,
  title={Demonstration of the trapped-ion quantum CCD computer architecture},
  author={Pino, J. M. and Dreiling, J. M. and Figgatt, C. and Gaebler, J. P. and Moses, S. A. and Allman, M. S. and Baldwin, C. H. and Foss-Feig, M. and Hayes, D. and Mayer, K. and Ryan-Anderson, C. and Neyenhuis, B.},
  journal={Nature},
  volume = {592},
  pages = {209-213},
  year = {2021},
  month = {April},
  doi = {10.1038/s41586-021-03318-4}
}

@article{ryananderson2021realization,
  title = {Realization of Real-Time Fault-Tolerant Quantum Error Correction},
  author = {Ryan-Anderson, C. and Bohnet, J. G. and Lee, K. and Gresh, D. and Hankin, A. and Gaebler, J. P. and Francois, D. and Chernoguzov, A. and Lucchetti, D. and Brown, N. C. and Gatterman, T. M. and Halit, S. K. and Gilmore, K. and Gerber, J. A. and Neyenhuis, B. and Hayes, D. and Stutz, R. P.},
  journal = {Phys. Rev. X},
  volume = {11},
  issue = {4},
  pages = {041058},
  numpages = {29},
  year = {2021},
  month = {Dec},
  publisher = {American Physical Society},
  doi = {10.1103/PhysRevX.11.041058}
}

@article{iverson2020coherence,
  author = {Joseph K. Iverson and John Preskill},
  title = {Coherence in logical quantum channels},
  journal = {New Journal of Physics},
  volume = {22},
  pages = {073066},
  year = {2020},
  month = {July},
  publisher = {IOP Publishing Ltd on behalf of the Institute of Physics and Deutsche Physikalische Gesellschaft},
  doi = {10.1088/1367-2630/ab8e5c},
  note = {Open access}
}

@article{bravyi2018correcting,
	author = {Bravyi, Sergey and Englbrecht, Matthias and K{\"o}nig, Robert and Peard, Nolan},
	date = {2018/10/31},
	doi = {10.1038/s41534-018-0106-y},
	id = {Bravyi2018},
	isbn = {2056-6387},
	journal = {npj Quantum Information},
	number = {1},
	pages = {55},
	title = {Correcting coherent errors with surface codes},
	volume = {4},
	year = {2018}}

@article{beale2018quantum,
  title = {Quantum Error Correction Decoheres Noise},
  author = {Beale, Stefanie J. and Wallman, Joel J. and Guti\'errez, Mauricio and Brown, Kenneth R. and Laflamme, Raymond},
  journal = {Phys. Rev. Lett.},
  volume = {121},
  issue = {19},
  pages = {190501},
  numpages = {6},
  year = {2018},
  month = {Nov},
  publisher = {American Physical Society},
  doi = {10.1103/PhysRevLett.121.190501},
  url = {https://link.aps.org/doi/10.1103/PhysRevLett.121.190501}
}

@misc{ceasura2022nonexponential,
      title={Non-Exponential Behaviour in Logical Randomized Benchmarking}, 
      author={Athena Ceasura and Pavithran Iyer and Joel J. Wallman and Hakop Pashayan},
      year={2022},
      eprint={2212.05488},
      archivePrefix={arXiv},
      primaryClass={quant-ph}
}

@misc{endo2025nonmarkovianity,
      title={Non-Markovianity in Quantum Information Processing: Interplay with Quantum Error Mitigation}, 
      author={Suguru Endo and Hideaki Hakoshima and Tomohiro Shitara},
      year={2025},
      eprint={2510.20224},
      archivePrefix={arXiv},
      primaryClass={quant-ph}
}

@misc{ziyad2025emergent,
      title={Emergent Non-Markovianity in Logical Qubit Dynamics}, 
      author={Jalan A. Ziyad and Robin Blume-Kohout and Kenneth Rudinger},
      year={2025},
      eprint={2512.08893},
      archivePrefix={arXiv},
      primaryClass={quant-ph}
}

@misc{kwiatkowski2025constructing,
      title={Constructing an approximate logical Markovian model of consecutive QEC cycles of a stabilizer code}, 
      author={Alex Kwiatkowski and Aaron J. Friedman and Shawn Geller and Jalan A. Ziyad and Scott Glancy and Emanuel Knill},
      year={2025},
      eprint={2509.16887},
      archivePrefix={arXiv},
      primaryClass={quant-ph}
}

@article{Greenbaum2017modeling,
doi = {10.1088/2058-9565/aa9a06},
year = {2017},
month = {dec},
publisher = {IOP Publishing},
volume = {3},
number = {1},
pages = {015007},
author = {Greenbaum, Daniel and Dutton, Zachary},
title = {Modeling coherent errors in quantum error correction},
journal = {Quantum Science and Technology}
}

@misc{witzel2026correcting,
      title={Correcting coherent quantum errors by going with the flow}, 
      author={Wayne M. Witzel and Anand Ganti and Tzvetan S. Metodi},
      year={2026},
      eprint={2602.21076},
      archivePrefix={arXiv},
      primaryClass={quant-ph}
}

@misc{chang2025taming,
      title={Taming coherent noise with teleportation}, 
      author={Kathleen Chang and Qile Su and Shruti Puri},
      year={2025},
      eprint={2508.04947},
      archivePrefix={arXiv},
      primaryClass={quant-ph},
      url={https://arxiv.org/abs/2508.04947}, 
}

@article{cheng2025emergent,
  title = {Emergent Unitary Designs for Encoded Qubits from Coherent Errors and Syndrome Measurements},
  author = {Cheng, Zihan and Huang, Eric and Khemani, Vedika and Gullans, Michael J. and Ippoliti, Matteo},
  journal = {PRX Quantum},
  volume = {6},
  issue = {3},
  pages = {030333},
  numpages = {22},
  year = {2025},
  month = {Aug},
  publisher = {American Physical Society},
  doi = {10.1103/bnld-2chd},
  url = {https://link.aps.org/doi/10.1103/bnld-2chd}
}

@misc{combes2017logical,
      title={Logical Randomized Benchmarking}, 
      author={Joshua Combes and Christopher Granade and Christopher Ferrie and Steven T. Flammia},
      year={2017},
      eprint={1702.03688},
      archivePrefix={arXiv},
      primaryClass={quant-ph}
}

@article{lacroix2025scaling,
  author = {Google Quantum AI and Collaborators},
  title = {Scaling and logic in the colour code on a superconducting quantum processor},
  journal = {Nature},
  volume = {645},
  pages = {614--619},
  year = {2025},
  doi = {10.1038/s41586-025-09061-4}
}

@misc{dasu2026computing,
      title={Computing with many encoded logical qubits beyond break-even}, 
      author={Shival Dasu and others},
      year={2026},
      eprint={2602.22211},
      archivePrefix={arXiv},
      primaryClass={quant-ph},
      url={https://arxiv.org/abs/2602.22211}, 
}

@misc{berthusen2026experimental,
      title={Experimental validation of a compact fault-tolerant architecture for trapped ions}, 
      author={Noah Berthusen and Ali Lavasani and Asmae Benhemou and M. S. Allman and Joan Dreiling and Brian Estey and Cameron Foltz and Trent Jacobs and Michael Mills and Annie Jihyun Park and Adam P. Reed and David Hayes and Tzvetan S. Metodi and Andrew C. Potter},
      year={2026},
      eprint={2609.03194},
      archivePrefix={arXiv},
      primaryClass={quant-ph},
      url={https://arxiv.org/abs/2609.03194}, 
}

@article{daguerre2025experimental,
  title = {Experimental Demonstration of High-Fidelity Logical Magic States from Code Switching},
  author = {Daguerre, Lucas and Blume-Kohout, Robin and Brown, Natalie C. and Hayes, David and Kim, Isaac H.},
  journal = {Phys. Rev. X},
  volume = {15},
  issue = {4},
  pages = {041008},
  numpages = {11},
  year = {2025},
  month = {Oct},
  publisher = {American Physical Society},
  doi = {10.1103/dck4-x9c2}}

@article{reinhold2020error-corrected,
	author = {Reinhold, Philip and Rosenblum, Serge and Ma, Wen-Long and Frunzio, Luigi and Jiang, Liang and Schoelkopf, Robert J.},
	date = {2020/08/01},
	doi = {10.1038/s41567-020-0931-8},
	id = {Reinhold2020},
	isbn = {1745-2481},
	journal = {Nature Physics},
	number = {8},
	pages = {822--826},
	title = {Error-corrected gates on an encoded qubit},
	url = {https://doi.org/10.1038/s41567-020-0931-8},
	volume = {16},
	year = {2020}}

@article{zhang2026universal,
	author = {Zhang, Chunhui and Xu, Feng and Zhang, Shihang and Duan, Mingchao and Zhong, Dupeng and Bai, Xuesong and Wang, Hao and Huang, Chao and Deng, Yi and Gao, Miao and Zhang, Yu-Ning and Liu, Jiaze and Li, Chunhui and Jiang, Yan and Zhao, Baolong and Shu, Huan and Wu, Kunrong and Shi, Keji and Ding, Qiming and Tian, Zhen and Wang, Guanyong and Yuan, Xiao and Xin, Tao and Hu, Guangchong and Liu, Song and Pan, Tianluo and Huang, Peihao and He, Yu and Yu, Dapeng},
	date = {2026/03/23},
	doi = {10.1038/s41565-026-02140-1},
	id = {Zhang2026},
	isbn = {1748-3395},
	journal = {Nature Nanotechnology},
	title = {Universal logical operations in a silicon quantum processor},
	year = {2026}}

@article{gutierrez2016errors,
  title = {Errors and pseudothresholds for incoherent and coherent noise},
  author = {Guti\'errez, Mauricio and Smith, Conor and Lulushi, Livia and Janardan, Smitha and Brown, Kenneth R.},
  journal = {Phys. Rev. A},
  volume = {94},
  issue = {4},
  pages = {042338},
  numpages = {14},
  year = {2016},
  month = {Oct},
  publisher = {American Physical Society},
  doi = {10.1103/PhysRevA.94.042338}}

@article{Fazio2025logical,
  doi = {10.22331/q-2025-06-24-1779},
  url = {https://doi.org/10.22331/q-2025-06-24-1779},
  title = {Logical {N}oise {B}ias in {M}agic {S}tate {I}njection},
  author = {Fazio, Nicholas and Harper, Robin and Bartlett, Stephen D.},
  journal = {{Quantum}},
  issn = {2521-327X},
  publisher = {{Verein zur F{\"{o}}rderung des Open Access Publizierens in den Quantenwissenschaften}},
  volume = {9},
  pages = {1779},
  month = jun,
  year = {2025}
}

@article{merkel2013selfconsistent,
  title = {Self-consistent quantum process tomography},
  author = {Merkel, Seth T. and Gambetta, Jay M. and Smolin, John A. and Poletto, Stefano and C\'orcoles, Antonio D. and Johnson, Blake R. and Ryan, Colm A. and Steffen, Matthias},
  journal = {Phys. Rev. A},
  volume = {87},
  issue = {6},
  pages = {062119},
  numpages = {9},
  year = {2013},
  month = {Jun},
  publisher = {American Physical Society},
  doi = {10.1103/PhysRevA.87.062119}}

@article{Nielsen2021gatesettomography,
  doi = {10.22331/q-2021-10-05-557},
  title = {Gate Set Tomography},
  author = {Nielsen, Erik and Gamble, John King and Rudinger, Kenneth and Scholten, Travis and Young, Kevin and Blume-Kohout, Robin},
  journal = {{Quantum}},
  issn = {2521-327X},
  publisher = {{Verein zur F{\"{o}}rderung des Open Access Publizierens in den Quantenwissenschaften}},
  volume = {5},
  pages = {557},
  month = {October},
  year = {2021}
  }

@article{blume-kohout2017,
	author = {Blume-Kohout, Robin and Gamble, John King and Nielsen, Erik and Rudinger, Kenneth and Mizrahi, Jonathan and Fortier, Kevin and Maunz, Peter},
	journal = {Nature Communications},
	number = {1},
	pages = {14485},
	title = {Demonstration of qubit operations below a rigorous fault tolerance threshold with gate set tomography},
	volume = {8},
	year = {2017}}

@article{Govia2020bootstrapping,
	author = {Govia, L. C. G. and Ribeill, G. J. and Rist{\`e}, D. and Ware, M. and Krovi, H.},
	date = {2020/02/27},
	doi = {10.1038/s41467-020-14873-1},
	id = {Govia2020},
	isbn = {2041-1723},
	journal = {Nature Communications},
	number = {1},
	pages = {1084},
	title = {Bootstrapping quantum process tomography via a perturbative ansatz},
	url = {https://doi.org/10.1038/s41467-020-14873-1},
	volume = {11},
	year = {2020}}

@article{brieger2023compressive,
  title = {Compressive Gate Set Tomography},
  author = {Brieger, Raphael and Roth, Ingo and Kliesch, Martin},
  journal = {PRX Quantum},
  volume = {4},
  issue = {1},
  pages = {010325},
  numpages = {33},
  year = {2023},
  month = {Mar},
  publisher = {American Physical Society},
  doi = {10.1103/PRXQuantum.4.010325},
  url = {https://link.aps.org/doi/10.1103/PRXQuantum.4.010325}
}

@article{cao2024efficient,
  title = {Efficient Characterization of Qudit Logical Gates with Gate Set Tomography Using an Error-Free Virtual $Z$ Gate Model},
  author = {Cao, Shuxiang and Lall, Deep and Bakr, Mustafa and Campanaro, Giulio and Fasciati, Simone D. and Wills, James and Chidambaram, Vivek and Shteynas, Boris and Rungger, Ivan and Leek, Peter J.},
  journal = {Phys. Rev. Lett.},
  volume = {133},
  issue = {12},
  pages = {120802},
  numpages = {6},
  year = {2024},
  month = {Sep},
  publisher = {American Physical Society},
  doi = {10.1103/PhysRevLett.133.120802},
  url = {https://link.aps.org/doi/10.1103/PhysRevLett.133.120802}
}

@misc{wysocki2026detailed,
      title={Detailed, interpretable characterization of mid-circuit measurement on a transmon qubit}, 
      author={Piper C. Wysocki and Luke D. Burkhart and Madeline H. Morocco and Corey I. Ostrove and Riley J. Murray and Tristan Brown and Jeffrey M. Gertler and David K. Kim and Nathan E. Miller and Bethany M. Niedzielski and Katrina M. Sliwa and Robin Blume-Kohout and Gabriel O. Samach and Mollie E. Schwartz and Kenneth M. Rudinger},
      year={2026},
      eprint={2602.03938},
      archivePrefix={arXiv},

}

@article{Madzik2022precision,
	author = {Madzik, Mateusz T. and Asaad, Serwan and Youssry, Akram and Joecker, Benjamin and Rudinger, Kenneth M. and Nielsen, Erik and Young, Kevin C. and Proctor, Timothy J. and Baczewski, Andrew D. and Laucht, Arne and Schmitt, Vivien and Hudson, Fay E. and Itoh, Kohei M. and Jakob, Alexander M. and Johnson, Brett C. and Jamieson, David N. and Dzurak, Andrew S. and Ferrie, Christopher and Blume-Kohout, Robin and Morello, Andrea},
	date = {2022/01/01},
	doi = {10.1038/s41586-021-04292-7},
	id = {Madzik2022},
	isbn = {1476-4687},
	journal = {Nature},
	number = {7893},
	pages = {348--353},
	title = {Precision tomography of a three-qubit donor quantum processor in silicon},
	volume = {601},
	year = {2022}}

@article{Xue2022quantum,
	author = {Xue, Xiao and Russ, Maximilian and Samkharadze, Nodar and Undseth, Brennan and Sammak, Amir and Scappucci, Giordano and Vandersypen, Lieven M. K.},
	date = {2022/01/01},
	doi = {10.1038/s41586-021-04273-w},
	id = {Xue2022},
	isbn = {1476-4687},
	journal = {Nature},
	number = {7893},
	pages = {343--347},
	title = {Quantum logic with spin qubits crossing the surface code threshold},
	url = {https://doi.org/10.1038/s41586-021-04273-w},
	volume = {601},
	year = {2022}}

@misc{jones2026midcircuit,
      title={Mid-circuit logic executed in the qubit layer of a quantum processor}, 
      author={Cameron Jones and others},
      year={2026},
      eprint={2512.12648},
      archivePrefix={arXiv},
      primaryClass={quant-ph},
      url={https://arxiv.org/abs/2512.12648}, 
}

@article{rozanov2026benchmarking,
  title = {Benchmarking single-qubit gates on a neutral-atom quantum processor},
  author = {Rozanov, A. and Bantysh, B. and Bobrov, I. and Struchalin, G. and Straupe, S.},
  journal = {Phys. Rev. A},
  volume = {113},
  issue = {4},
  pages = {042603},
  numpages = {10},
  year = {2026},
  month = {Apr},
  publisher = {American Physical Society},
  doi = {10.1103/vbwh-sk7s},
  url = {https://link.aps.org/doi/10.1103/vbwh-sk7s}
}

@article{bartling2025universal,
  title = {Universal high-fidelity quantum gates for spin qubits in diamond},
  author = {Bartling, H.P. and Yun, J. and Schymik, K.N. and van Riggelen, M. and Enthoven, L.A. and van Ommen, H.B. and Babaie, M. and Sebastiano, F. and Markham, M. and Twitchen, D.J. and Taminiau, T.H.},
  journal = {Phys. Rev. Appl.},
  volume = {23},
  issue = {3},
  pages = {034052},
  numpages = {26},
  year = {2025},
  month = {Mar},
  publisher = {American Physical Society},
  doi = {10.1103/PhysRevApplied.23.034052}
}

@article{hughes2020benchmarking,
  title = {Benchmarking a High-Fidelity Mixed-Species Entangling Gate},
  author = {Hughes, A. C. and Sch\"afer, V. M. and Thirumalai, K. and Nadlinger, D. P. and Woodrow, S. R. and Lucas, D. M. and Ballance, C. J.},
  journal = {Phys. Rev. Lett.},
  volume = {125},
  issue = {8},
  pages = {080504},
  numpages = {7},
  year = {2020},
  month = {Aug},
  publisher = {American Physical Society},
  doi = {10.1103/PhysRevLett.125.080504},
  url = {https://link.aps.org/doi/10.1103/PhysRevLett.125.080504}
}

@article{proctor2020detecting,
	author = {Proctor, Timothy and Revelle, Melissa and Nielsen, Erik and Rudinger, Kenneth and Lobser, Daniel and Maunz, Peter and Blume-Kohout, Robin and Young, Kevin},
	date = {2020/10/26},
	doi = {10.1038/s41467-020-19074-4},
	id = {Proctor2020},
	isbn = {2041-1723},
	journal = {Nature Communications},
	number = {1},
	pages = {5396},
	title = {Detecting and tracking drift in quantum information processors},
	url = {https://doi.org/10.1038/s41467-020-19074-4},
	volume = {11},
	year = {2020}}

@article{rudinger2021experimental,
  author = {Rudinger, Kenneth and Blume-Kohout, Robin and Gamble, John K. and Nielsen, Erik and Proctor, T. and Young, Kevin},
  title = {Experimental Characterization of Crosstalk Errors with Simultaneous Gate Set Tomography},
  journal = {PRX Quantum},
  volume = {2},
  pages = {040338},
  year = {2021},
  publisher = {American Physical Society},
  doi = {10.1103/PRXQuantum.2.040338},
}

@misc{piliouras2026highperformance,
      title={High-performance gates on trapped ion qubits using counterpropagating pulse-shaped laser beams}, 
      author={Evangelos Piliouras and Hisham Amer and Susan M. Clark and Melissa C. Revelle and Edward C. Tortorici and Matthew N. H. Chow and Brandon Ruzic and Daniel S. Lobser and Brian K. McFarland and Christopher G. Yale and Edwin Barnes and Sophia E. Economou},
      year={2026},
      eprint={2606.15672},
      archivePrefix={arXiv},
      primaryClass={quant-ph},
      url={https://arxiv.org/abs/2606.15672}, 
}

@article{Steane1996multiple,
  author = {Andrew Steane},
  title = {Multiple-particle interference and quantum error correction},
  journal = {Proceedings of the Royal Society A},
  volume = {452},
  number = {1954},
  pages = {2551--2577},
  year = {1996},
  month = {November},
  doi = {10.1098/rspa.1996.0136}
}

@article{Steane1996error,
  author = {A. M. Steane},
  title = {Error Correcting Codes in Quantum Theory},
  journal = {Physical Review Letters},
  volume = {77},
  pages = {793},
  year = {1996},
  month = {July},
  doi = {10.1103/PhysRevLett.77.793}
}

@article{Reichardt_2021,
doi = {10.1088/2058-9565/abc6f4},
url = {https://doi.org/10.1088/2058-9565/abc6f4},
year = {2020},
month = {nov},
publisher = {IOP Publishing},
volume = {6},
number = {1},
pages = {015007},
author = {Reichardt, Ben W},
title = {Fault-tolerant quantum error correction for Steane’s seven-qubit color code with few or no extra qubits},
journal = {Quantum Science and Technology},
}

@article{chamberland2019fault-tolerant,
  doi = {10.22331/q-2019-05-20-143},
  url = {https://doi.org/10.22331/q-2019-05-20-143},
  title = {Fault-tolerant magic state preparation with flag qubits},
  author = {Chamberland, Christopher and Cross, Andrew W.},
  journal = {{Quantum}},
  issn = {2521-327X},
  publisher = {{Verein zur F{\"{o}}rderung des Open Access Publizierens in den Quantenwissenschaften}},
  volume = {3},
  pages = {143},
  month = {may},
  year = {2019}
}

@software{Pecos,
author = {Ciaran Ryan-Anderson},
title = {Pecos: Performance estimator of codes on surfaces},
year = {2018},
url = {https://github.com/PECOS-packages/PECOS}}

@article{rahn2002exact,
  title = {Exact performance of concatenated quantum codes},
  author = {Rahn, Benjamin and Doherty, Andrew C. and Mabuchi, Hideo},
  journal = {Phys. Rev. A},
  volume = {66},
  issue = {3},
  pages = {032304},
  numpages = {13},
  year = {2002},
  month = {Sep},
  publisher = {American Physical Society},
  doi = {10.1103/PhysRevA.66.032304},
  url = {https://link.aps.org/doi/10.1103/PhysRevA.66.032304}
}

@INPROCEEDINGS{ostrove2023near,
author={Ostrove, Corey and Rudinger, Kenneth and Seritan, Stefan and Young, Kevin and Blume-Kohout, Robin},
  booktitle={2023 IEEE International Conference on Quantum Computing and Engineering (QCE)}, 
  title={Near-Minimal Gate Set Tomography Experiment Designs}, 
  year={2023},
  volume={01},
  number={},
  pages={1422-1432},
  doi={10.1109/QCE57702.2023.00161}}

@misc{osti_code-28250,
  author       = {Nielsen, Erik and others},
  title        = {Python GST Implementation (PyGSTi) v. 0.9},
  doi          = {10.11578/dc.20190722.2},
  place        = {United States},
  year         = {2019},
  month        = {07}}

@article{Nielsen2020,
doi = {10.1088/2058-9565/ab8aa4},
year = {2020},
month = {jul},
publisher = {IOP Publishing},
volume = {5},
number = {4},
pages = {044002},
author = {Nielsen, Erik and Rudinger, Kenneth and Proctor, Timothy and Russo, Antonio and Young, Kevin and Blume-Kohout, Robin},
title = {Probing quantum processor performance with pyGSTi},
journal = {Quantum Science and Technology}
}

@misc{rudinger2023two,
      title={Two-Qubit Gate Set Tomography with Fewer Circuits}, 
      author={Kenneth M. Rudinger and Corey I. Ostrove and Stefan K. Seritan and Matthew D. Grace and Erik Nielsen and Robin J. Blume-Kohout and Kevin C. Young},
      year={2023},
      eprint={2307.15767},
      archivePrefix={arXiv},
      primaryClass={quant-ph}
}

@article{hashim2025practical, title = {Practical Introduction to Benchmarking and Characterization of Quantum Computers}, author = {Hashim, Akel and others}, journal = {PRX Quantum}, volume = {6}, issue = {3}, pages = {030202}, numpages = {132}, year = {2025}, month = {Aug}, publisher = {American Physical Society}, doi = {10.1103/PRXQuantum.6.030202}}

@online{emulator_docs,
  author       = {{Quantinuum}},
  title        = {Emulator User Guide},
  year         = {2025},
  url          = {https://docs.quantinuum.com/systems/user_guide/emulator_user_guide/introduction.html},
  urldate      = {2026-09-10},
}

@article{moses2023a,
  title = {A Race-Track Trapped-Ion Quantum Processor},
  author = {Moses, S. A. and others},
  journal = {Phys. Rev. X},
  volume = {13},
  issue = {4},
  pages = {041052},
  numpages = {25},
  year = {2023},
  month = {Dec},
  publisher = {American Physical Society},
  doi = {10.1103/PhysRevX.13.041052},
  url = {https://link.aps.org/doi/10.1103/PhysRevX.13.041052}
}

@article{ransford2026a,
	author = {Ransford, Anthony and others},
	date = {2026/07/01},
	doi = {10.1038/s41586-026-10676-4},
	id = {Ransford2026},
	isbn = {1476-4687},
	journal = {Nature},
	number = {8121},
	pages = {81--86},
	title = {A 98-qubit trapped-ion quantum computer with all-to-all connectivity},
	url = {https://doi.org/10.1038/s41586-026-10676-4},
	volume = {655},
	year = {2026}}

@article{wilksthm,
author = {S. S. Wilks},
title = {{The Large-Sample Distribution of the Likelihood Ratio for Testing Composite Hypotheses}},
volume = {9},
journal = {The Annals of Mathematical Statistics},
number = {1},
publisher = {Institute of Mathematical Statistics},
pages = {60 -- 62},
year = {1938},
doi = {10.1214/aoms/1177732360},
URL = {https://doi.org/10.1214/aoms/1177732360}
}

@misc{blumekohout2020wildcard,
      title={Wildcard error: Quantifying unmodeled errors in quantum processors}, 
      author={Robin Blume-Kohout and Kenneth Rudinger and Erik Nielsen and Timothy Proctor and Kevin Young},
      year={2020},
      eprint={2012.12231},
      archivePrefix={arXiv},
      primaryClass={quant-ph}}

@article{hashim2023benchmarking,
  author = {Akel Hashim and others},
  title = {Benchmarking quantum logic operations relative to thresholds for fault tolerance},
  journal = {npj Quantum Information},
  volume = {9},
  pages = {109},
  year = {2023},
  publisher = {Nature Publishing Group},
  doi = {10.1038/s41534-023-00764-y}
}

@article{blume-kohout2022a,
  title = {A Taxonomy of Small Markovian Errors},
  author = {Blume-Kohout, Robin and da Silva, Marcus P. and Nielsen, Erik and Proctor, Timothy and Rudinger, Kenneth and Sarovar, Mohan and Young, Kevin},
  journal = {PRX Quantum},
  volume = {3},
  issue = {2},
  pages = {020335},
  numpages = {21},
  year = {2022},
  month = {May},
  publisher = {American Physical Society},
  doi = {10.1103/PRXQuantum.3.020335}
}

@article{Nielsen2021efficient,
   title={Efficient flexible characterization of quantum processors with nested error models},
   volume={23},
   ISSN={1367-2630},
   DOI={10.1088/1367-2630/ac20b9},
   number={9},
   journal={New Journal of Physics},
   publisher={IOP Publishing},
   author={Nielsen, Erik and Rudinger, Kenneth and Proctor, Timothy and Young, Kevin and Blume-Kohout, Robin},
   year={2021},
   month=sep, pages={093020} }

@article{PORTET2020111,
title = {A primer on model selection using the Akaike Information Criterion},
journal = {Infectious Disease Modelling},
volume = {5},
pages = {111-128},
year = {2020},
issn = {2468-0427},
doi = {https://doi.org/10.1016/j.idm.2019.12.010},
url = {https://www.sciencedirect.com/science/article/pii/S2468042719300508},
author = {Stéphanie Portet}
}

@mastersthesis{sinha2026efficient,
  author = {Sinha, Savar Dayal},
    title = {Efficient and SPAM-Robust Ansatz-Free Lindbladian Learning},
    school = {California Institute of Technology},
    year = {2026},
    url = {https://resolver.caltech.edu/CaltechTHESIS:04142026-214248334},
    type={Bachelor's Thesis},
    doi = {10.7907/HPRR-FQ68},
}

@misc{FOGI,
      title={A gauge-invariant theory of small Markovian errors in quantum gate sets}, 
      author={Juan Gonzalez De Mendoza and Corey Ostrove and Timothy Proctor and Kevin Young and Erik Nielsen and Robin Blume-Kohout},
      year={2026},
      eprint={2608.14891},
      archivePrefix={arXiv},
      primaryClass={quant-ph},
      url={https://arxiv.org/abs/2608.14891}, 
}

@article{Stemp2024scalable,
	author = {Stemp, Holly G. and others},
	date = {2024/09/28},
	doi = {10.1038/s41467-024-52795-4},
	id = {Stemp2024},
	isbn = {2041-1723},
	journal = {Nature Communications},
	number = {1},
	pages = {8415},
	title = {Tomography of entangling two-qubit logic operations in exchange-coupled donor electron spin qubits},
	volume = {15},
	year = {2024}}

@article{Gaebler2021,
  title = {Suppression of midcircuit measurement crosstalk errors with micromotion},
  author = {Gaebler, J. P. and Baldwin, C. H. and Moses, S. A. and Dreiling, J. M. and Figgatt, C. and Foss-Feig, M. and Hayes, D. and Pino, J. M.},
  journal = {Phys. Rev. A},
  volume = {104},
  issue = {6},
  pages = {062440},
  numpages = {12},
  year = {2021},
  month = {Dec},
  publisher = {American Physical Society},
  doi = {10.1103/PhysRevA.104.062440}
}

@article{Hayes2020,
   title={Eliminating Leakage Errors in Hyperfine Qubits},
   volume={124},
   ISSN={1079-7114},
   DOI={10.1103/physrevlett.124.170501},
   number={17},
   journal={Physical Review Letters},
   publisher={American Physical Society (APS)},
   author={Hayes, D. and Stack, D. and Bjork, B. and Potter, A. C. and Baldwin, C. H. and Stutz, R. P.},
   year={2020},
   month={April}}

@article{ryananderson2024teleport,
  title   = {High-fidelity teleportation of a logical qubit using transversal gates and lattice surgery},
  author  = {Ryan-Anderson, C. and others},
  journal = {Science},
  volume  = {385},
  pages   = {1327--1331},
  year    = {2024},
  doi     = {10.1126/science.adp6016}
}

@article{dehollain2016optimization,
   title={Optimization of a solid-state electron spin qubit using gate set tomography},
   volume={18},
   ISSN={1367-2630},
   url={http://dx.doi.org/10.1088/1367-2630/18/10/103018},
   DOI={10.1088/1367-2630/18/10/103018},
   number={10},
   journal={New Journal of Physics},
   publisher={IOP Publishing},
   author={Dehollain, Juan P and Muhonen, Juha T and Blume-Kohout, Robin and Rudinger, Kenneth M and Gamble, John King and Nielsen, Erik and Laucht, Arne and Simmons, Stephanie and Kalra, Rachpon and Dzurak, Andrew S and Morello, Andrea},
   year={2016},
   month=Oct, pages={103018} }

@article{lucero_high-fidelity_2008,
	title = {High-{Fidelity} {Gates} in a {Single} {Josephson} {Qubit}},
	volume = {100},
	copyright = {http://link.aps.org/licenses/aps-default-license},
	issn = {0031-9007, 1079-7114},
	url = {https://link.aps.org/doi/10.1103/PhysRevLett.100.247001},
	doi = {10.1103/PhysRevLett.100.247001},
	number = {24},
	urldate = {2026-08-28},
	journal = {Physical Review Letters},
	author = {Lucero, Erik and Hofheinz, M. and Ansmann, M. and Bialczak, Radoslaw C. and Katz, N. and Neeley, Matthew and O’Connell, A. D. and Wang, H. and Cleland, A. N. and Martinis, John M.},
	month = {June},
	year = {2008},
	pages = {247001}
}

@article{babu_state_2021,
	title = {State leakage during fast decay and control of a superconducting transmon qubit},
	volume = {7},
	issn = {2056-6387},
	url = {https://www.nature.com/articles/s41534-020-00357-z},
	doi = {10.1038/s41534-020-00357-z},
	number = {1},
	urldate = {2026-08-28},
	journal = {npj Quantum Information},
	author = {Babu, Aravind Plathanam and Tuorila, Jani and Ala-Nissila, Tapio},
	month = feb,
	year = {2021},
	pages = {30}
}

@article{yoshihara_decoherence_2006,
	title = {Decoherence of {Flux} {Qubits} due to 1 / f {Flux} {Noise}},
	volume = {97},
	copyright = {http://link.aps.org/licenses/aps-default-license},
	issn = {0031-9007, 1079-7114},
	url = {https://link.aps.org/doi/10.1103/PhysRevLett.97.167001},
	doi = {10.1103/PhysRevLett.97.167001},
	number = {16},
	urldate = {2026-08-28},
	journal = {Physical Review Letters},
	author = {Yoshihara, F. and Harrabi, K. and Niskanen, A. O. and Nakamura, Y. and Tsai, J. S.},
	month = oct,
	year = {2006},
	pages = {167001}
}

@article{slichter_measurement-induced_2012,
	title = {Measurement-{Induced} {Qubit} {State} {Mixing} in {Circuit} {QED} from {Up}-{Converted} {Dephasing} {Noise}},
	volume = {109},
	copyright = {http://link.aps.org/licenses/aps-default-license},
	issn = {0031-9007, 1079-7114},
	url = {https://link.aps.org/doi/10.1103/PhysRevLett.109.153601},
	doi = {10.1103/PhysRevLett.109.153601},
	number = {15},
	urldate = {2026-08-28},
	journal = {Physical Review Letters},
	author = {Slichter, D. H. and Vijay, R. and Weber, S. J. and Boutin, S. and Boissonneault, M. and Gambetta, J. M. and Blais, A. and Siddiqi, I.},
	month = oct,
	year = {2012},
	pages = {153601}
}

@article{taylor_relaxation_2007,
	title = {Relaxation, dephasing, and quantum control of electron spins in double quantum dots},
	volume = {76},
	copyright = {http://link.aps.org/licenses/aps-default-license},
	issn = {1098-0121, 1550-235X},
	url = {https://link.aps.org/doi/10.1103/PhysRevB.76.035315},
	doi = {10.1103/PhysRevB.76.035315},
	number = {3},
	urldate = {2026-08-28},
	journal = {Physical Review B},
	author = {Taylor, J. M. and Petta, J. R. and Johnson, A. C. and Yacoby, A. and Marcus, C. M. and Lukin, M. D.},
	month = jul,
	year = {2007},
	pages = {035315}
}

@article{wellstood_low-frequency_1987,
	title = {Low-frequency noise in dc superconducting quantum interference devices below 1 {K}},
	volume = {50},
	issn = {0003-6951, 1077-3118},
	url = {https://pubs.aip.org/apl/article/50/12/772/52092/Low-frequency-noise-in-dc-superconducting-quantum},
	doi = {10.1063/1.98041},
	number = {12},
	urldate = {2026-08-28},
	journal = {Applied Physics Letters},
	author = {Wellstood, Frederick C. and Urbina, Cristian and Clarke, John},
	month = {March},
	year = {1987},
	pages = {772--774},
}

@article{steffen_accurate_2003,
	title = {Accurate control of {Josephson} phase qubits},
	volume = {68},
	copyright = {http://link.aps.org/licenses/aps-default-license},
	issn = {0163-1829, 1095-3795},
	url = {https://link.aps.org/doi/10.1103/PhysRevB.68.224518},
	doi = {10.1103/PhysRevB.68.224518},
	number = {22},
	urldate = {2026-08-28},
	journal = {Physical Review B},
	author = {Steffen, Matthias and Martinis, John M. and Chuang, Isaac L.},
	month = {December},
	year = {2003},
	pages = {224518},
}

@article{medford_self-consistent_2013,
	title = {Self-consistent measurement and state tomography of an exchange-only spin qubit},
	volume = {8},
	copyright = {http://www.springer.com/tdm},
	issn = {1748-3387, 1748-3395},
	url = {https://www.nature.com/articles/nnano.2013.168},
	doi = {10.1038/nnano.2013.168},
	number = {9},
	urldate = {2026-08-28},
	journal = {Nature Nanotechnology},
	author = {Medford, J. and Beil, J. and Taylor, J. M. and Bartlett, S. D. and Doherty, A. C. and Rashba, E. I. and DiVincenzo, D. P. and Lu, H. and Gossard, A. C. and Marcus, C. M.},
	month = sep,
	year = {2013},
	pages = {654--659}}

@article{fogarty_nonexponential_2015,
	title = {Nonexponential fidelity decay in randomized benchmarking with low-frequency noise},
	volume = {92},
	copyright = {http://link.aps.org/licenses/aps-default-license},
	issn = {1050-2947, 1094-1622},
	url = {https://link.aps.org/doi/10.1103/PhysRevA.92.022326},
	doi = {10.1103/PhysRevA.92.022326},
	number = {2},
	urldate = {2026-08-28},
	journal = {Physical Review A},
	author = {Fogarty, M. A. and Veldhorst, M. and Harper, R. and Yang, C. H. and Bartlett, S. D. and Flammia, S. T. and Dzurak, A. S.},
	month = {August},
	year = {2015},
	pages = {022326}
}

@article{ryan_randomized_2009,
	title = {Randomized benchmarking of single- and multi-qubit control in liquid-state {NMR} quantum information processing},
	volume = {11},
	issn = {1367-2630},
	url = {https://iopscience.iop.org/article/10.1088/1367-2630/11/1/013034},
	doi = {10.1088/1367-2630/11/1/013034},
	number = {1},
	urldate = {2026-08-28},
	journal = {New Journal of Physics},
	author = {Ryan, C A and Laforest, M and Laflamme, R},
	month = jan,
	year = {2009},
	pages = {013034}}

@article{ozeri_errors_2007,
	title = {Errors in trapped-ion quantum gates due to spontaneous photon scattering},
	volume = {75},
	copyright = {http://link.aps.org/licenses/aps-default-license},
	issn = {1050-2947, 1094-1622},
	url = {https://link.aps.org/doi/10.1103/PhysRevA.75.042329},
	doi = {10.1103/PhysRevA.75.042329},
	number = {4},
	urldate = {2026-08-28},
	journal = {Physical Review A},
	author = {Ozeri, R. and Itano, W. M. and Blakestad, R. B. and Britton, J. and Chiaverini, J. and Jost, J. D. and Langer, C. and Leibfried, D. and Reichle, R. and Seidelin, S. and Wesenberg, J. H. and Wineland, D. J.},
	month = apr,
	year = {2007},
	pages = {042329}
}

@article{rudinger2019probing,
  title = {Probing Context-Dependent Errors in Quantum Processors},
  author = {Rudinger, Kenneth and Proctor, Timothy and Langharst, Dylan and Sarovar, Mohan and Young, Kevin and Blume-Kohout, Robin},
  journal = {Phys. Rev. X},
  volume = {9},
  issue = {2},
  pages = {021045},
  numpages = {12},
  year = {2019},
  month = {June},
  publisher = {American Physical Society},
  doi = {10.1103/PhysRevX.9.021045},
  url = {https://link.aps.org/doi/10.1103/PhysRevX.9.021045}
}

@article{holm1979simple,
  title={A simple sequentially rejective multiple test procedure},
  author={Holm, Sture},
  journal={Scandinavian journal of statistics},
  pages={65--70},
  year={1979},
  publisher={JSTOR}
}

@misc{zheng2026efficient,
      title={Efficient learning of logical noise from syndrome data}, 
      author={Han Zheng and Chia-Tung Chu and Senrui Chen and Argyris Giannisis Manes and Su-un Lee and Sisi Zhou and Liang Jiang},
      year={2026},
      eprint={2601.22286},
      archivePrefix={arXiv},
      primaryClass={quant-ph},
      url={https://arxiv.org/abs/2601.22286}}

@misc{xiao2026insitu,
      title={In-situ benchmarking of fault-tolerant quantum circuits. I. Clifford circuits}, 
      author={Xiao Xiao and Dominik Hangleiter and Dolev Bluvstein and Mikhail D. Lukin and Michael J. Gullans},
      year={2026},
      eprint={2601.21472},
      archivePrefix={arXiv},
      primaryClass={quant-ph}}

@article{wagner2023learning,
  title = {Learning Logical Pauli Noise in Quantum Error Correction},
  author = {Wagner, Thomas and Kampermann, Hermann and Bru\ss{}, Dagmar and Kliesch, Martin},
  journal = {Phys. Rev. Lett.},
  volume = {130},
  issue = {20},
  pages = {200601},
  numpages = {7},
  year = {2023},
  month = {May},
  publisher = {American Physical Society},
  doi = {10.1103/PhysRevLett.130.200601}
}

@article{Derks2025designing,
  doi = {10.22331/q-2025-11-06-1905},
  title = {Designing fault-tolerant circuits using detector error models},
  author = {Derks, Peter-Jan H.S. and Townsend-Teague, Alex and Burchards, Ansgar G. and Eisert, Jens},
  journal = {{Quantum}},
  issn = {2521-327X},
  publisher = {{Verein zur F{\"{o}}rderung des Open Access Publizierens in den Quantenwissenschaften}},
  volume = {9},
  pages = {1905},
  month = {November},
  year = {2025}
}

@misc{blumekohout2025estimating,
      title={Estimating detector error models from syndrome data}, 
      author={Robin Blume-Kohout and Kevin Young},
      year={2025},
      eprint={2504.14643},
      archivePrefix={arXiv},
      primaryClass={quant-ph} 
}

@article{remm2026experimentallyinformeddecoding,
  title = {Experimentally informed decoding of stabilizer codes based on syndrome correlations},
  author = {Remm, Ants and Lacroix, Nathan and B\"odeker, Lukas and Genois, Elie and Hellings, Christoph and Swiadek, Fran\ifmmode \mbox{\c{c}}\else \c{c}\fi{}ois and Norris, Graham J. and Eichler, Christopher and Blais, Alexandre and M\"uller, Markus and Krinner, Sebastian and Wallraff, Andreas},
  journal = {Phys. Rev. Res.},
  volume = {8},
  issue = {1},
  pages = {013044},
  numpages = {21},
  year = {2026},
  month = {Jan},
  publisher = {American Physical Society},
  doi = {10.1103/z1ng-wg3k},
  url = {https://link.aps.org/doi/10.1103/z1ng-wg3k}
}

@misc{arms2026estimating,
      title={Estimating Detector Error Models on Google's Willow}, 
      author={Kregg Elliot Arms and Martin James McHugh and Joseph Edward Nyhan and William Frederick Reus and James Loudon Ulrich},
      year={2026},
      eprint={2512.10814},
      archivePrefix={arXiv},
      primaryClass={quant-ph},
      url={https://arxiv.org/abs/2512.10814}, 
}

@misc{takou2025estimating,
      title={Estimating and decoding coherent errors of QEC experiments with detector error models}, 
      author={Evangelia Takou and Kenneth R. Brown},
      year={2025},
      eprint={2510.23797},
      archivePrefix={arXiv},
      primaryClass={quant-ph}
}

@misc{blumekohout2013robust,
      title={Robust, self-consistent, closed-form tomography of quantum logic gates on a trapped ion qubit}, 
      author={Robin Blume-Kohout and John King Gamble and Erik Nielsen and Jonathan Mizrahi and Jonathan D. Sterk and Peter Maunz},
      year={2013},
      eprint={1310.4492},
      archivePrefix={arXiv},
      primaryClass={quant-ph},
      url={https://arxiv.org/abs/1310.4492}, 
}

@article{pla_high-fidelity_2013,
	title = {High-fidelity readout and control of a nuclear spin qubit in silicon},
	volume = {496},
	copyright = {http://www.springer.com/tdm},
	issn = {0028-0836, 1476-4687},
	url = {https://www.nature.com/articles/nature12011},
	doi = {10.1038/nature12011},
	number = {7445},
	urldate = {2026-08-28},
	journal = {Nature},
	author = {Pla, Jarryd J. and Tan, Kuan Y. and Dehollain, Juan P. and Lim, Wee H. and Morton, John J. L. and Zwanenburg, Floris A. and Jamieson, David N. and Dzurak, Andrew S. and Morello, Andrea},
	month = apr,
	year = {2013},
	pages = {334--338}
}

@misc{proctor2026benchmarking,
      title={Benchmarking the computational power of quantum computers}, 
      author={Timothy Proctor and others},
      year={2026},
      eprint={2609.12146},
      archivePrefix={arXiv},
      primaryClass={quant-ph},
      url={https://arxiv.org/abs/2609.12146}, 
}

\appendix 
\onecolumngrid

\section{Hardware overview}\label{appendix:h1-1}
Quantinuum H1-1 is a 20-qubit linear QCCD trapped-ion quantum computer, whose architecture was first introduced in Ref.~\cite{Pino2020} with further upgrades described in Ref.~\cite{ryananderson2021realization}. H1-1 encodes qubits into the hyperfine ground manifold of $^{171}$Yb$^+$ ions, which are transported by stationary dc electrodes into 5 ``gate zones" to perform parallel quantum operations. The electrode control allows physical swaps of ions to create arbitrary rearrangements and pairings inside gate zones enabling low-error all-to-all connectivity. Each $^{171}$Yb$^+$ ion is co-trapped with a $^{138}$Ba$^+$ sympathetic coolant ion. Coolant ions do not encode qubits but receive Doppler cooling during transport to keep motional excitation low and sympathetic cooling inside gate zones prior to two-qubit gates to reach high fidelities. Qubit measurement is also done inside the gate zones with ion fluorescence of $^{171}$Yb$^+$ and can be implemented mid-circuit with low-crosstalk by micro-motion hiding~\cite{Gaebler2021}. H1-1 also enables leakage repump to move population from non-qubit hyperfine ground manifold states to computational states~\cite{Hayes2020}.

H1-1's performance is regularly measured with a suite of component-level benchmarks. Typical one-qubit gate average infidelity is around $3\times 10^{-5}$, two-qubit gate average infidelity is around $9\times 10^{-4}$, and transport average infidelity is around $2.5\times10^{-4}$, each measured with Clifford randomized benchmarking. Spontaneous emission during gates also causes incoherent leakage of population outside of the qubit subspace at a rate of around $1\times10^{-5}$ for single-qubit gates and $2\times10^{-4}$ for two-qubit gates. The other major sources of errors are state-preparation and measurement errors, bit-flip probabilities around $2.5\times10^{-3}$, and mid-circuit measurement and reset crosstalk, single-qubit average infidelity around $4\times10^{-5}$. Drift of these values is measured and controlled by periodic calibrations of system control parameters.

\section{Steane $[\![7,1,3]\!]$ code}\label{app:steane}
Logical-level programs in this work were constructed using the Simple Logical
Representation (SLR) framework~\cite{Pecos,mayer2024benchmarking}, a Python-embedded
language for expressing quantum circuits at the level of encoded qubits rather
than physical operations. In SLR, a logical program is composed from a library
of QEC primitives (encoded state preparation, transversal and logical gates,
syndrome extraction, and decode-and-correct blocks) together with
code-agnostic classical control-flow constructs (e.g., \texttt{If}/\texttt{Then})
that condition logical operations on mid-circuit measurement outcomes. 
The explicit gate construction and details can be found in ~\cite{ryananderson2021realization}.
 
SLR is distributed as part of the open-source PECOS package~\cite{Pecos} and
compiles logical programs to OpenQASM~2.0 augmented with Quantinuum's
real-time classical extensions, allowing the identical program to be executed
in simulation or submitted directly to Quantinuum hardware. This workflow
underpinned earlier demonstrations of real-time fault-tolerant error
correction~\cite{ryananderson2021realization}, logically encoded three-qubit
algorithms~\cite{mayer2024benchmarking}, and fault-tolerant logical qubit
teleportation~\cite{ryananderson2024teleport}.

\section{H1-1E noise model parameters}\label{app:noise_model}
As previously described in Sec.~\ref{sec:testing}, the H1-1E noise model includes error sources in one- and two-qubit gates, initialization, measurement, and memory. The emulator uses the same transport compilations as H1-1 to provide a schedule of operations including parallel operations and routing. It then simulates the quantum operations and errors with Monte-Carlo state vector simulation. Gate errors are modeled primarily as Pauli stochastic error channels with a leakage component due to spontaneous-emission. Memory error is treated as a coherent phase error accumulated during transport and cooling \cite{moses2023a,ransford2026a}. Table~\ref{tab:emulator-error-parameters} gives the values and names of the noise model parameters in the context of the four H1-1E datasets. In the main text, we discuss the effects of leakage and coherent errors, further details of each are given below.

Leakage errors are part of several modeled error sources from different quantum operations (with different fractions of errors causing leakage in parenthesis): (1) State initialization ($2/3$ \texttt{p\_init}), (2) single-qubit gates ($1/2\; \texttt{p1} \times \texttt{p1\_emission\_ratio}$), (3) two-qubit gates ($1/2\; \texttt{p2} \times \texttt{p2\_emission\_ratio}$ on either ion), (4) measurement crosstalk (0 if state is zero, $2/3\; \texttt{p\_crosstalk\_meas}$ if 1 state) and (4) initialization crosstalk (0 if state is zero, $1/3\; \texttt{p\_crosstalk\_meas}$, if 1 state). Most of these values are derived from detailed atomic models and different emulator parameters are measured regularly on H1-1. If a qubit is in a leaked state when it is measured then the measurement returns a one outcome.

Coherent errors only occur during qubit idling either during ion transport or delays for cooling or gating, which we call memory errors. Each memory error is modeled as an $R_z(\theta)$ rotation for $\theta = 2 \pi \sqrt{2/3} \, t \,\texttt{quadratic\_dephasing\_rate}$, where $t$ is idle time during transport, cooling, other gates, and \texttt{quadratic\_dephasing\_rate} is given in Table~\ref{tab:emulator-error-parameters}.

\begin{table*}
\centering
\begin{tabular}{llll}
\hline
Parameter or mechanism & Setting name & Recorded value & Status \\
\hline
Single-qubit gate error & \texttt{p1} & $2.97\times10^{-5}$ & Enabled in all datasets \\
Two-qubit gate error & \texttt{p2} & $1.38\times10^{-3}$ & Enabled in all datasets \\
Measurement error for state 0 & \texttt{p\_meas[0]} & $1.32\times10^{-3}$ & Enabled in all datasets \\
Measurement error for state 1 & \texttt{p\_meas[1]} & $3.98\times10^{-3}$ & Enabled in all datasets \\
Initialization error & \texttt{p\_init} & $3.62\times10^{-5}$ & Enabled in all datasets \\
Measurement crosstalk & \texttt{p\_crosstalk\_meas} & $8.73\times10^{-6}$ & Enabled in all datasets \\
Initialization crosstalk & \texttt{p\_crosstalk\_init} & $5.02\times10^{-6}$ & Enabled in all datasets \\
Single-qubit emission fraction & \texttt{p1\_emission\_ratio} & $0.350$ & Enabled in all datasets \\
Two-qubit emission fraction & \texttt{p2\_emission\_ratio} & $0.547$ & Enabled in all datasets \\
Quadratic dephasing coefficient & \texttt{quadratic\_dephasing\_rate} & $0.1224744871$ & Active only when memory is ON \\
Memory errors & \texttt{memory\_scale} &  0 or 1 & Varied OFF or ON \\
Leakage errors & \texttt{leakage\_scale} &   0 or 1 & Varied OFF or ON \\
Coherent memory error & \texttt{coherent\_dephasing} & True & Effective only when memory is ON \\
\hline
\end{tabular}
\caption{\label{tab:emulator-error-parameters}Error parameters and model components used in the H1-1E simulations. The memory and leakage mechanisms were independently enabled or disabled to produce the four simulated datasets.}
\end{table*}

\section{Metrics for the simulated data}\label{app:simulated_data}
\begin{figure}[h]
    \centering
    \includegraphics[width=1\linewidth]{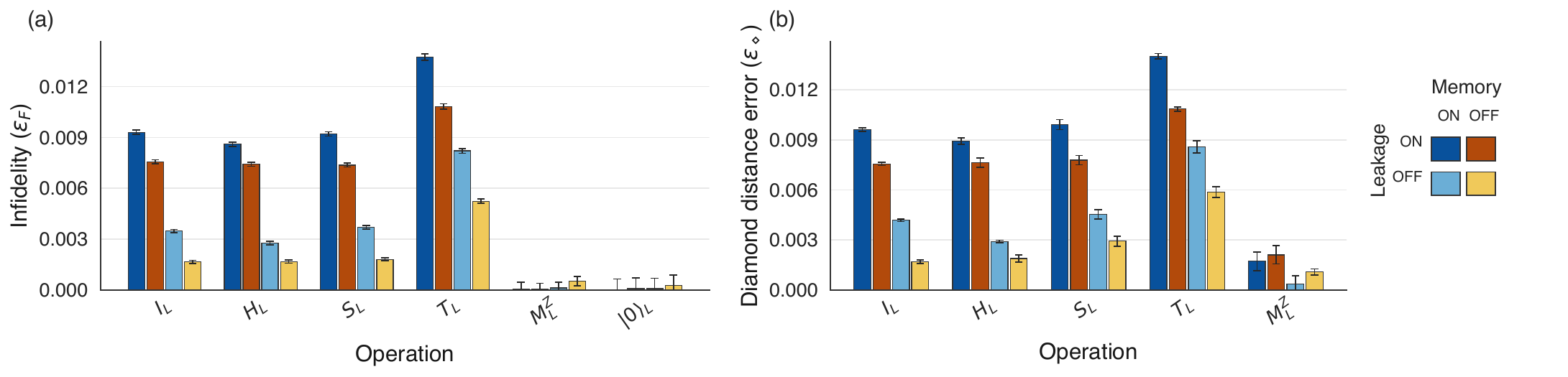}
    \caption{Logical error metrics for the simulated data, analogous to those reported for the experimental data in Table~\ref{tab:gate_metrics}. Blue (orange) shows memory error enabled (disabled); dark (light) colors show leakage enabled (disabled).  Panels show (a) entanglement (state) infidelity $\epsilon_F$ and (b) diamond-distance error $\epsilon_\diamond$ \cite{hashim2025practical}. For Pauli-stochastic noise, the two metrics coincide. For $|0\rangle_L$, $\epsilon_\diamond$ is undefined; for $M_L^Z$ errors are computed for the associated POVM map. Error bars denote $2\sigma$ confidence intervals.}
    \label{fig:error_metrics_simulated}
\end{figure}

\begin{figure}[h]
    \centering
    \includegraphics[width=0.75\linewidth]{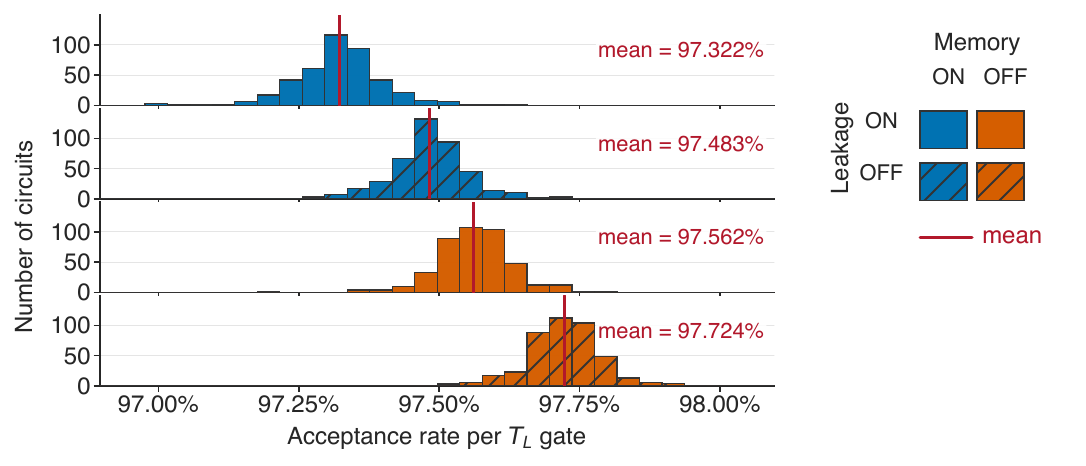}
    \caption{Histogram of post-selection acceptance rate per $T_L$ gate for each of the four simulated datasets with leakage and memory error toggled on and off. Blue (orange) shows memory error enabled (disabled); dark (light) colors show leakage enabled (disabled). As error modes are removed the mean acceptance rate (red line) increases, and memory error has the largest impact on the acceptance rate.}
    \label{fig:post_selection_simulated}
\end{figure}

This appendix collects additional metrics and diagnostics for the logical GST datasets simulated on Quantinuum's device emulator, H1-1E. These datasets were discussed in Sec.~\ref{sec:testing}. Figures~\ref{fig:error_metrics_simulated} and \ref{fig:post_selection_simulated} show, respectively, the logical error metrics and post-selection acceptance-rate distributions for the simulated datasets with memory error and leakage independently enabled or disabled. In these simulations, memory error has the largest effect on the $T_L$-gate post-selection acceptance rate, while leakage has the largest effect on the logical error metrics. Table~\ref{tab:nsigma_comparison} gives in tabular form the model-validation diagnostics reported in Fig.~\ref{fig:overrotation_ramsey}(a)-(c); the hardware result is included in the table for completeness. Additionally, the table reports the number of circuits and shots per circuit in each experiment to allow comparison of the experiment size. We also simulated linearly-spaced Ramsey experiments out to depth $p_{\mathrm{max}}=128$, for a total of $N_{\mathrm{circuits}}=128$ circuits with $N_{\mathrm{shots}} =14100$ shots per circuit; these results are reported in Fig.~\ref{fig:overrotation_ramsey}(d). 

\begin{table*}[h]
    \centering
    \begin{tabular}{lrrrlrl}
        \hline
        \textbf{Dataset} & \textbf{Circuits} & \textbf{Shots per circuit} & $N_\sigma^{\text{CPTP}}$ &$\alpha_{\text{CPTP}}$ (\%)& $N_\sigma^{\text{S}}$   &$\gamma\mathrm{(CPTP, S)}$ \\
        \hline
        Hardware (H1-1) &  160& 50 & 1.7  &--& 0.9   &0.7 \\
        Emulator (Both ON)& 931 & 14100 & 4.9  &0.23& 115.4   &153.9 \\
        Emulator (Memory OFF)  & 931 & 14100 & 2.6  &0.07& 3.6   &2.4 \\
        Emulator (Leakage OFF) & 931& 14100&2.3 &0.16& 172.8  &236.9 \\
        Emulator (Both OFF) & 931 & 14100 & 0.9 & -- &2.2 &2.9\\
        \hline
    \end{tabular}
    \caption{\label{tab:nsigma_comparison} Logical GST circuit-sampling details and model-validation diagnostics for the hardware and simulated logical GST datasets. The simulated datasets independently toggle memory error and leakage; the hardware result is included only for completeness. The table reports total number of circuits, shots per circuit, $N_\sigma$ [Eq.~\eqref{eq:n_sigma}] for the CPTP and S models, unmodeled wildcard error $\alpha$ for the CPTP model, and the evidence ratio $\gamma(\mathrm{CPTP,S})$ [Eq.~\eqref{eq:evidence_ratio}] comparing the predictive power of the CPTP and S model. The CPTP and S models have 43 and 12 non-gauge parameters, respectively. Effective shots per circuit will be lower for some circuits due to post-selection on the $T_L$ gate and the hardware experiment included two repeated circuits at the start and end of the experiment to assess temporal drift.}
\end{table*}

\section{Logical non-Markovianity}\label{app:non_markovianity}
Prior GST experiments on physical qubits \cite{blume-kohout2017, rudinger2021experimental, blumekohout2013robust, Madzik2022precision, hashim2023benchmarking, dehollain2016optimization, Stemp2024scalable} have provided extensive evidence that physical qubit systems can, and almost always do, display non-Markovian behavior. This is unsurprising and predicted by device modeling \cite{Hayes2020, lucero_high-fidelity_2008, babu_state_2021, yoshihara_decoherence_2006, slichter_measurement-induced_2012, taylor_relaxation_2007, pla_high-fidelity_2013, wellstood_low-frequency_1987, steffen_accurate_2003, medford_self-consistent_2013, fogarty_nonexponential_2015, ryan_randomized_2009, ozeri_errors_2007}. It is reasonable to expect that logical qubits implemented within non-Markovian physical qubits may also display non-Markovian behavior — e.g., slow drift in the error rates of the physical qubits (a very common cause of non-Markovianity) would probably yield slow drift in logical error rates as well. However, a less obvious possibility, predicted by Ref.~\cite{ziyad2025emergent}, is that logical non-Markovianity can emerge from completely Markovian dynamics at the physical-qubit level. Ref.~\cite{ziyad2025emergent} demonstrated that emergent logical non-Markovianity can result when the syndrome degrees of freedom (the complementary subsystem of the logical qubit within the data qubits) act as a persistent environment for the logical qubit, but other mechanisms may also be possible.  We sought evidence for logical non-Markovianity in the experimental logical GST data, and the simulated logical GST data, generated for this paper.

GST is known ~\cite{Nielsen2021gatesettomography, rudinger2019probing} to provide a very sensitive detector of logical non-Markovianity in both experiments and simulations, because it prescribes a set of experiments that are (collectively) very sensitive to all aspects of the gate set model, and then performs statistically rigorous inference to find the best-fit Markovian model. A byproduct of this fitting procedure is a quality-of-fit metric called the log-likelihood ratio statistic (LLRS), $\lambda$, whose empirical value can be compared to its distribution under the null hypothesis “the data were generated by a Markovian process”. If the LLRS for a GST fit to a given dataset is statistically inconsistent with its null distribution — e.g., under the null hypothesis we expect $\lambda = 100 \pm 10$, but we observe $\lambda = 170$ — then this constitutes evidence for some non-Markovianity in that data. Additional qualitative conclusions may be drawn from the behavior of the LLRS, but not rigorously and only with careful analysis, because the LLRS is a witness to the statistical evidence for non-Markovianity, not for the amount or impact of non-Markovianity. One useful (and provable) property of the LLRS is the following: if we consider multiple datasets generated by the same non-Markovian process whose only significant difference is the total number of shots $N$, then the expected difference between the LLRS $\lambda$ and its expected null value $\expect{\lambda}_0$ is proportional to $N$. 

For notational compactness, we denote this difference as $\delta$:
\begin{equation}
    \llrdiff := \lambda - \expect{\lambda}_0.
\end{equation}
If we analyze a set of several datasets which are putatively “equivalent” except for variation in $N$, we can distinguish four potentially relevant cases according to how $\llrdiff$ varies with $N$:

\begin{enumerate}
    \item $\llrdiff$ is statistically consistent with 0 for all N: no evidence for non-Markovianity of any kind.
    \item $\llrdiff$ is inconsistent with 0 for small N, but does not increase with N: the various experiments are not actually equivalent. There is evidence for non-Markovianity, but the additional shots added as N increases provide no additional evidence (e.g., they are consistent with Markovian behavior).
    \item $\llrdiff$ is consistent with linear growth proportional to N: evidence for non-Markovianity, and consistency with the expectation that the datasets are equivalent.
    \item $\llrdiff$ is inconsistent with linear growth, but indicates superlinear growth as a function of N: evidence for non-Markovianity, and evidence that the datasets are not equivalent, and that the additional shots added as N increases are increasingly inconsistent with Markovian behavior.
\end{enumerate}

In our analysis of the experimental data, we found no compelling evidence for violation of Markovianity. We found $\llrdiff= 24.2197$ which is only a $1.59\sigma$ (p-value = $0.062479$) deviation from the null hypothesis when a $\chi^2$ approximation of the distribution is assumed. This is unsurprising and does not imply that the logical dynamics were perfectly Markovian, just that the experiment was not sensitive enough to detect any non-Markovianity. A good heuristic is that non-Markovian effects whose impact on the data is less than the size of the error bars in the GST-estimated gate errors are likely to go undetected. The median of the error bars on the logical gate error generator rates is $.02$, so we conclude that if logical non-Markovian effects were present, their magnitude did not exceed $.02$. No theory led us to expect non-Markovianity on this scale, and we conclude that our experiment did not perform enough shots of each GST circuit (N=50) to probe scientifically interesting (small) non-Markovian effects. If we had observed non-Markovianity, we do not expect that we could have confidently attributed it either to underlying physical-qubit non-Markovianity or to emergent effects.

In simulations on the Quantinuum emulator, however, we have access to (1) far more shots and (2) a guarantee that the underlying physical-qubit model is perfectly Markovian. The emulator’s model normally includes a very small amount of physical-qubit leakage, which is a non-Markovian effect that cannot be correctly modeled by standard GST. However, this can be turned off. We used the emulator to generate GST datasets using four distinct noise settings, enumerated by turning leakage on and off, and independently turning coherent memory errors on the physical qubits on and off. For each of these four noise settings, we simulated a $14100$-shot experiment.  The overall results of these experiments are discussed in the main text and Table II. 
With respect to detecting logical non-Markovianity, the fit for the leakage ON, memory ON dataset yields $N_\sigma=4.9$, yielding reasonable confidence about the presence of non-Markovian noise.  The fits for the leakage ON, memory OFF and leakage OFF, memory ON datasets yield $N_\sigma=2.6$ and $N_\sigma=2.3$, respectively, indicating weak but non-zero evidence thereof.  The fit for the leakage OFF, memory OFF dataset yields $N_\sigma=0.9$, indicating no detection of non-Markovian noise.

However, instead of just looking at the aggregate $N_\sigma$ values, we can, as discussed above, examine the trend of the excess log-likelihood ratio as the GST dataset iteratively grows in size. This can give us deeper insight into whether or not the noise models are generating non-Markovian logical dynamics, and if so, what the nature of those non-Markovian dynamics are.

\begin{figure}[h]
    \centering
    \begin{subfigure}{0.49\textwidth}
        \centering
        \includegraphics[width=\linewidth]{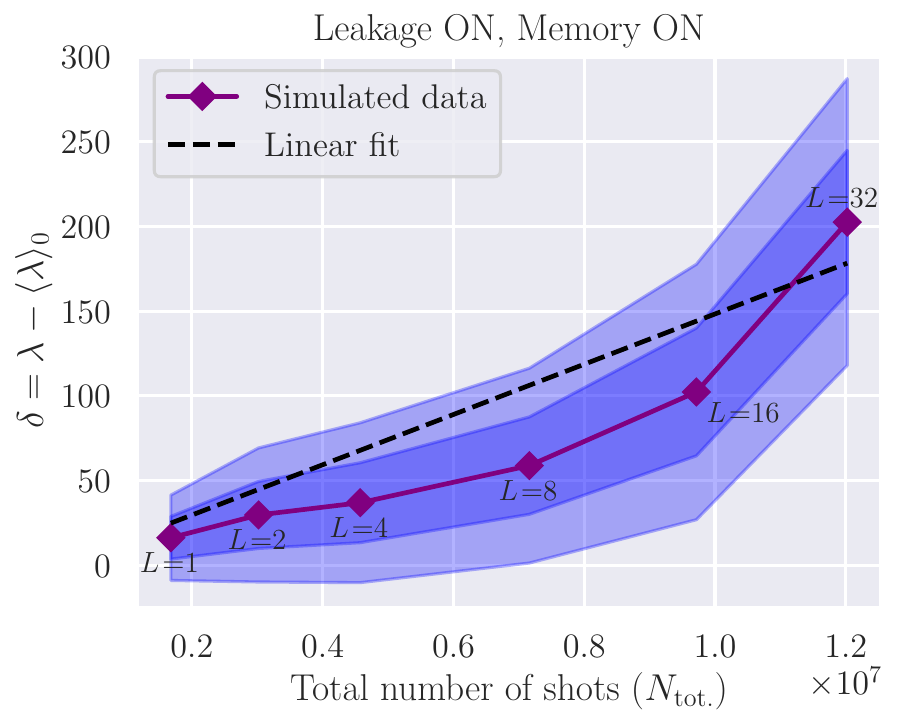}
        \caption{}
    \end{subfigure}%
    \hfill
    \begin{subfigure}{0.49\textwidth}
        \centering
        \includegraphics[width=\linewidth]{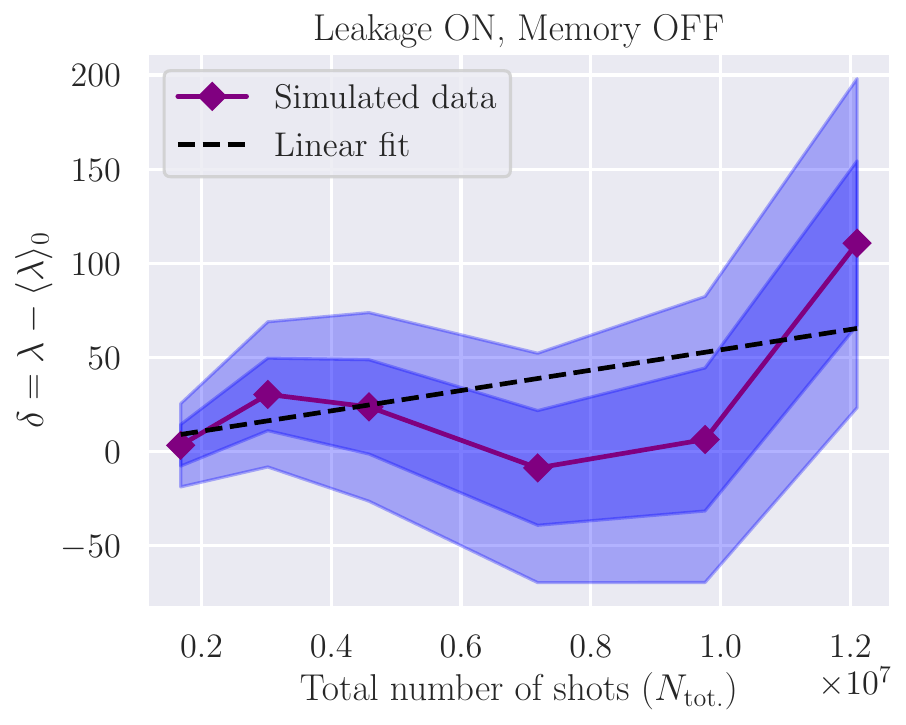}
        \caption{}
    \end{subfigure}
    \hfill
        \begin{subfigure}{0.49\textwidth}
        \centering
        \includegraphics[width=\linewidth]{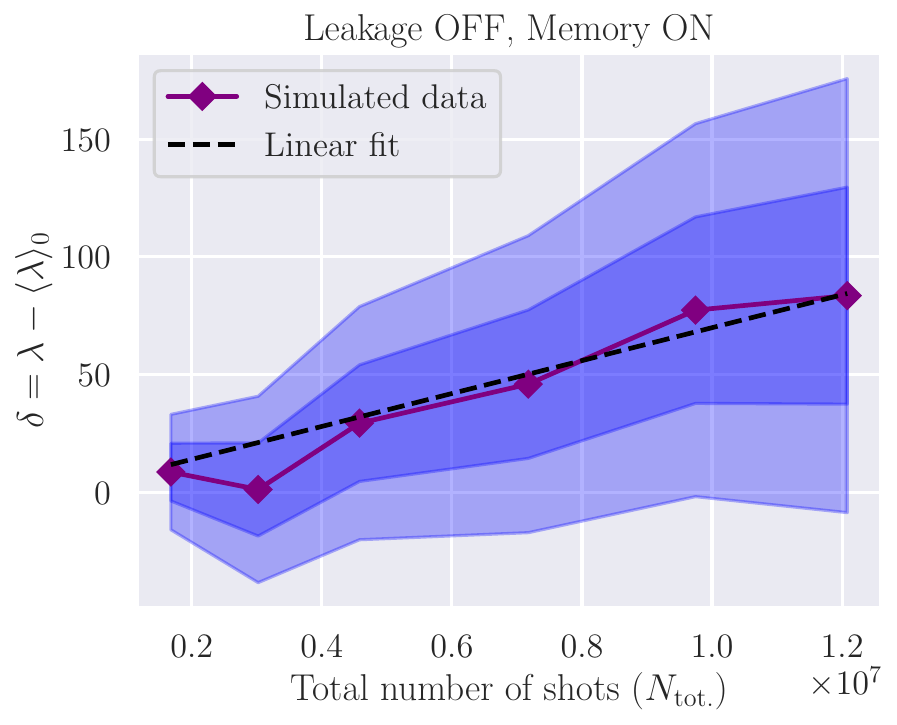}
        \caption{}
    \end{subfigure}
    \hfill
    \begin{subfigure}{0.49\textwidth}
        \centering
        \includegraphics[width=\linewidth]{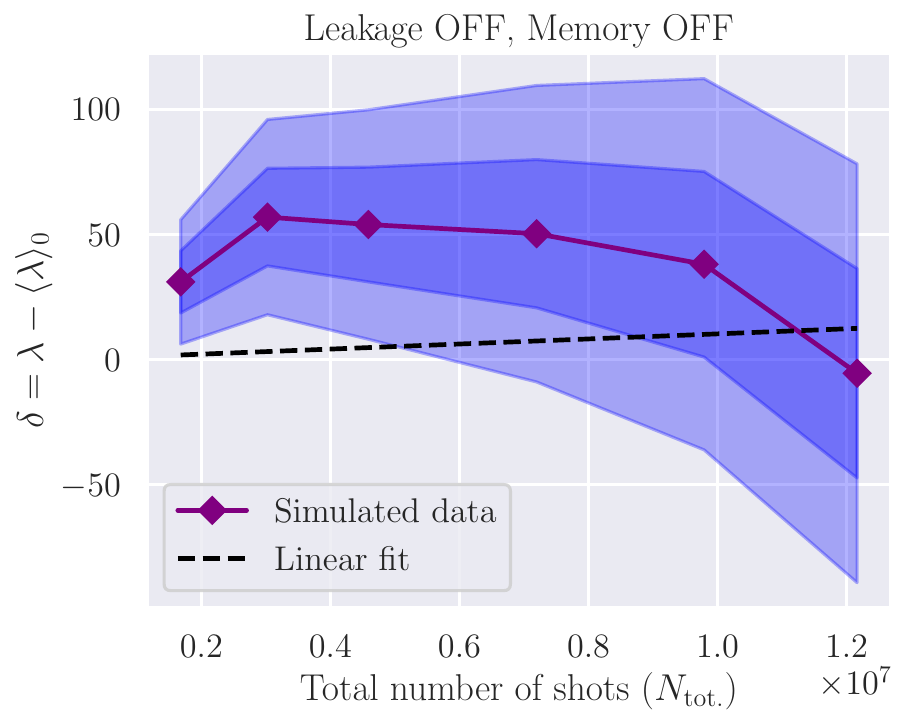}
        \caption{}
    \end{subfigure}
    \caption{Log-likelihood ratio statistic $\lambda$ of simulated data minus $\expect{\lambda}_0$ (purple diamonds), versus total number of shots ($N_\text{tot.}$). The value of $\expect{\lambda}_0$ is the mean LLRS of $100$ (bootstrap) datasets generated from the null (Markovian) hypotheses.  corresponds to datasets of varying $L$, a proxy for the maximum length of circuits. The values of $L$ are labeled on (a). The standard deviation, $\sigma$, of the distribution of the null hypothesis is estimated from the bootstrap datasets. The shaded region represents $\pm 2\sigma$ sigma significance level, meaning if the shaded region is above $0$, the corresponding dataset violates the Markovian assumption with $95\%$ confidence. If the datasets at varying $L$ are equivalent, then $\lambda$ will be roughly linear with respect to the number of circuits. Each plot has a black dashed line corresponding to the generalized least squares fit; fit quality is analyzed in Table~\ref{tab:holm}.}
    \label{fig:numcircs}
\end{figure}

For each dataset, we define several sub-datasets, labeled by an integer $L>0$, containing all the circuits whose base sequence length (including germ powers but \textit{not} including fiducials) is less than or equal to $L$. \footnote{Elsewhere in this paper we use the symbol $p$ to denote base sequence length; here it collides with $p$-value, so we instead use the label $L$.} For each sub-dataset, we perform GST analysis to find a best (maximum likelihood) fit gate set, we compute the log-likelihood ratio $\lambda$ quantifying how much that best fit \textit{failed} to fit the data, and then we compare $\lambda$ to its expected null behavior by generating an ensemble of 100 datasets from the best-fit (Markovian) model and computing $\lambda$ for each one. The mean and standard deviation of the collection of $\{\lambda\}$ values are used, respectively, as the estimates of $\expect{\lambda}_0$ and the 1$\sigma$ confidence interval thereon.  Results are shown in Fig.~\ref{fig:numcircs}.  For each value of $L$, the value of $\llrdiff$ (i.e., $\lambda - \expect{\lambda}_0$) is plotted against the \textit{total} number shots ($N_\text{tot.}$) in the sub-dataset (which increases with $L$); 1$\sigma$ and 2$\sigma$ confidence intervals are shown as well, the latter indicating which deviations of $\llrdiff$ from 0 are statistically significant.  

For each physical error model simulated, we can statistically rigorously test if $\delta$ grows linearly with $N_\mathrm{tot.}$, i.e., to see if $\delta=a\,N_\mathrm{tot.}$ for some slope $a$.  We use generalized least squares to obtain an estimate of $a$ and to determine if indeed the data are consistent with the obtained linear fit. \footnote{Because datasets at successive values of $L$ contain preceding datasets, the values of $\delta$ are not uncorrelated, requiring the use of generalized least squares, as opposed to weighted least squares.}  For each fit we can compute a $\chi^2$ value and a corresponding $p$-value.  To avoid p-hacking, we use the Holm method \cite{holm1979simple} for multiple comparisons, allowing us to retain a 95\% confidence level without subjecting all $p$-values to the same more stringent threshold.  The results of our fits are summarized in Table~\ref{tab:holm}, and the linear fits are displayed in the plots in Fig.~\ref{fig:numcircs}.  

We reject the null hypothesis ($\delta=a\,N_\mathrm{tot.}$) for all noise models except for the leakage OFF, memory ON case.  When both error mechanisms are present (Fig.~\ref{fig:numcircs}(a)), $\llrdiff$ grows super-linearly with $\Ntot$, indicating the deeper circuits added to the experiment design are more sensitive to the induced logical non-Markovianity; this is consistent with some manner of ``logical leakage'' induced by each logical cycle.  Interestingly, if physical leakage is turned off, but memory errors remain (Fig.~\ref{fig:numcircs}(c)), $\llrdiff$ \textit{does} grow linearly.  This indicates that, for that error model, any evidence for logical non-Markovianity unearthed by deeper circuits is no stronger that (and is essentially equivalent to) the evidence unearthed by shallower circuits.  On the other hand, turning off memory error but leaving leakage on (Fig.~\ref{fig:numcircs}(b)) yields non-linear and non-monotonic growth in $\llrdiff$; the sharp growth in $\llrdiff$ at $L=32$ (which is statistically significantly different from 0) indicates that physical leakage does not induce easy-to-detect logical non-Markovianity at low circuit depths, but the addition of $L=32$ circuits is sufficient to produce an unambiguous signal of logical non-Markovianity.  Understanding precisely how physical leakage (which itself is non-Markovian at the physical level) and physical memory errors (which are not) contribute to logical non-Markovianity, both separately and jointly, is an intriguing venue for future research.

Perhaps \textit{most} interesting, however, is the case in which both leakage and memory effects are turned off. In this instance, $\llrdiff$ is not consistent with linear growth in $\Ntot$ but \textit{also} only displays detectable non-Markovianity for $L\leq4$.  In this instance, there are still residual physical error mechanisms present (all of which are Markovian).  Therefore we witness here evidence of \textit{transient, emergent} logical non-Markovianity.  Transient, emergent logical non-Markovianity was predicted, for certain circumstances, in Refs. \cite{ziyad2025emergent,kwiatkowski2025constructing}. In \cite{ziyad2025emergent} it was shown that errors during syndrome extraction can become correlated with logical errors in subsequent rounds of error correction; in other words, non-Markovian dynamics emerge due to correlations across rounds of error correction between logical and syndrome qubit states.  The impact of those correlations are predicted to fall quickly as more rounds of error correction are performed.  Here we also have errors occurring during error correction (due to both gate and readout errors). Additionally, no non-Markovian dynamics are detected by the models fit to $L\geq8$ circuits, indicating that whatever impact non-Markovian dynamics \textit{have} on short circuits, that impact is heavily diluted by the data of long-depth Markovian (or approximately Markovian) circuits.  It is therefore \textit{possible} that we are witnessing dynamics akin to what was predicted by Refs. \cite{ziyad2025emergent,kwiatkowski2025constructing}.  However, those works studied circuits that contain \textit{only} rounds of error correction (as opposed to non-trivial logic operations studied in this work). Additionally, Ref. \cite{ziyad2025emergent} showed how logical error rate changes per round of error correction; while GST analysis reconstructs Pauli transfer matrices for each value of $L$, each increment of $L$ adds a collection of circuits with a range of depths (owing to variable depths for germs as wells as the fiducial sub-circuits used for state preparation and measurement), so we cannot reconstruct error rates of logic operations occurring at particular depths (and therefore cannot see if those error rates follow the trend predicted in \cite{ziyad2025emergent}).  Further careful study is required to firmly conclude if the dynamics we are witnessing here are manifestations of the predictions of Refs. \cite{ziyad2025emergent,kwiatkowski2025constructing}.

\begin{table}[h!]

\begin{tabular}{@{}c|l|r@{\,$\pm$\,}l|r|r|r|c@{}}
\toprule
$i$ & Noise model & \multicolumn{2}{c|}{$a$ $[10^{-6}]$} & $\chi^2$ & $p$
 & $\alpha/(m{-}i{+}1)$ & Reject $H_0$? \\
\midrule
\hline
1 & Leakage ON,\phantom{f} memory OFF
  &  5.40 & 3.39 & 27.886 & $3.8\times10^{-5}$ & 0.0125 & yes \\
2 & Leakage ON,\phantom{f} memory ON
  & 14.83 & 3.33 & 15.020 & 0.0103 & 0.0167 & yes \\
3 & Leakage OFF, memory OFF
  &  1.01 & 3.32 & 14.506 & 0.0127 & 0.0250 & yes \\
4 & Leakage OFF, memory ON
  &  6.99 & 3.62 &  3.084 & 0.687  & 0.0500 & no \\
\bottomrule
\end{tabular}
\caption{%
\label{tab:holm}
Summary of our analysis of whether or not $\llrdiff=a\,\Ntot$
for the emulator error models considered. Fitting of the data presented in Fig.~\ref{fig:numcircs}
was performed using generalized least squares, providing us with values for the fit parameter $a$
and goodness-of-fit metric $\chi^2$, which in turn yields a $p$-value. To account for the
fact that we are performing multiple tests, we use the Holm method. Noise models are sorted by
ascending $p$-value with the corresponding Holm threshold listed in the $\alpha/(m-i+1)$ column.
Here $\alpha$ is 0.05 (for $95\%$ confidence level), $m$ is four (number of models being tested) and $i$
is the index labeling the $p$-sorted model.  The procedure stops at the first row where
$p>\alpha/(m-i+1)$; the null hypothesis $H_0$ (that $\llrdiff=a\,\Ntot$) is rejected for 
all rows above, but is not for that row or any subsequent one.  We reject $H_0$ for
all error models except leakage OFF, memory ON (Fig.~\ref{fig:numcircs}).}

\end{table}

\begin{figure}
    \centering
    \includegraphics[width=0.5\linewidth]{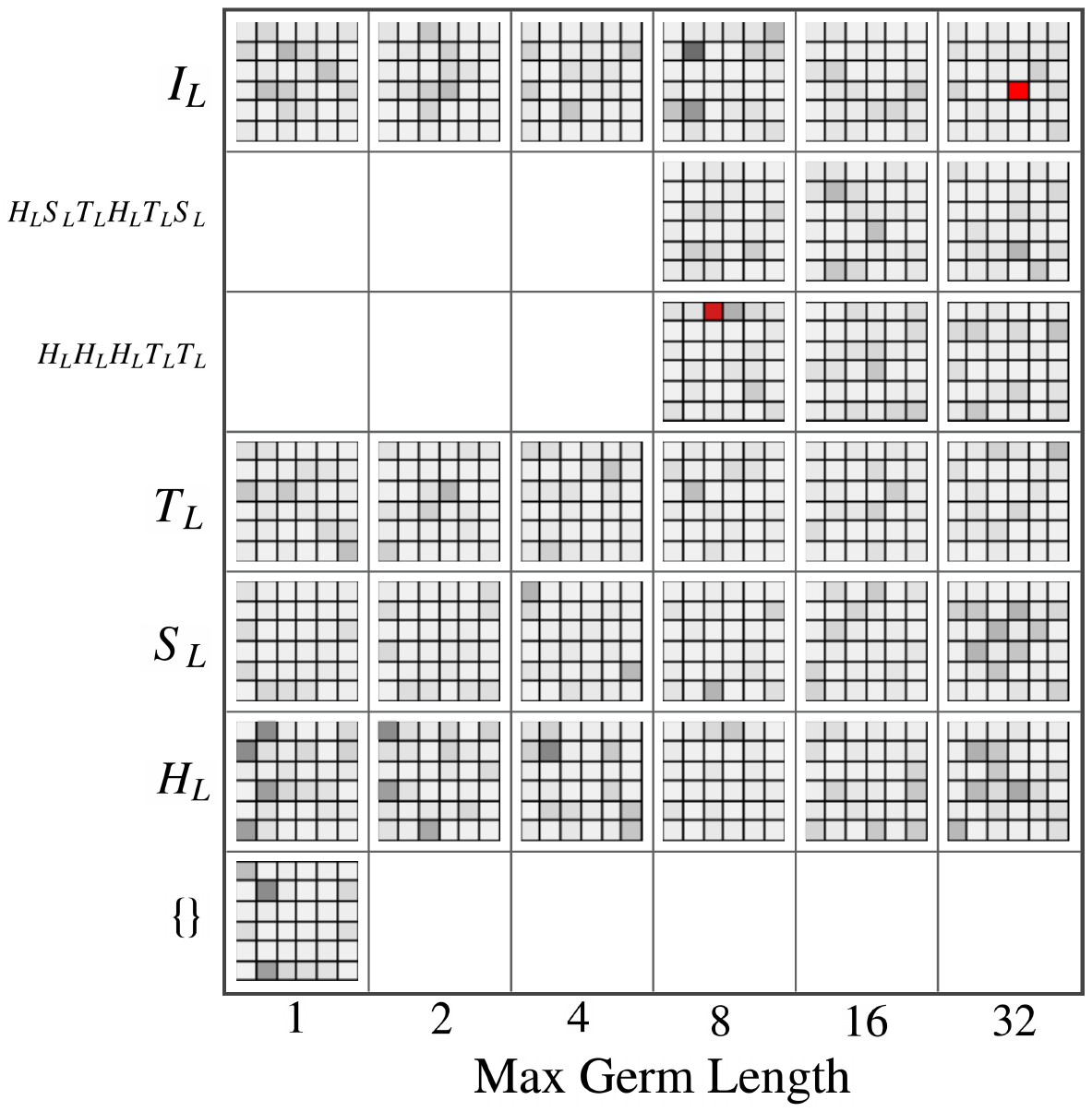}
    \caption{Visualization of each circuits LLRS for the leakage ON, memory ON noise model. Each box represents a circuit in the GST protocol. They are arranged into “plaquettes” according to germ (the vertical labels) and maximum germ length (the horizontal labels) where "germ" refers the the gate sequence being repeated in the circuit. The circuits are labelled using the \texttt{pyGSTi} convention of time running left to right. Each plaquette is a $5\times 5$ group of boxes corresponding to the various initial (vertical axis) and final (horizontal axis) fiducial SPAM operations. The darker the color of the box, the more a circuits data deviates from the null hypothesis. Any one of the circuits being red implies that the entire dataset passes $2\sigma$ threshold of significance for violating the Markovian hypothesis.}
    \label{fig:boxplot}
\end{figure}

Finally, we also disaggregate $\lambda$ into the contributions of every individual circuit (see Fig. \ref{fig:boxplot}), where each circuit’s $\lambda$ value is indicated by color; statistically significant outliers are colored red, grayscale is used otherwise.  The boxes representing circuits are arranged into “plaquettes” by length and germ, and then within plaquettes by initial and final fiducial), to seek evidence for a particular mechanism of non-Markovianity (e.g. whether greater LLRS values were associated with a particular germ or fiducial pair). We do not find any clear evidence or pattern suggesting a mechanism. Notably, only the leakage ON, memory ON setting (shown in  Fig. \ref{fig:boxplot}) had statistical outliers (i.e., circuits whose corresponding boxes were red), meaning that those circuits' $\lambda$ values are large enough by themselves to push the entire dataset past the $2\sigma$ threshold of significance. This means that the other noise settings demonstrated statistically significant non-Markovianity only when their circuits were considered in aggregate. The red boxes in the figure correspond to the circuits $H_L P_L^4T_L^{2}$ and $H_L P_L^2 I_L^{32} T_L^{2}$.  The circuits are labeled using the \texttt{pyGSTi} convention of time running left to right.
For each circuit, there is not an immediately obvious single noise mechanism that explains its statistically significantly large $\lambda$ value.

This analysis in this appendix is not sufficient to support confident scientific conclusions about the presence, or nature, of emergent non-Markovianity in the logical-qubit system that we studied here. It is strongly suggestive that emergent non-Markovianity exists and is detectable in simulations, but (1) the statistical significance of the data is borderline and (2) a clear mechanism has not been hypothesized and confirmed. The most physically realistic noise model (leakage ON, memory ON) generates evidence of a form of logical non-Markovianity that does not follow the pattern predicted by Refs. \cite{ziyad2025emergent, kwiatkowski2025constructing}, because we observe the evidence to grow with $L$, whereas the mechanism analyzed in Refs. \cite{ziyad2025emergent, kwiatkowski2025constructing} suggests non-Markovian effects that vanish at longer $L$. However, as stated above, this is not a rigorous connection because Refs. \cite{ziyad2025emergent,kwiatkowski2025constructing} only analyzed idle gates in detail, and the heterogeneous circuits used in GST may yield different behavior and/or enable different mechanisms of non-Markovianity. Additionally, physical leakage itself is non-Markovian, so any resulting logical non-Markovianity is not ``emergent''.  We note that $\delta$ is very frequently observed to grow superlinearly with $L$ in physical-qubit GST experiments, and this behavior is often attributed to either a) slowly-varying coherent errors, i.e., “non-Markovian dephasing” or b) leakage, both of which get amplified by long circuits. The prospect that either of these phenomena might also manifest in logical GST is intriguing, and a compelling question for future investigation.

\section{Model selection details}\label{app:model_selection}

We considered a variety of different models to fit GST data with the underlying goal of inferring logical error mechanisms that generated the data. In order to do any kind of inference from a statistical model, one must first ensure that the model accurately describes the logical error mechanisms in the system. As a consequence we aim to find models that avoid (1) underfitting, i.e. inadequately describes the data, and (2) overfitting, i.e. do not capture statistical errors which are uncorrelated with future data points. 

The ideal model selection procedure to obtain a parsimonous model require considering every possible candidate model and selecting one that maximizes or minimizes some score like the Akaike Information Criterion (AIC) \cite{PORTET2020111}. Unfortunately, the number of candidate models for Markovian dynamics grows combinatorially, making this approach intractable. In Sec. \ref{sec:pauli_stochastic_model} we present the definition for the evidence ratio between two models $\mathcal{M}_A$ and $\mathcal{M}_B$, $\gamma(\mathcal{M}_A,\mathcal{M}_B)$. We will assume by convention that $\mathcal{M}_A$'s parameters are a superset of $\mathcal{M}_B$'s (i.e., the models are nested models). The evidence ratio provides a way to quantify how much more expressive power per extra parameter $\mathcal{M}_A$ has over $\mathcal{M}_B$. The AIC suggests that an evidence ratio of $\gamma(\mathcal{M}_A,\mathcal{M}_B) < 2$ implies that $\mathcal{M}_B$ is likely to have greater predictive accuracy than $\mathcal{M}_A$, even though it is a strictly less expressive model in general. Here and throughout we utilize the evidence ratio to compare pairs of models and ``accept" a smaller of two models only if it satisfies $\gamma < 2$.

Some of the model selection techniques we use are hindered by gauge-freedom, which is an inherent property of the CPTP models considered in the main text \cite{Nielsen2021gatesettomography}. To circumvent this we utilize first-order gauge-invariant (FOGI) parameters \cite{FOGI}. In addition to ameliorating gauge-related problems, FOGI parameters can be more amenable to interpretation and inference, especially in the case of coherent errors where they naturally correspond to on-axis overrotations and axis-misalignment errors. Tables \ref{tab:FOGI-error-generator-combinations-short} and \ref{tab:FOGI-error-generator-combinations-long}  contain a list of 43 FOGI parameters represented as linear combinations of the elementary error generators which forms a basis for all the observable errors of the gate set in question. 

%
%
%

\begingroup
\scriptsize
\renewcommand{\arraystretch}{1.18}
\setlength{\tabcolsep}{3pt}

\newcommand{\FOGIexprleft}[1]{\parbox[t]{0.25\textwidth}{\raggedright $#1$}}
\newcommand{\FOGIinterpleft}[1]{\parbox[t]{0.12\textwidth}{\raggedright #1}}
\newcommand{\FOGIexprright}[1]{\parbox[t]{0.34\textwidth}{\raggedright $#1$}}
\newcommand{\FOGIexprlong}[1]{\parbox[t]{0.86\textwidth}{\raggedright $#1$}}


\begin{table}[tb]
\centering
\caption{%
All of the intrinsic FOGI parameters for the gate set $\{\ket{0}_L, M_L^Z,H_L,S_L,I_L,T_L \}$. As dicussed in Sec.~\ref{app:model_selection}, the first 13 parameters (shown on the left table) have the highest per-circuit TVD between all four simulated datasets. Thus, the parameters on the left correspond to those plotted in Figure~\ref{fig:AMS}.}
\label{tab:FOGI-error-generator-combinations-short}

\vspace{0.5em}

\begin{center}
\begin{tabular}{|r|l|l|}
\hline
\# & Expression & Interpretation \\
\hline

1  & \FOGIexprleft{\Hg{Z}{\TL}}
   & \multirow[t]{4}{*}{\FOGIinterpleft{Overrotation errors}} \\
\cline{1-2}

2  & \FOGIexprleft{\pfac{2}\bigl[\Hg{X}{\HL}+\Hg{Z}{\HL}\bigr]}
   & \\
\cline{1-2}

3  & \FOGIexprleft{\Hg{Z}{\IL}}
   & \\
\cline{1-2}

4  & \FOGIexprleft{\Hg{Z}{\SL}}
   & \\
\hline

5  & \FOGIexprleft{\Sg{X}{\TL}+\Sg{Y}{\TL}}
   & \multirow[t]{4}{*}{\FOGIinterpleft{Generalized bitflip errors}} \\
\cline{1-2}

6  & \FOGIexprleft{\Sg{X}{\HL}+\Sg{Z}{\HL}}
   & \\
\cline{1-2}

7  & \FOGIexprleft{\Sg{X}{\IL}+\Sg{Y}{\IL}}
   & \\
\cline{1-2}

8  & \FOGIexprleft{\Sg{X}{\SL}+\Sg{Y}{\SL}}
   & \\
\hline

9  & \FOGIexprleft{\Sg{X}{\IL}-\Sg{Y}{\IL}}
   & \FOGIinterpleft{Bitflip anisotropy} \\
\hline

10 & \FOGIexprleft{\Sg{Z}{\TL}}
   & \multirow[t]{4}{*}{\FOGIinterpleft{Dephasing errors}} \\
\cline{1-2}

11 & \FOGIexprleft{\Sg{Y}{\HL}}
   & \\
\cline{1-2}

12 & \FOGIexprleft{\Sg{Z}{\IL}}
   & \\
\cline{1-2}

13 & \FOGIexprleft{\Sg{Z}{\SL}}
   & \\
\hline

\end{tabular}
\hspace{6pt}
\begin{tabular}{|r|l|}
\hline
\# & Expression \\
\hline

14 & \FOGIexprright{\Ag{X}{Y}{\TL}} \\
\hline

15 & \FOGIexprright{\Cg{Z}{X}{\HL}} \\
\hline

16 & \FOGIexprright{\pfac{2}\bigl[-\Cg{Y}{X}{\HL}+\Cg{Z}{Y}{\HL}\bigr]} \\
\hline

17 & \FOGIexprright{\pfac{2}\bigl[\Ag{X}{Y}{\HL}+\Ag{Y}{Z}{\HL}\bigr]} \\
\hline

18 & \FOGIexprright{\Hg{X}{\IL}} \\
\hline

19 & \FOGIexprright{\Hg{Y}{\IL}} \\
\hline

20 & \FOGIexprright{\Cg{Y}{X}{\IL}} \\
\hline

21 & \FOGIexprright{\Ag{X}{Y}{\IL}} \\
\hline

22 & \FOGIexprright{\Cg{Z}{X}{\IL}} \\
\hline

23 & \FOGIexprright{\Ag{X}{Z}{\IL}} \\
\hline

24 & \FOGIexprright{\Cg{Z}{Y}{\IL}} \\
\hline

25 & \FOGIexprright{\Ag{Y}{Z}{\IL}} \\
\hline

26 & \FOGIexprright{\Ag{X}{Y}{\SL}} \\
\hline

\end{tabular}
\end{center}

\vspace{1em} 


\caption{%
A list of FOGI parameters which together with those listed in Table~\ref{tab:FOGI-error-generator-combinations-short} form a basis for the space of observable errors of the gate set $\{\ket{0}_L, M_L^Z,H_L,S_L,I_L,T_L \}$.}
\label{tab:FOGI-error-generator-combinations-long}

\vspace{0.5em}

\begin{center}
\begin{tabular}{|r|l|}
\hline
\# & Expression \\
\hline

27 & \FOGIexprlong{\pfac{5+2\rt}\bigl[-\Ag{X}{Z}{\TL}+\onept\,\Ag{Y}{Z}{\TL}+\Ag{X}{Z}{\HL}\bigr]} \\
\hline

28 & \FOGIexprlong{\pfac{5+2\rt}\bigl[\onept\,\Hg{X}{\TL}+\Hg{Y}{\TL}-\Hg{Y}{\HL}\bigr]} \\
\hline

29 & \FOGIexprlong{\pfac{30+8\rt}\bigl[\Sg{X}{\TL}-\Sg{Y}{\TL}+2\,\Cg{Y}{X}{\TL}+2\onept\,\Cg{Z}{X}{\TL}+2\,\Cg{Z}{Y}{\TL}-2\,\Cg{Y}{X}{\HL}-2\,\Cg{Z}{Y}{\HL}\bigr]} \\
\hline

30 & \FOGIexprlong{\pfac{6+2\rt}\bigl[-\Hg{X}{\TL}+\onept\,\Hg{Y}{\TL}+\Hg{X}{\SL}-\Hg{Y}{\SL}\bigr]} \\
\hline

31 & \FOGIexprlong{\pfac{6+2\rt}\bigl[-\Cg{Z}{X}{\TL}+\onept\,\Cg{Z}{Y}{\TL}+\Cg{Z}{X}{\SL}-\Cg{Z}{Y}{\SL}\bigr]} \\
\hline

32 & \FOGIexprlong{\pfac{6+2\rt}\bigl[\onept\,\Ag{X}{Z}{\TL}+\Ag{Y}{Z}{\TL}-\Ag{X}{Z}{\SL}-\Ag{Y}{Z}{\SL}\bigr]} \\
\hline

33 & \FOGIexprlong{\pfac{6+2\rt}\bigl[\onept\,\Cg{Z}{X}{\TL}+\Cg{Z}{Y}{\TL}-\Cg{Z}{X}{\SL}-\Cg{Z}{Y}{\SL}\bigr]} \\
\hline

34 & \FOGIexprlong{\pfac{190}\bigl[-6\,\Sg{X}{\TL}+8\,\Cg{Y}{X}{\TL}+6\,\Sg{Y}{\TL}+5\,\Sg{X}{\SL}+2\,\Cg{Y}{X}{\SL}-5\,\Sg{Y}{\SL}\bigr]} \\
\hline

35 & \FOGIexprlong{\pfac{6+2\rt}\bigl[-\Ag{X}{Z}{\TL}+\onept\,\Ag{Y}{Z}{\TL}+\Ag{X}{Z}{\SL}-\Ag{Y}{Z}{\SL}\bigr]} \\
\hline

36 & \FOGIexprlong{\pfac{6+2\rt}\bigl[\onept\,\Hg{X}{\TL}+\Hg{Y}{\TL}-\Hg{X}{\SL}-\Hg{Y}{\SL}\bigr]} \\
\hline

37 & \FOGIexprlong{\pfac{10}\bigl[-\Sg{X}{\TL}-2\,\Cg{Y}{X}{\TL}+\Sg{Y}{\TL}+2\,\Cg{Y}{X}{\SL}\bigr]} \\
\hline

38 & \FOGIexprlong{\pfac{56+14\rt}\bigl[2\rt\,\Hg{Y}{\rz}-4\,\Ag{Y}{Z}{\rz}+2\,\Cg{Z}{X}{\rz}-(2+\rt)\,\Hg{X}{\TL}-\rt\,\Hg{Y}{\TL}+2\onept\,\Ag{X}{Z}{\TL}+2\,\Ag{Y}{Z}{\TL}-\Cg{Z}{X}{\TL}+\onept\,\Cg{Z}{Y}{\TL}\bigr]} \\
\hline

39 & \FOGIexprlong{\pfac{56+14\rt}\bigl[-2\rt\,\Hg{X}{\rz}+4\,\Ag{X}{Z}{\rz}+2\,\Cg{Z}{Y}{\rz}+\rt\,\Hg{X}{\TL}-(2+\rt)\,\Hg{Y}{\TL}-2\,\Ag{X}{Z}{\TL}+2\onept\,\Ag{Y}{Z}{\TL}-\onept\,\Cg{Z}{X}{\TL}-\Cg{Z}{Y}{\TL}\bigr]} \\
\hline

40 & \FOGIexprlong{\pfac{24+6\rt}\bigl[-2\rt\,\Hg{X}{\Md}-2\,\Cg{Z}{Y}{\Md}-\rt\,\Hg{X}{\TL}+(2+\rt)\,\Hg{Y}{\TL}-\onept\,\Cg{Z}{X}{\TL}-\Cg{Z}{Y}{\TL}\bigr]} \\
\hline

41 & \FOGIexprlong{\pfac{24+6\rt}\bigl[2\rt\,\Hg{Y}{\Md}-2\,\Cg{Z}{X}{\Md}+(2+\rt)\,\Hg{X}{\TL}+\rt\,\Hg{Y}{\TL}-\Cg{Z}{X}{\TL}+\onept\,\Cg{Z}{Y}{\TL}\bigr]} \\
\hline

42 & \FOGIexprlong{\frac{1}{2\sqrt{3}}\bigl[\Sg{X}{\rz}+2\,\Ag{X}{Y}{\rz}+\Sg{Y}{\rz}+\Sg{X}{\Md}+2\,\Ag{X}{Y}{\Md}+\Sg{Y}{\Md}\bigr]} \\
\hline

43 & \FOGIexprlong{\pfac{10+2\rt}\bigl[2\,\Ag{X}{Y}{\Md}+\onept\,\Ag{X}{Z}{\TL}+\Ag{Y}{Z}{\TL}+\Ag{X}{Y}{\HL}-\Ag{Y}{Z}{\HL}\bigr]} \\
\hline

\end{tabular}
\end{center}

\end{table}

\endgroup

In the case of the hardware data, we found that the logical dynamics are well modeled by a few-parameter Pauli-stochastic model (S). Since our main interest lies in exploring logical coherent dynamics, we probe individual coherent errors by constructing augmented models each consisting of all the Pauli-stochastic degrees of freedom plus a single coherent FOGI parameter. A coherent FOGI parameter is any FOGI parameter that contains only Hamiltonian error generators i.e. $h_X, h_Y \text{ and } h_Z$. These correspond to parameters $f_{1-4}, f_{18-19}, f_{28}, f_{30}, \text{ and } f_{36}$ in Tables \ref{tab:FOGI-error-generator-combinations-short} \& \ref{tab:FOGI-error-generator-combinations-long}. This yields a total of 9 candidate models with a single FOGI parameter, out of which only three are recommended by the AIC ($\gamma > 2$) as discussed in Sec.  \ref{sec:pauli_stochastic_model}, with the following evidence ratios:
 \begin{itemize}
     \item $\gamma(S,S + f_{28})=3.598$,
     \item $\gamma(S,S + f_{30})=3.090$,
     \item $\gamma(S,S + f_{36})=3.038$.
 \end{itemize}

We then continue adding individual FOGI parameters to each of the three models identified above to probe combinations of coherent errors. Only one model with two FOGI parameters is accepted by the AIC with evidence ratio $\gamma(S,S + f_{30}+f_{36})=2.115$. 

A FOGI parameter can be either \emph{intrinsic} to a single operation or \emph{relational} between multiple operations. The former describes errors which can be isolated to a single operation, while the latter describes errors with support fundamentally on multiple gates. An example of an intrinsic FOGI property is a gate's over-rotation. An example of a relational FOGI property is the axis-misalignment between a pair of gates\cite{FOGI}. All of the FOGI parameters identified as statistically significant in the above analysis quantify axis-misalignment errors, with the final model only suggesting axis-misalignment between $T_L$ and $S_L$. 

We now turn our attention to the analysis of the four simulated datasets discussed in Sec. \ref{sec:model_selection}. Unlike the hardware data, the simulated dataset dynamics are not captured adequately by a Pauli-stochastic model. To find a parsimonious model we start with a model containing the entire FOGI basis. We then consider all candidate models with a single parameter removed and compute their evidence ratios with respect to the full model. We choose the model with the highest evidence ratio only if it is accepted by the AIC. If the model is not accepted by the AIC, the procedure terminates, otherwise it recursively continues by removing an extra parameter from the model chosen in the previous step. We call this procedure automated model selection (AMS).

\begin{figure}
    \centering
    \includegraphics[width=1\linewidth]{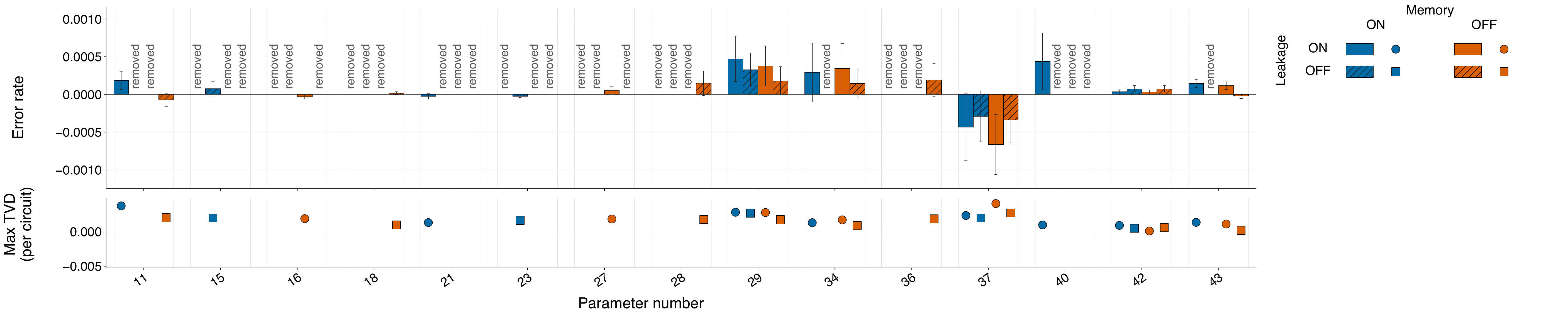}
    \caption{A complementary figure to Fig. \ref{fig:AMS} containing the remaining parameters of the reduced models found by AMS for the four simulated datasets from H1-1E. Some parameters were removed from all datasets and omitted. The labels correspond to the index of the FOGI parameters found in Tables~\ref{tab:FOGI-error-generator-combinations-short} and \ref{tab:FOGI-error-generator-combinations-long}.}
    \label{fig:AMS_remaining_params}
\end{figure}

The models returned by the AMS procedure removed more than half of the parameters on all four datasets. 13 of the parameters were kept, i.e. considered statistically significant, in all datasets, while 15 of them were removed from all datasets.
In addition, we compute the ``max TVD loss" of every parameter kept in the resulting models as an alternative heuristic measure of the explanatory loss associated with the removal of a parameter. First, we calculate the TVD between the AMS model predictions and the empirical frequencies for the dataset for every circuit. Then, we create a further reduced model by removing the parameter and question, run GST analysis on it to fit the data, and calculate the TVD between the predicted outcome probabilities and empirical frequences for every circuit. The difference between the two TVD values is the ``TVD loss" per circuit, it quantifies how much TVD each circuit would lose if the corresponding parameter was not included in the model. For every parameter we isolate the circuit with the maximum TVD loss across all four simulated datasets and plot the top 13 in Fig. \ref{fig:AMS} in the main text. Here, in Fig. \ref{fig:AMS_remaining_params}, we present the remaining parameters that were not removed for each of the four datasets.

As previously mentioned, some FOGI parameters are intrinsic to a single gate. This implies the error can be isolated to a single gate regardless of gauge choice. Intrinsic FOGI parameters are simple to write down, and quite amenable to interpretation. Luckily, the most important parameters -- in terms of TVD loss-- for the simulated datasets are all intrinsic. These fall into one of four categories:
\begin{enumerate}
    \item \emph{Overrotation} quantifies coherent rotations aligned with the gate's target axis. 
    \item \emph{Generalized bitflip} quantifies the symmetric mixing of the target gate's eigenstates. For example, in the case of a $Z$ gate it quantifies the mixing of the $\ket{0}$ and $\ket{1}$ state.
    \item \emph{Dephasing} quantifies the decoherence of a state along its eigenbasis. For example, in the case of a $Z$ gate it quantifies the shrinking of the transverse plane of the Bloch sphere perpendiculat to its eigenstates i.e. the equator of the Bloch sphere.
    \item \emph{Bitflip anisotropy} determines the \emph{shape} of the dephasing in the transverse plane perpendicular to the target gate's eigenstates. For example, in the case of a $Z$ gate it determines if the eigenstates of the $X$ and the $Y$ gate are equally mixed (symmetric along the equator), or otherwise (squeezed along the $X$ or $Y$ axis).
\end{enumerate}
For more details about FOGI parameters and their interpretation, including figures to help visualize some of their effects on the Bloch sphere, see Ref. \cite{FOGI}.

\section{Emergent logical coherent error on the $T_L$ gate}\label{sec: appendix Kevin's argument}
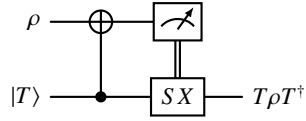
\begin{figure}[h]
    \begin{quantikz}[]
        \lstick{$\rho$} & \targ{} & \meter{}\wire[d][1]{c}  \\
        \lstick{$\ket{T}$} & \ctrl{-1} & \gate{SX} & \rstick{$T\rho T^{\dagger}$}
    \end{quantikz}
    \caption{Logical teleportation gadget for performing a $T$ gate.}
    \label{fig:T-gate-teleportation}
\end{figure}

In this appendix we provide an analytic argument for how a logical-level measurement error in the $\ket{T}_L$ state injection gadget can  produce a coherent error component in the $T_L$ gate.
Let $\rho_L$ be the logical state before the gadget. As shown in Fig.~\ref{fig:T-gate-teleportation},
the gadget involves a logical CNOT between an ancilla code block prepared in the $\ket{T}_L$ state and the data code block,
followed by a logical measurement on the data code block and a conditional $S_L\cdot X_L$ gate. A logical error in the decoded measurement outcome leads to either an $S_L\cdot X_L$ or $X_L\cdot S_L^{\dagger}$ error,
depending on whether the conditional gate is erroneously applied or not applied.
As an error channel, suppressing the logical subscript, this is given by
\begin{equation}
\rho \mapsto(1-p)\rho+\frac{p}{2}SX\rho XS^{\dagger} + \frac{p}{2}XS^{\dagger}\rho SX,
\end{equation}
where $p$ is the probability of an error in the decoded measurement.
Writing $S=\frac{1}{\sqrt{2}}(I-iZ)$ up to an overall phase, we have
$SX=\frac{1}{\sqrt{2}}(X+Y)$, and therefore
\begin{align}
\rho&\mapsto (1-p)\rho + \frac{p}{2}(X+Y)\rho(X+Y)\notag\\
&= (1-p)\rho + \frac{p}{2}(X\rho X + Y\rho Y ) +  \frac{p}{2}( X\rho Y +Y\rho X)\notag\\
&= (1-p)\rho + \frac{p}{2}(X\rho X + Y\rho Y ) +  \frac{p}{2}( iX\rho XZ -iZX\rho X)\notag\\
&= (1-p)\rho + \frac{p}{2}(X\rho X + Y\rho Y) -\frac{p}{2}i[Z, X\rho X].
\end{align}
The commutator term to lowest order in $p$ matches the action of first applying $X$ and then applying a $Z$ rotation by angle $p$.

The logical measurement includes decoding that corrects any weight-1 $X$ error on the data qubits, based on the values of the $Z$-type stabilizers.
If a weight-2 or higher $X$-type error on the data qubit has occurred, this may result in an error in the decoded measurement outcome. Such an error could result from a combination of errors in the $\ket{T}_L$ state prep, CNOT, and logical measurement.

\section{Syndrome statistics}\label{app:syndrome_statistics}

Syndrome data carries rich information about the underlying physical error mechanisms experiences within logical-qubit gadgets \cite{zheng2026efficient, xiao2026insitu, wagner2023learning}. In this appendix, we discuss our analysis of syndrome data from our logical GST experiment.

The structure of the syndrome extraction cycle in our experiment has implications for our analysis. We used active error correction during our experiments, implying that after each round of decoding (which occurs immediately after syndrome extraction), the state of the system is such that all syndromes should be 0. As a result, each syndrome measurement is itself a detector \cite{Derks2025designing}---i.e., each syndrome measurement is definite outcome in the absence of errors. The conditional structure of our syndrome extraction protocol cannot be captured in a detector error model, and our limited experiment data makes reliably estimating a series of conditional detector error models to describe the syndrome extraction cycle \cite{blumekohout2025estimating, remm2026experimentallyinformeddecoding, arms2026estimating, takou2025estimating} infeasible. We therefore focus on aggregate syndrome statistics rather than a complete model of detection histories. 

\begin{figure}
    \centering
    \includegraphics{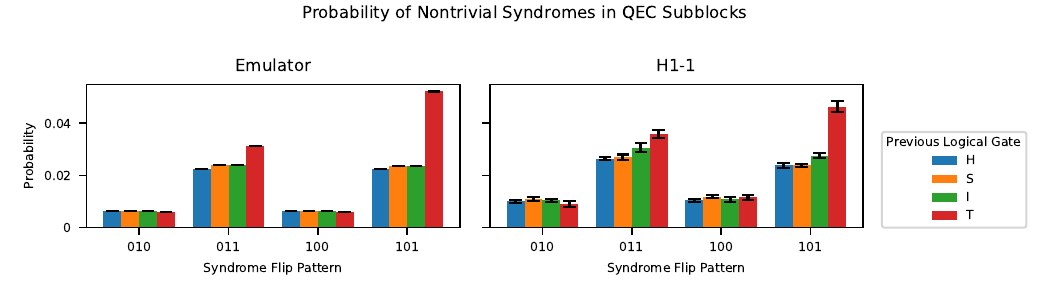}
    \caption{\textbf{Coarse-grained syndrome probabilities.} We coarse-grained syndrome data from each round of syndrome extraction into 3 bits $b_0b_1b_2$, where $b_0=1$ if the first block of flagged syndrome extraction yields at least one 1, $b_1=1$ iff the second block of flagged syndrome extraction yields at least one 1, and $b_2=1$ if unflagged syndrome extraction yields at least one 1. Rounds of syndrome extraction immediately after a $T_L$ gate see the coarse-grained syndromes $101$ and $011$ (patterns indicative of data-qubit errors) with a much higher probability than the Clifford gates, whereas the probability of $010$ and $101$ (patterns indicative of measurement errors) occur with approximately the same probability for all gates. In the H1-1 experiment, we observed higher probabilities of all nontrivial syndrome flip patterns than predicted by the emulator (with memory error enabled and leakage error disabled), with the largest difference being in the $101$ pattern.}
    \label{fig:coarse_syndromes}
\end{figure}

We first look at how syndromes vary dependent on the logical gate implemented immediately prior. The $T_L$ gate exhibits higher rates of nontrivial rounds of syndrome extraction than the $I_L$, $S_L$, and $H_L$ gates (Table~\ref{tab:hardware_syndromes})---consistent with the higher logical infidelity and non-transversality of the gate. The higher fraction of non-trivial syndromes after $T_L$ gates can be attributed to errors that cause a nontrivial syndrome in both the flagged and unflagged portions of syndrome extraction (Fig.~\ref{fig:coarse_syndromes}), which we call \emph{data-like errors} (because this pattern is typical of single-data-qubit errors). All gates have approximately the same rate of syndrome patterns with non-trivial syndromes only during flagged syndrome extraction, which we call \emph{measurement-like errors}. The T gate has a significantly higher rate of data-like errors than the three Clifford gates---in particular, we find that 101 occurs more than $2\times$ as often after a $T_L$ gate than a Clifford gate, and 011 occurs approximately $1.3\times$ more often after a $T_L$ gate than after a Clifford gate. 

\begin{figure}
    \centering
    \includegraphics{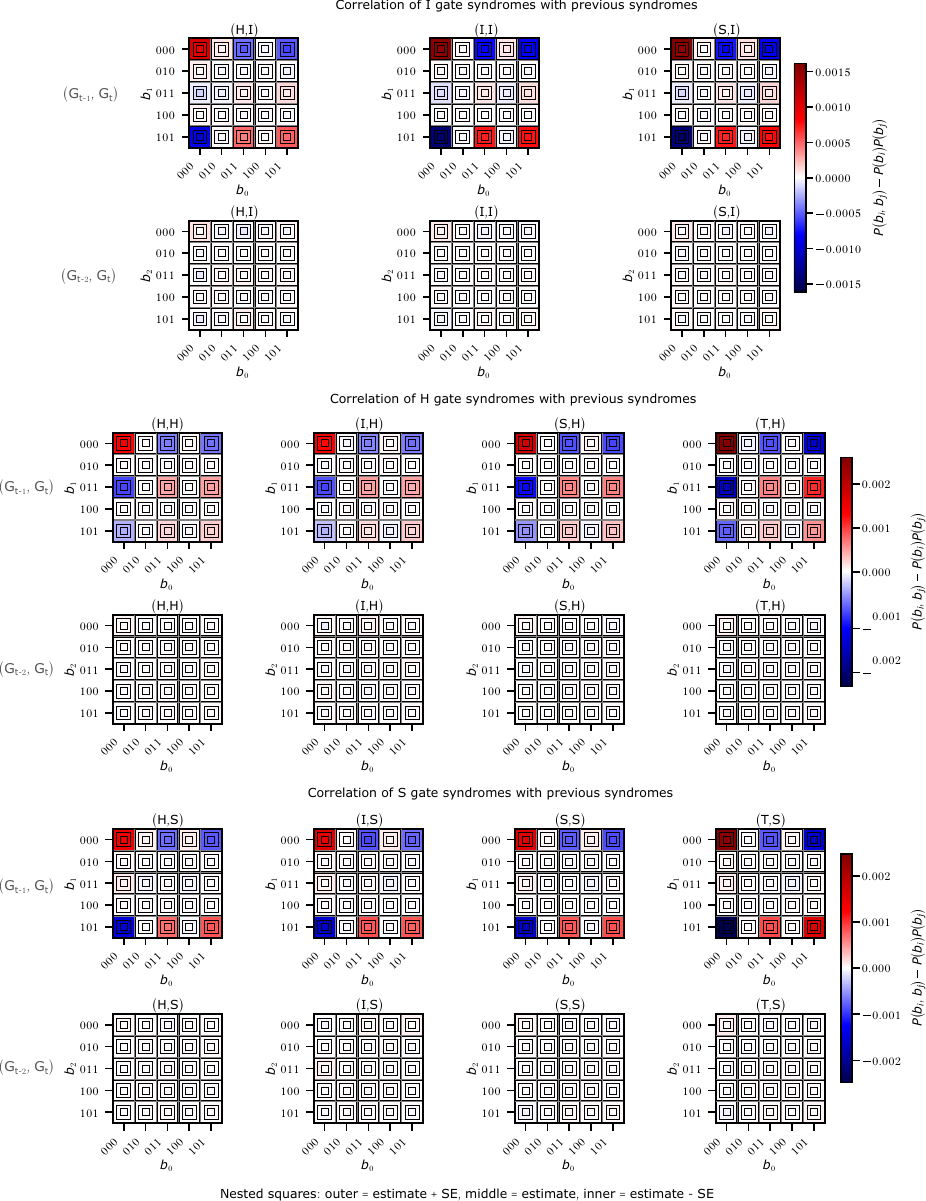}
    \caption{\textbf{Correlations in syndrome data across rounds.} Using emulated GST data (with memory error enabled and leakage error disabled), we looked at correlations between 3-bit coarse-grained syndromes across rounds of syndrome extraction. For gate pairs $(G_{t-1}, G_t)$ and $(G_{t-2},G_t)$, we estimate the probability distributions over coarse-grained syndromes from our emulator data and compute $P(b_{t-1}, b_t) - P(b_{t-1})P(b_t)$ (top row) and $P(b_{t-2}, b_t) - P(b_{t-2})P(b_t)$ (bottom row) for each gate $G_t$. For all gates, we see significant correlation between syndromes $b_t$ and the previous syndromes $b_{t-1}$, but not between $b_t$ and $b_{t-2}$. }
    \label{fig:correlation_emulator}
\end{figure}

\begin{figure}
    \centering
    \includegraphics{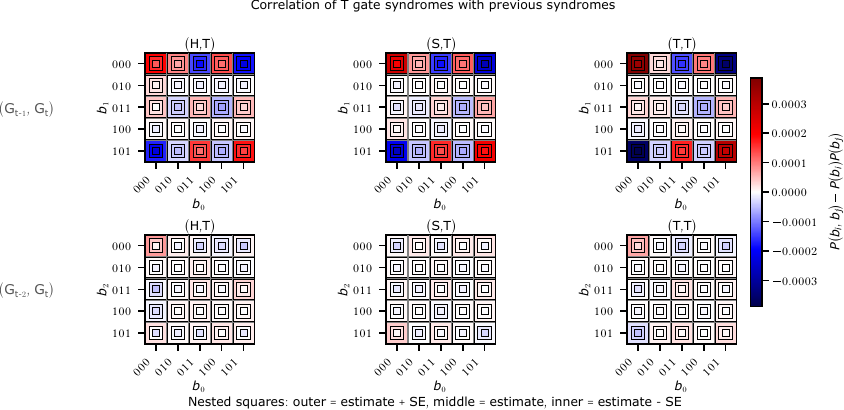}
    \caption{\textbf{Correlations in syndrome data across rounds for $T_L$ gates.} Correlations between 3-bit coarse-grained syndromes across rounds of syndrome extraction in emulated data for $G_t = T_L$, with memory error enabled and leakage error disabled. Correlation strength is roughly 4 times smaller for $G_t = T_L$ than for $G_t = S_L, H_L, I_L$. }
    \label{fig:correlation_emulator_t}
\end{figure}

While detectors in our experiment are contained within individual rounds of syndrome extraction, it is still possible for detectors to be correlated across time. Using the H1-1E simulated data, where we have a large number of samples available, we explored correlations between syndromes in local time windows. We focus on the dataset with memory enabled and leakage disabled, because this H1-1E configuration most likely matches the experiment (which inclues leakage repump) closest. Figs.~\ref{fig:correlation_emulator}-~\ref{fig:correlation_emulator_t} compares the probability of coarse-grained syndromes conditioned on the logical gates performed previously, where we have excluded pairs that do not occur together in the GST germs ($I_L$ and $T_L$). In this simulated data, we found that seeing a data-error-like syndrome pattern after an $H_L$ or $S_L$ gate is significantly correlated with the probability of having seen a data-error-like syndrome pattern in the round of syndrome extraction prior to the gate. However, we did not see significant correlation between syndromes in rounds $t$ and $t+2$. These results suggest that while our circuits have detectors contained within a single round of syndrome extraction, syndromes exhibit context-dependence that spans two rounds.

\begin{figure}
    \centering
    \includegraphics[width=1\linewidth]{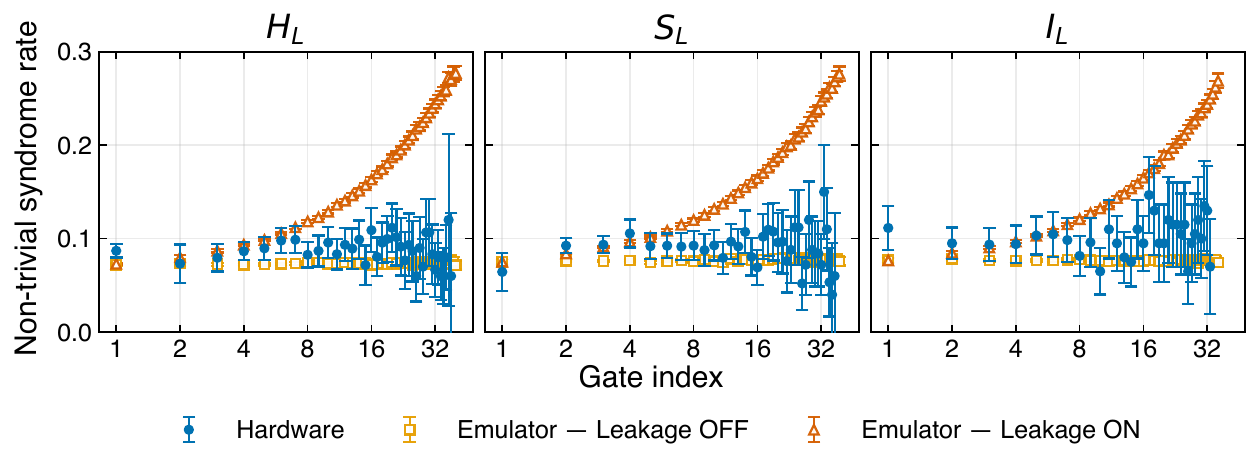}
    \caption{\textbf{Signatures of leakage in syndrome data.} The probability of non-trivial syndromes in circuits consisting of a single repeated logical Clifford gate. In simulations with leakage and without leakage removal, the probability of non-trivial syndromes increases dramatically with time within a circuit due to leakage buildup. In simulations without leakage, the rate of non-trivial syndromes remains approximately constant in time. In each dataset, memory error is enabled. Performing this analysis on our H1-1 experiment data, we see no evidence of an increase in non-trivial syndromes with time within these circuits.}
    \label{fig:leakage}
\end{figure}

Syndrome data can also exhibit signatures of leakage. Persistent leakage causes the probability of observing non-trivial syndromes to increase with time within each shot. Figure~\ref{fig:leakage} shows this effect in an emulated logical GST experiment without leakage removal and the probability of non-trivial syndromes in our experiment on H1-1. We do not see a statistically significant increase in the non-trivial syndrome extraction rounds in the H1-1 experiment data, as expected since leakage repump was enabled. However, we note that the limited number of shots severely limits the leakage rates we can reliably detect from the data. 

\begin{table}[htbp]
\centering
\caption{Number of syndromes and percentage of non-trivial syndromes for each gate using H1-1 hardware data.}
\label{tab:hardware_syndromes}
\small
\begin{tabular}{lcc}
\hline
\textbf{Gate}
& \shortstack{\textbf{Total number of}\\\textbf{syndromes}}
& \shortstack{\textbf{Percentage}\\\textbf{non-trivial}} \\
\hline
Hadamard (H) &  $34265$ & $(8.9 \pm 0.3)\%$ \\
Phase (S)   & $31455$ & $(9.2\pm 0.3)\%$ \\
Idle (I)    & $13400$  &  $(9.9\pm0.5)\%$ \\
T           & $7832$ & $(12.8 \pm 0.7)\%$ \\
\hline
\end{tabular}
\end{table}

\end{document}